\documentclass{jfm}

\usepackage{float} 
\usepackage{placeins}
\usepackage{graphicx}
\usepackage{gensymb} 
\usepackage{newtxtext}
\usepackage{newtxmath}
\usepackage{natbib,amsmath,amssymb}
\usepackage{relsize} 
\usepackage{hyperref}
\usepackage{makecell} 
\hypersetup{
    colorlinks = true,
    allcolors  = blue,
    urlcolor    = blue,
    citecolor  = blue,
}
\usepackage[absolute,overlay]{textpos}
\newcommand\Ec{\mbox{\textit{Ec}}}
\newcommand\Ma{\mbox{\textit{Ma}}}

\newcommand{\RomanNumeralCaps}[1]
\linenumbers

\usepackage[dvipsnames]{xcolor}
\definecolor{myred}{RGB}{162,22,52}       

\newcommand\B{$[b]$}
\newcommand\E{[\mbox{{EoS}}]}
\newcommand\V{$[\mu]$}
\newcommand\T{$[\kappa]$}
\renewcommand\spadesuit{[b]}
\renewcommand\clubsuit{[\mbox{EoS}]}
\renewcommand\diamondsuit{[\mu]}
\renewcommand\heartsuit{[\kappa]}

\title{Sensitivity of three-dimensional boundary-layer stability to intrinsic uncertainties of fluid properties: a study on supercritical CO2}
\author{
	  Jie Ren\aff{1,2,3} \corresp{\email{jie.ren@bit.edu.cn}}, 
	  Yongxiang Wu\aff{2},
	  Xuerui Mao\aff{1,3,4}, 
	  Cheng Wang\aff{1,3},
	  \vspace{1mm}
	  \and 
  	  Markus Kloker\aff{2} 
	   }
\affiliation{
\aff{1} State Key Laboratory of Explosion Science and Safety Protection, Beijing Institute of Technology, Beijing 100081, China
\aff{2} Institute of Aerodynamics and Gas Dynamics, University of Stuttgart, Pfaffenwaldring 21, D-70569 Stuttgart, Germany
\aff{3} Yangtze Delta Region Academy of Beijing Institute of Technology, Jiaxing 314003, China
\aff{4} Advanced Research Institute of Multidisciplinary Sciences, Beijing Institute of Technology, Beijing 100081, China}

\graphicspath{{Figures/}}
\begin{document}

\begin{textblock}{0.4}(0.45,0.14) 
\centering
\textblockcolour{green!20} 
\vspace{1mm}
Accepted by \emph{J. Fluid Mech.} \\ \vspace{1mm}doi: \textrm{10.1017/jfm.2025.100}
\vspace{1mm}
\end{textblock}

\maketitle
\begin{abstract}
The intrinsic uncertainty of fluid properties, including the equation of state, viscosity, and thermal conductivity, on boundary layer stability has scarcely been addressed. When a fluid is operating in the vicinity of the Widom line (defined as the maximum of isobaric specific heat) in supercritical state, its properties exhibit highly non-ideal behavior, which is an ongoing research field leading to refined and more accurate fluid property databases. Upon crossing the Widom line, new mechanisms of flow instability emerge, feasibly leading to changes in dominating modes that yield turbulence. The present work investigates the sensitivity of three-dimensional boundary-layer modal instability to these intrinsic uncertainties in fluid properties. The uncertainty, regardless of its source and the fluid regimes, gives rise to distortions of all profiles that constitute the inputs of the stability operator. The effect of these distortions on flow stability is measured by sensitivity coefficients, which are formulated with the adjoint operator and validated against linear modal stability analysis. The results are presented for carbon dioxide at a representative supercritical pressure of about 80 bar. The sensitivity to different inputs of the stability operator across various thermodynamic regimes show an immense range of sensitivity amplitude. A balancing relationship between the density gradient and its perturbation leads to a quadratic effect across the Widom line, provoking significant sensitivity to distortions of the second derivative of the pressure with respect to the density, $\partial^2 p/\partial \rho^2$. From an application-oriented point of view, one important question is whether the correct baseflow profiles can be meaningfully analyzed by the simplified ideal-fluid model. The integrated modal disturbance growth — the N factor calculated with different partly idealized models — indicates that the answer depends strongly on the thermodynamic regime investigated.
\end{abstract}


\section{Introduction}\label{S1}

Non-ideal compressible fluids are increasingly used in various applications for improved efficiency and reduced pollution \citep{Guardone2024}. Examples include power generation \citep{white2021review}, heat exchangers \citep{chai2020review}, and fuel injections \citep{bellan2020high}. ``Non-ideal" describes fluids that do not conform to the ideal gas equation-of-state and exhibit unique phenomena such as pseudo-boiling \citep{banuti2015crossing}, heat transfer deterioration \citep{pizzarelli2018status} and nonclassical rarefaction shock waves \citep{alferez2017one}. These non-ideal characteristics pose significant challenges to the traditional ideal-gas framework used for predicting flows subject to distortions and laminar-turbulent transition \citep{li2024microscale}.

Besides the complexity of a non-ideal fluid, the difficulty of predicting flow transition is essentially due to its high sensitivity and the multi-fold path from laminar to turbulence \citep{reshotko2008transition}, which depends not only on the flow configuration but also the form and amplitude of external perturbations present in the environment. Consequently, the dominant mechanisms are varied. In the linear regime, well-known examples are Tollmien-Schlichting waves due to eigenmodal growth of instabilities, the second mode in hypersonic boundary layer flows \citep{mack1984boundary}, cross-flow waves resulting from a three-dimensional swept flow \citep{saric2003stability}, centrifugal instabilities due to the presence of concave surfaces \citep{saric1994gortler}, and streamwise velocity streaks following non-modal growth \citep{trefethen1993hydrodynamic,schmidstability}, among others.

In a typical linear stability analysis, the growth rate and dispersion relations are obtained for a pre-calculated laminar baseflow. However, actual flows are inevitably affected by numerous extraneous factors that are not thoroughly accounted for by the theoretical model. Examples include freestream turbulence \citep{hunt1978free}, particles \citep{browne2021numerical}, noise \citep{schneider2001effects}, and leading-edge contamination \citep{spalart1989direct}, to name a few. To connect to realistic configurations, a key question is how robust the analytical results are and to what extent the growth rate will change when certain distortions are present. Additionally, determining the appropriate distortion that leads to desired transition promotion or delay is crucial for controling purposes.

The above requirement aligns with the operator perturbation theory \citep{kato2013perturbation}, a well-developed field in mathematics \citep{bottaro2003effect}. In the context of flow instability, seminal works were performed by \citet{pralits2000sensitivity}, \citet{bottaro2003effect}, \citet{marquet2008sensitivity}, \citet{bagheri2009input} and \citet{brandt2011effect} for parabolised stability equations, local \& global modal stability analyses, feedback control design, and non-modal growth respectively. By adopting the adjoint equations \citep[see reviews by][]{luchini2014adjoint}, a measure of the system's response to input variations is formulated. The computed sensitivity field indicates the regions where flow distortions most effectively modify the growth rate, thereby pointing to optimal control strategies.

Recent research on sensitivity analysis has extended to account for high-speed boundary layer flows. \citet{park2019sensitivity} investigated a Mach 4.5 flat-plate boundary layer, focusing on the sensitivity properties of the fast and slow modes, whose synchronization gives rise to Mack's second mode \citep{fedorov2011high}. \citet{guo2021sensitivity} recognized two routes for sensitivity: one where distortion influences the baseflow, leading to variations in the linear stability operator, and another where the stability changes directly. \citet{chen2024non} found that for an inclined blunt cone, the structural sensitive region is located on the windward side, just downstream of the inlet. \citet{Poulain_Content_Rigas_Garnier_Sipp_2024} formulated the sensitivity based on global instability and resolvent analysis. They identified the optimal locations for steady wall blow/suction and heating/cooling, some of which were shown to successfully damp Mack's first/second modes and boundary-layer streaks simultaneously.

In relation to the current study, \citet{Brynjell-Rahkola_Shahriari_Schlatter_Hanifi_Henningson_2017} conducted a meaningful analysis on the sensitivity of a 3-D Falkner-Skan-Cooke boundary-layer flow to numerical details. Despite significant knowledge gained regarding the sensitivity of boundary-layer stability, studies so far have been mostly limited to ideal gases. The idea here aligns with \citet{juniper2018sensitivity}, who emphasized that ``\emph{the systematic approach in adjoint methods requires an accurate thermoacoustic model}.''  Recent efforts on supercritical fluids have discovered new inviscid instabilities \citep[see][for a short review]{robinet2019instabilities} occurring during pseudo-boiling, where the fluid shifts from liquid-like to gas-like behavior \citep{simeoni2010widom}. Pseudo-boiling represents significant non-ideal thermodynamic regions of a supercritical fluid where the phase change vanishes and is replaced by substantial mutations in thermodynamic and transport properties. These highly non-ideal regions are recognized by the Widom line \citep{banuti2015crossing}, typically defined as the maximum of the isobaric-specific heat ($C_p$).

Linear stability analyses on canonical flows of supercritical fluids explored so far have demonstrated commonalities --- the presence of a new inviscid instability that dominates. For example, the binary mixing layer was shown to be destabilized by a new thermodynamically induced instability \citep{Nguyen2022}, which can affect the performance of supercritical fuel injection systems. Plane Poiseuille and Couette flows can both become inviscidly unstable upon crossing the Widom line \citep{ren2019linear, Bugeat_Boldini_Hasan_Pecnik_2024}. Under similar conditions, two-dimensional (2D) boundary layers are subject to dual-mode instability, where the flow is dominated by new inviscid instability in addition to the conventional viscous Tollmien-Schlichting waves \citep{ren2019boundary}. Moreover, in three-dimensional boundary layers of accelerating flows with wall cooling, the dominating cross-flow (CF) modes are replaced by the 2D inviscid mode, which has a growth rate significantly more prominent than that of the CF modes, despite the strong favorable pressure gradient \citep{Ren2022}.


The above correlative phenomena have motivated recent efforts to develop novel solvers  and to understand the fundamental mechanisms.  To the best of the authors' knowledge, \citet{boldini2025cubens} has recently introduced the first open-source high-order solver `CUBENS' for single-phase non-ideal fluids in canonical geometries.\citet{Bugeat_Boldini_Hasan_Pecnik_2024} showed that for stratified plane Couette flow, a minimum in the kinematic viscosity of the baseflow profile produces a generalized inflection point that fulfills Fjørtoft's generalized inviscid instability criterion \citep{fjvarphirtoft1950application}. They further extended Rayleigh's criterion \citep{rayleigh1880stability} to stratified flows, demonstrating that the excess of density-weighted vorticity (attributed to shear and inertial baroclinic effects) relative to its spatial thickness is responsible for the inviscid instability. Specifically, different fluid models featuring an extremum of the kinematic viscosity were devised to verify the generalized stability model.

In contrast to Couette flow, boundary-layer flows are non-parallel, requiring further theoretical work to better understand the dual-mode instability \citep{bugeat2022new}. By analyzing the boundary-layer equation, \citet{Ren2022} concluded that in a three-dimensional boundary layer, the tremendous negative near-wall viscosity gradient $\partial\mu/\partial y$ is responsible for the inflectional shape of the streamwise velocity profile. This viscosity gradient, upon wall cooling, can be mathematically expressed as
\begin{equation}
\left.\frac{\partial\mu}{\partial y}\right|_{\left(-\right)}=\left.\frac{\partial\mu}{\partial T}\right|_{\left(+\right)}\left.\frac{\partial T}{\partial y}\right|_{\left(+\right)}+\left.\frac{\partial\mu}{\partial\rho}\right|_{\left(+\right)}\left.\frac{\partial\rho}{\partial y}\right|_{\left(-,\:\mathrm{dominating}\right).}
\end{equation}
Along with the increase of $\partial \rho/\partial y$ in the pseudo-boiling regime, an inflection point is established (note that the first term on the right-hand side is positive for a gas or gas-like fluid, cf. figure \ref{fig2}). Meanwhile, the cross-flow component $w_s$, responsible for the CF instability, is influenced by the boundary layer's density distributions through the balancing relationship of the centripetal and centrifugal forces on a curved streamline.

A rational equation of state (EOS) and laws of transport properties are essential  for both the baseflow and the stability operator, to capture the behaviour correctly. Numerically, look-up tables are used to obtain non-ideal fluid properties during the integration of flow equations.  Widely used databases include the NIST (National Institute of Standards and Technology) Reference Fluid Thermodynamic and Transport Properties Database (RefProp) \citep{huber2022nist} and the open-source library CoolProp \citep{doi:10.1021/ie4033999}. 
In both Refprop and Coolprop, the Helmholtz energy form for a fundamental EOS is used:
\begin{equation}\label{eq1}
\alpha(\tau, \delta) = \alpha^{\text{id}} + \alpha^{\text{r}} 
= \alpha^{\text{id}} + \sum_{k} N_k \delta^{d_k} \tau^{t_k} + \sum_{k} N_k \delta^{d_k} \tau^{t_k} \exp(-\delta^{l_k}).
\end{equation}
where $\tau = T/T_c$ and $\delta = \rho/\rho_c$ are the reduced temperature and density (by critical values), $\alpha$ is the reduced molar Helmholtz energy, $\alpha^{\text{id}}$ is the ideal gas contribution, and $\alpha^{\text{r}}$ is the real-fluid contribution. $N_k$ is coefficients obtained by fitting experimental data, and the exponents $d_k$, $t_k$, and $l_k$ are also determined by regression. To recover the  state equation $p=f(\rho, T)$, the following thermodynamic relationship is used:
\begin{equation}\label{eq2}
\frac{p}{\rho RT} = 1 + \delta \left( \frac{\partial \alpha^{\text{r}}}{\partial \delta} \right)_{\tau}.
\end{equation}
However, it is important to note that to better represent properties in the critical region, additional terms are necessary in equation \eqref{eq1}, which can make the expression extremely complex (sometimes with more than 50 terms). Due to the empirical nature of the equation, RefProp and CoolProp retain the equation of state in an implicit form, but one that closely aligns with physical values—achieving uncertainties that approach the level of the underlying experimental data, in line with the objectives of such libraries. Therefore, the constraints on EOS imposed by thermodynamic stability \citep[see Section II.C of][]{menikoff1989riemann} shall be satisfied.  
Accurately describing transport properties, especially in non-ideal regimes, is a challenging and ongoing task that depends heavily on precise experimental measurements. These libraries incorporate various empirical and theoretical (fluid-specific) models for transport properties \citep[see][for an example of thermal conductivity in CO2]{huber2016reference}. 

\begin{figure}
\centerline{
\includegraphics[scale=0.35]{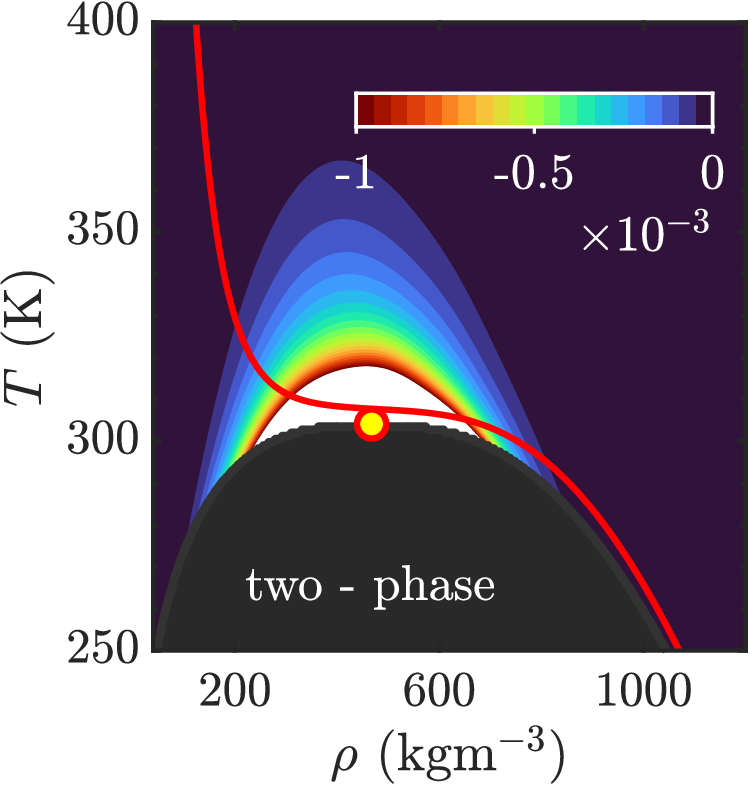}\hspace{2mm}
\includegraphics[scale=0.35]{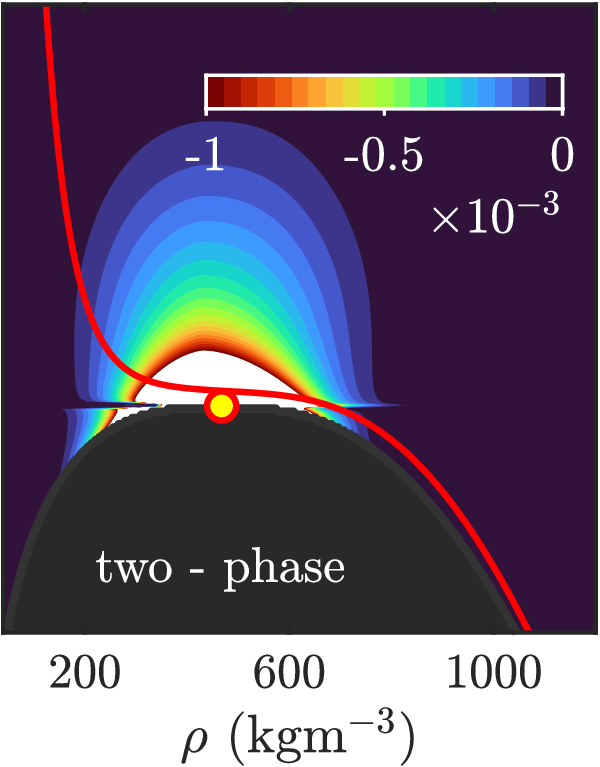}\hspace{2mm}
\includegraphics[scale=0.35]{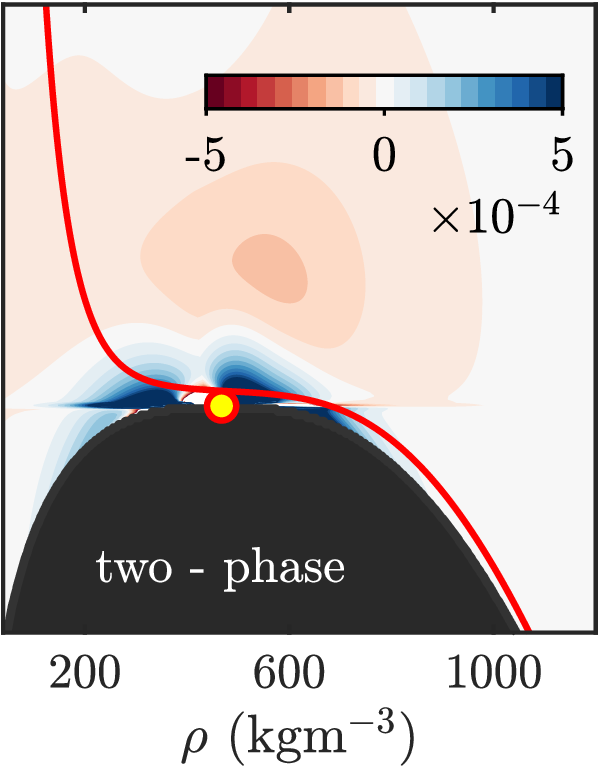}
\put(-308,120) {\textcolor{white}{({\it a})}}
\put(-199,120) {\textcolor{white}{({\it b})}}
\put(-090,120) {\textcolor{black}{({\it c})}}
}
\caption{Comparison of the fluid property $\partial \kappa/\partial T$ (in dimension Wm$^{-1}$K$^{-2}$) for carbon dioxide. The values are generated using RefProp in panel (a) and CoolProp in panel (b). Differences between the two are highlighted in panel (c). The red circle marks the critical point, and the red line represents the isobar at 80.}
\label{fig_table}
\end{figure}

Figure \ref{fig_table} compares the gradient of thermal conductivity with respect to temperature, one of the inputs for the flow stability operator. Data were generated using RefProp (version 8.0.4) and CoolProp (version 6.4.1), respectively. As shown, panel (b) suffers from model imperfections and non-smooth behavior near the critical temperature. Even far from the critical point, the values differ significantly between the two databases. The overall difference (see panel c) is generally of a similar order as the values, reflecting considerable uncertainties. \citet{ren2022non} demonstrates that even slight adjustments to the stability operator can substantially influence modal growth, emphasizing the importance of considering non-ideal gas behaviors. However, the root causes of this sensitivity variation are still unexplored, pointing to the need for a formalized sensitivity framework to gain deeper understanding. To date, the sensitivity characteristics of boundary-layer stability in relation to these fluid properties remain unclear.

We aim to characterize and quantify the behaviour by examining the sensitivity of linear stability to each of the inputs of the stability operator. This study will highlight the importance of accurately including the thermodynamic and transport properties of the non-ideal fluid, which are often missed in conventional hydrodynamic stability theory when a temperature gradient crosses the Widom line. Additionally, the possibility of simplifying the fluid model in other regimes will be discussed. The organization of the paper is as follows: Section \ref{S2} defines the problem, clarifies the coupling between fluid properties, the laminar baseflow, and the linear instability, followed by the derivation of sensitivity profiles; Section \ref{S3} discusses the results for sensitivity; and Section \ref{S4} presents the conclusions.




\section{Problem definition and sensitivity}\label{S2}
\subsection{Governing equations and flow conditions}\label{S2a}
The flow satisfies the conservation laws for mass, momentum and energy for a generic fluid (Navier-Stokes equations). In cartesian coordinates and dimensionless form this reads 
\begin{subeqnarray}\label{NS1}
\frac{\partial\rho}{\partial t}+\frac{\partial\left(\rho u_{j}\right)}{\partial x_{j}}	&=& 0, \\[0pt]
\frac{\partial\left(\rho u_{i}\right)}{\partial t}+\frac{\partial\left(\rho u_{i}u_{j}-\sigma_{ij}\right)}{\partial x_{j}}&=&0, \\[0pt]
\frac{\partial\left(\rho e\right)}{\partial t}+\frac{\partial\left(\rho eu_{j}+q_{j}\right)}{\partial x_{j}}-\sigma_{ij}\frac{\partial u_{i}}{\partial x_{j}}	&=&0.
\end{subeqnarray}
Here the subscripts $i, j$ denote vector/matrix components in a three-dimensional space and follow Einstein's summation rule. With this denotation, $(x_1,x_2,x_3) =(x,y,z)$ are coordinated along the streamwise, wall-normal, and spanwise directions. Alike, $(u_1,u_2,u_3) =(u,v,w)$ stand for velocity components along $(x,y,z)$.  The stress tensor $\sigma_{ij}$ and heat flux $q_j$ are given by
\begin{subeqnarray}\label{NS2}
\sigma_{ij}&=&\frac{\mu}{Re}\left(\frac{\partial u_{i}}{\partial x_{j}}+\frac{\partial u_{j}}{\partial x_{i}}\right)+\frac{\lambda}{Re}\delta_{ij}\frac{\partial u_{k}}{\partial x_{k}}-p\delta_{ij},\\
q_{j}&=&-\frac{\kappa}{RePrEc}\frac{\partial T}{\partial x_{j}},
\end{subeqnarray}
where $\Rey$, $\Pran$ and $\Ec$ are the Reynolds, Prandtl and Eckert numbers, $\delta_{ij}$ stands for the Kronecker delta, $\mu$ for viscosity, $\kappa$ for thermal conductivity, $p$ for pressure, and $\lambda$ is the Lame's constant (set to $-2\mu/3$ in this study). The equations \eqref{NS1} and \eqref{NS2} are closed with the relations for thermodynamic and transport properties. For a genetical fluid of pure substance, these relations are written as binary functions of density and temperature,
\begin{equation}
\left[p,e,\mu,\kappa\right]=\left[p,e,\mu,\kappa\right]\left(\rho,T\right).
\end{equation}

\begin{figure}
\centerline{
\includegraphics[scale=0.35]{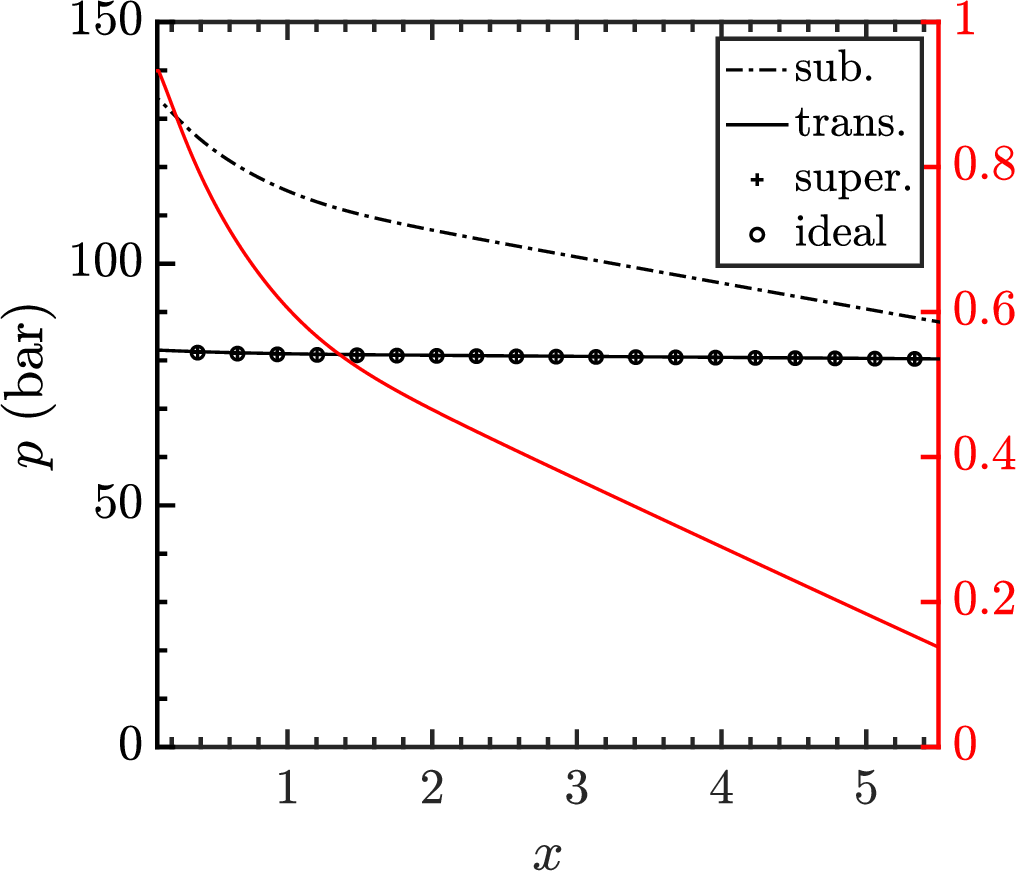}
\hspace{3mm}
\includegraphics[scale=0.35]{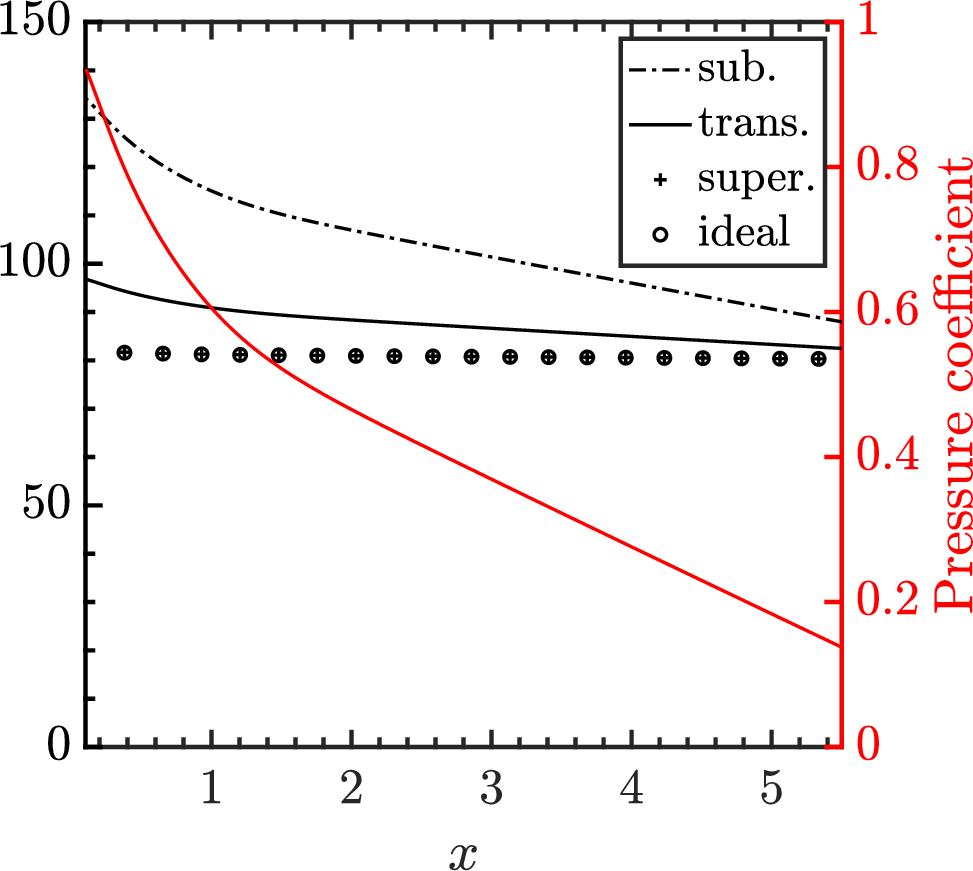}
\put(-316,132){({\it a})}
\put(-142,132){({\it b})}
\put(-40,48) {\textcolor{red}{$\longrightarrow$}}
\put(-210,48) {\textcolor{red}{$\longrightarrow$}}
\vspace{5mm}
}
\centerline{
\includegraphics[scale=0.35]{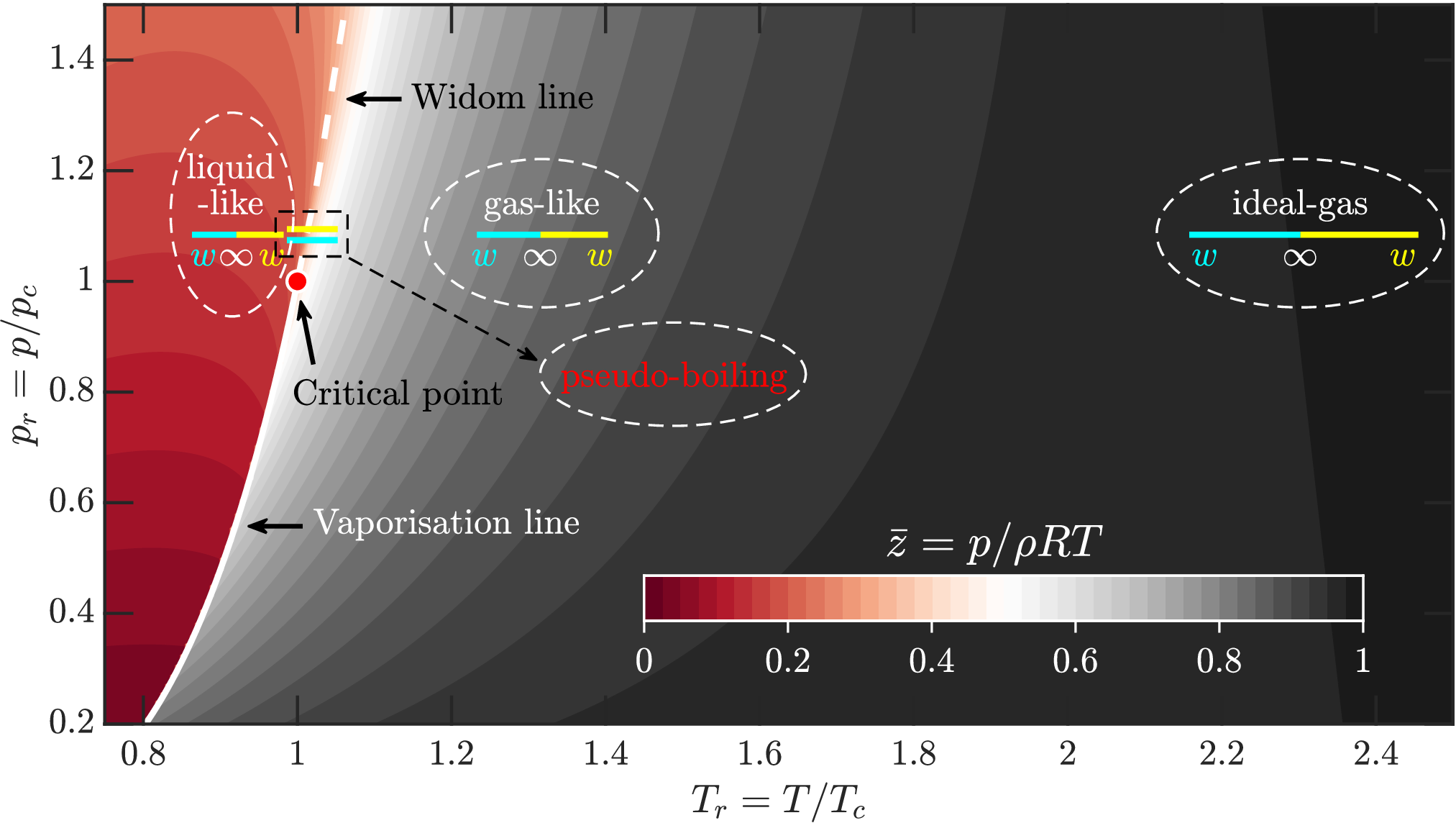}
\put(-310,180) {\textcolor{white}{({\it c})}}
}
\caption{Distribution of pressure (left axis) and pressure coefficient (right axis) for wall cooling (a) and heating cases (b). Panel (c) provides pressure$-$temperature ($P-T$) diagram of carbon dioxide. The contours represent the compressibility factor $\bar{z}=p/(\rho RT)$. Above the critical point (red dot), the Widom line is plotted using a white dashed line. The fluid regimes are considered along the isobar of 80 ($p/p_c=1.0844$). Four groups of cases are shown with yellow (wall-heating) and cyan (wall-cooling) lines. These lines characterize the distribution of flow temperature, with wall and freestream values denoted by $w$ and $\infty$, respectively. }
\label{fig1}
\end{figure}

We consider three-dimensional laminar boundary layers with a favorable pressure gradient. The pressure coefficient distribution matches the redesigned DLR experiment \citep{barth2018redesigned} on CF instability, leading to an established $p(x)$ as shown in Figure~\ref{fig1}(a,b). In all cases, the static pressure $p_\infty$ is fixed at 80 bar, and the pressure thus follows Bernoulli's relation. We prescribe various freestream and wall temperatures $(T_\infty, T_w)$ to explore the physics of four representative regimes: liquid-like, pseudo-boiling, gas-like, and ideal-gas, as shown in Figure~\ref{fig1}(c). Specifically, relative to the pseudo-critical (pseudo-boiling) temperature $T_{pc}=307.7$ K, the boundary temperatures are summarized in Table \ref{table1}. The temperature ratios are maintained at $T_w/T_\infty=16/15$ and $15/16$ for wall heating and cooling, respectively. An identical Reynolds number ($\Rey=\rho_\infty U_\infty L_{\rm ref}/\mu_\infty=1.4687\times10^5$) as in the experiment and a low Mach number ($\Ma=U_\infty/c_\infty=0.2$) have been used. The choice of these parameters ensures that the flow has exact comparability with typical cross-flow instabilities in the ideal gas regime, in which the same dimensionless baseflow and neutral curve is obtained \citep{dorr2017crossflow, Ren2022}.

\begin{table}
\begin{center}\def~{\hphantom{0}}
\begin{tabular}{c@{\hskip 5mm}c@{\hskip 5mm}c@{\hskip 5mm}c@{\hskip 5mm}c}
acronym & \makecell{relative to the\\ pseudo-critical point \\(Widom line)} & \makecell{fluid \\ characteristics}  & \makecell{$T_\infty,T_w$ \\ (wall heating) }& \makecell{$T_\infty,T_w$ \\ (wall cooling) }   \\ [5mm]
sub 		& subcritical	& liquid-like		& (280.0, 298.7)    K  & (280.0, 262.5) K  \\
trans 	& transcritical 	&pseudo-boiling 	& (300.0, 320.0)    K  & (320.0, 300.0) K  \\
super 	& supercritical  	&gas-like             	& (400.0, 426.7)    K  & (400.0, 375.0) K  \\
ideal 	& supercritical 	&ideal-gas            	& (700.0, 746.7)    K  & (700.0, 656.3) K   
\end{tabular}
\caption{A summary of flow cases investigated.}
\label{table1}                               
\end{center}
\end{table}

\subsection{The laminar baseflow}\label{S2b}

Matching the pressure coefficient distribution of the DLR experiments results in a non-self-similar laminar baseflow where the shape factor (ratio of displacement to momentum-loss thickness) is not constant. This laminar baseflow is obtained by solving the parabolized Navier-Stokes (PNS) equations. In steady and non-separating boundary-layer flows, the streamwise viscous gradient is notably smaller than the wall-normal component. Therefore, the PNS equations are derived from the full Navier-Stokes equations by neglecting the streamwise gradient in the viscous terms. The resulting equations in their 2-D form are given by
\begin{equation}\label{eq_PNS}
\mathsfbi{A}\dfrac{\partial\boldsymbol{Q}}{\partial 
x}+\mathsfbi{B}\dfrac{\partial\boldsymbol{Q}}{\partial y}=\textrm{RHS}.
\end{equation}
\begin{equation}\label{eq_PNS_A}
\mathsfbi{A}=\left(\begin{array}{ccccc}
u & \rho & 0 & 0 & 0\\
0 & \rho u & -\dfrac{1}{Re}\dfrac{\partial\mu}{\partial y} & 0 & 0\\
0 & -\dfrac{1}{Re}\dfrac{\partial\lambda}{\partial y} & \rho u & 0 & 0\\
0 & 0 & 0 & \rho u & 0\\
\rho u\dfrac{\partial e}{\partial\rho} & p & 0 & 0 & \rho u\dfrac{\partial 
e}{\partial T}
\end{array}\right),
\end{equation}
\begin{equation}\label{eq_PNS_B}
\mathsfbi{B}=\left(\begin{array}{ccccc}
v & 0 & \rho & 0 & 0\\
0 & b_{2,2} & 0 & 0 & 0\\
\dfrac{\partial p}{\partial\rho} & 0 & b_{3,3} & 0 & \dfrac{\partial p}{\partial 
T}\\
0 & 0 & 0 & b_{4,4} & 0\\
\rho v\dfrac{\partial e}{\partial\rho} & -\dfrac{\mu}{Re}\dfrac{\partial 
u}{\partial y} & p-\dfrac{2\mu+\lambda}{Re}\dfrac{\partial v}{\partial y} & 
-\dfrac{\mu}{Re}\dfrac{\partial w}{\partial y} & b_{5,5}
\end{array}\right),
\end{equation}
\begin{equation}\label{eq_PNS_B2}
\left.\begin{array}{ll}
b_{2,2} = & \rho v-\dfrac{1}{Re}\dfrac{\partial\mu}{\partial 
y}-\dfrac{\mu}{Re}D\\
b_{3,3} = & \rho 
v-\dfrac{1}{Re}\dfrac{\partial\left(2\mu+\lambda\right)}{\partial 
y}-\dfrac{2\mu+\lambda}{Re}D\\
b_{4,4} = & \rho v-\dfrac{1}{Re}\dfrac{\partial\mu}{\partial 
y}-\dfrac{\mu}{Re}D\\
b_{5,5} = & \rho v\dfrac{\partial e}{\partial 
T}-\dfrac{1}{RePrEc}\dfrac{\partial\kappa}{\partial y}-\dfrac{\kappa}{RePrEc}D,
\end{array}\right\} 
\end{equation}
\begin{equation}
\textrm{RHS}=(0,-dp/dx,0,0,0)^T. 
\end{equation}
The Prandtl ($\Pran={\mu_{\infty}C_{p\infty}}/{\kappa_\infty}$) and Eckert ($\Ec={u_{\infty}^{2}}/{C_{p\infty}T_\infty}$) numbers are not independent and can be calculated based on $\Rey$, $\Ma$, and the temperature conditions prescribed in Table \ref{table1}. The symbol $e$ denotes the internal energy. The operator $D$ in \eqref{eq_PNS_B2} stands for the wall-normal derivative. In numerically solving the PNS equations by an implicit Euler scheme, the system is linearized by ``lagging'' the coefficients $\mathsfbi{A}$ and $\mathsfbi{B}$ relative to the solution vector $\boldsymbol{Q}=\left(\rho,u,v,w,T\right)^{T}$ in an iterative procedure. Namely, sub-iterations are carried out to update $\mathsfbi{A}$ and $\mathsfbi{B}$ from $\boldsymbol{q}$ at each station of the streamwise marching to obtain the correct, fully non-linear values.
When external perturbations are not present, the boundary conditions are
\begin{subeqnarray}
	y=y_e:\;\frac{\partial u}{\partial y}=\frac{\partial w}{\partial y}=0,\;\rho=\rho_{e}\left(x\right),\;T=T_{e}\left(x\right);\\
	y=0:\;u=v=w=0,\;T=T_{w}.
\end{subeqnarray}
We employ the subscript $e$ to denote local boundary-layer edge values. At the upper edge of the boundary layer, both velocity components $u$ and $w$ are subject to Neumann conditions, while the gradient $\partial v/\partial y$ does not vanish due to the presence of streamwise pressure gradients. Thus, $v_e$ is calculated based on the continuity equation. The potential-flow values $\rho_{e}(x)$ and $T_{e}(x)$ are given by the isentropic relations (where $S$ stands for entropy):
\begin{equation}
S\left(\rho_{e}\left(x\right),p\left(x\right)\right)=S\left(\rho_{\infty},p_{\infty}\right),\;S\left(T_{e}\left(x\right),p\left(x\right)\right)=S\left(T_{\infty},p_{\infty}\right),
\end{equation} 
At the wall, no-slip velocity and a specified wall temperature apply (see Table \ref{table1}). The wall density is not prescribed; instead, it is allowed to vary to ensure that the pressure gradient at the wall is zero, thereby satisfying the momentum equation in the wall-normal direction. The PNS equations are integrated downstream, starting from an initial profile at $x=x_0$. In this study, we specify the streamwise and spanwise velocities $u(x_0,y)$ and $w(x_0,y)$ using the Falkner-Skan-Cooke (FSC) solution \citep{cooke1950boundary}, with $v(x_0,y)=0$. The thermodynamic variables ($\rho$, $T$) are either given as the potential-flow values (applying isentropic relations) or extrapolated from existing downstream data, ensuring that the influence of the initial profiles is insignificant.

\subsection{Linear instability and the sensitivity framework}\label{S2c}
\begin{figure}
\centering
\includegraphics[scale=0.26]{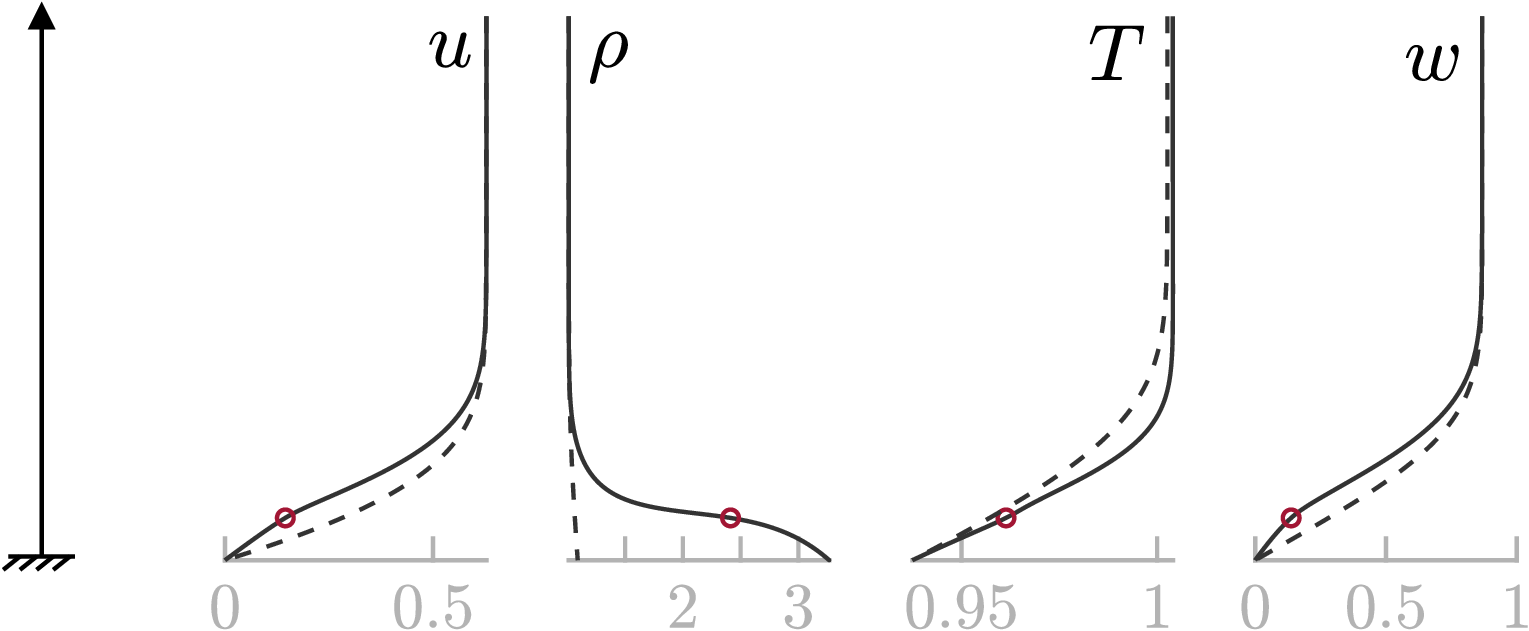} 
\put(-205,85){({\it a})}
\put(-182,12){wall}
\put(-182,68){$\infty$}
\put(-195,72){$y$}
\put(-165,49){\textcolor{myred}{pseudo-}}
\put(-165,39){\textcolor{myred}{critical}}
\put(-165,29){\textcolor{myred}{point}}
\put(-8,14){ideal}
\put(-38,44){pseudo-}
\put(-38,34){boiling}
\put(-28,23){$\uparrow$}
\put(-18,14){$\rightarrow$}
\put(-157.8,19){\textcolor{myred}{$\downarrow$}}
\put(-116,68){$\underbrace{\qquad\qquad\qquad\;\;\;\;}$}
\put(-86.5,51){\textcolor{myred}{$\Downarrow$}}
\put(-108,38){$\overbrace{\qquad\qquad\;\;\;\;\;}$}
\put(-109,32){\E~\V~\T}
\put(-80,51){\textcolor{myred}{influence}}
\put(-157,80){$\overbrace{\qquad\qquad\qquad\qquad\qquad\qquad\qquad\qquad\;}$}
\put(-90,91){\B}
\\ [2mm]
\includegraphics[scale=0.26]{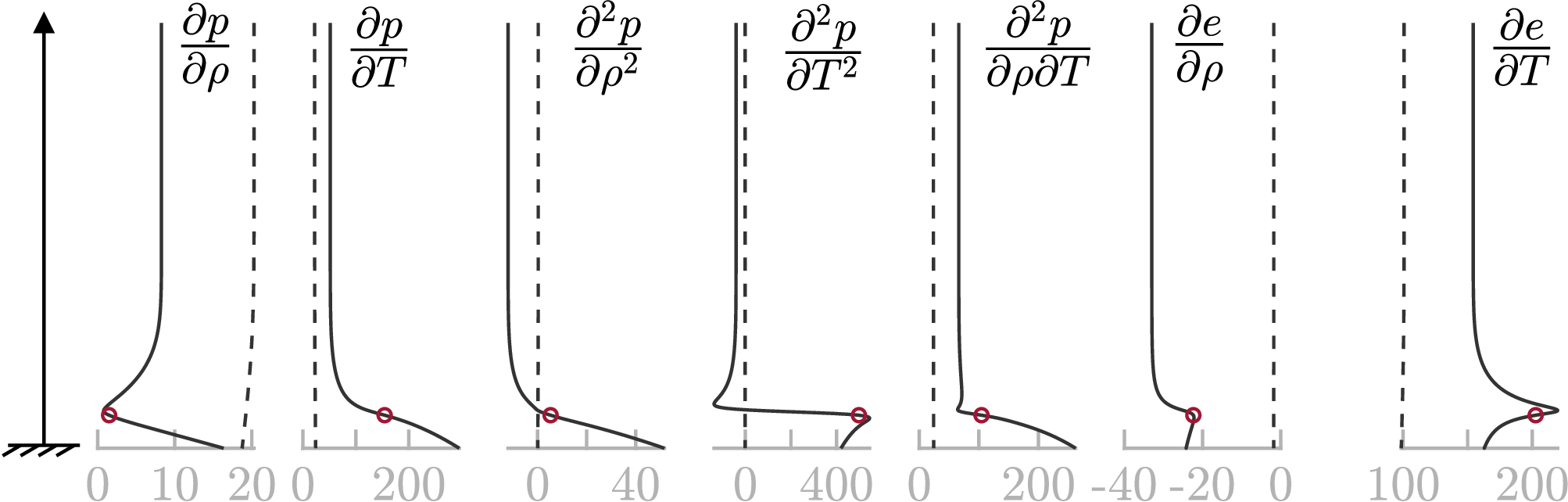} 
\put(-265,85){({\it b})}
\put(-253,72){$y$}
\put(-126,91){\E}
\put(-222,80){$\overbrace{\hspace{7.5cm}}$}
\\ [5mm]
\includegraphics[scale=0.26]{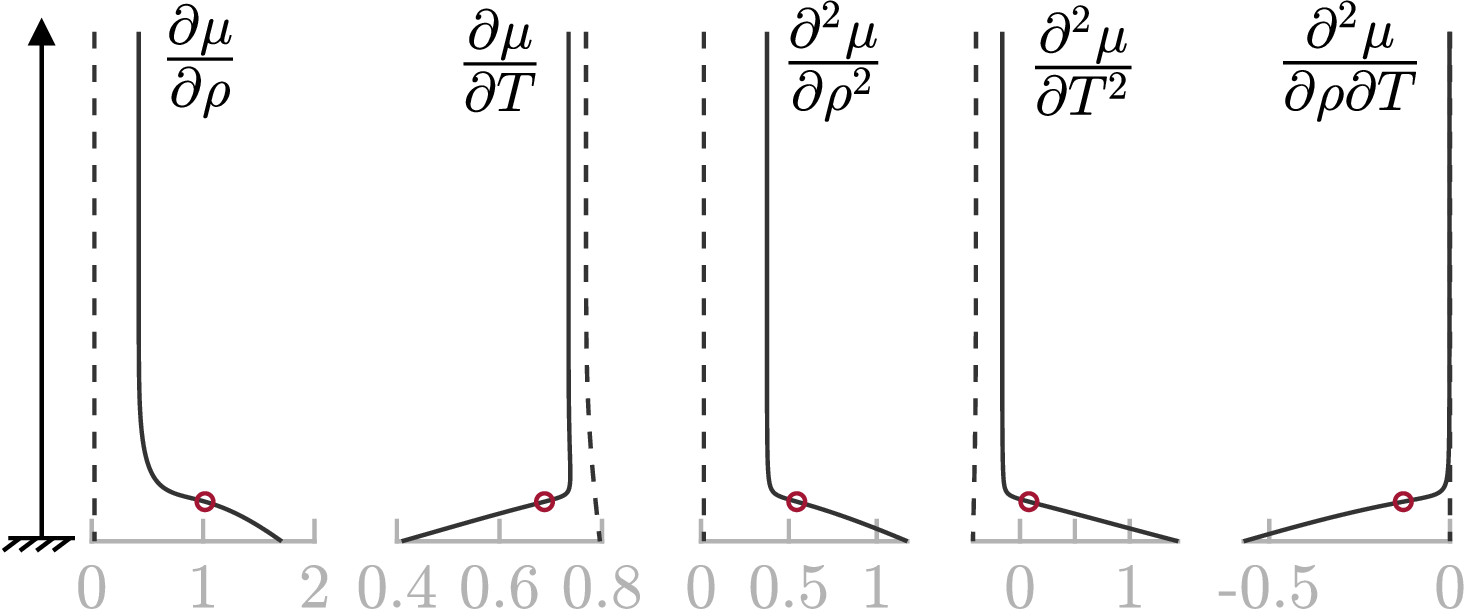} 
\includegraphics[scale=0.26]{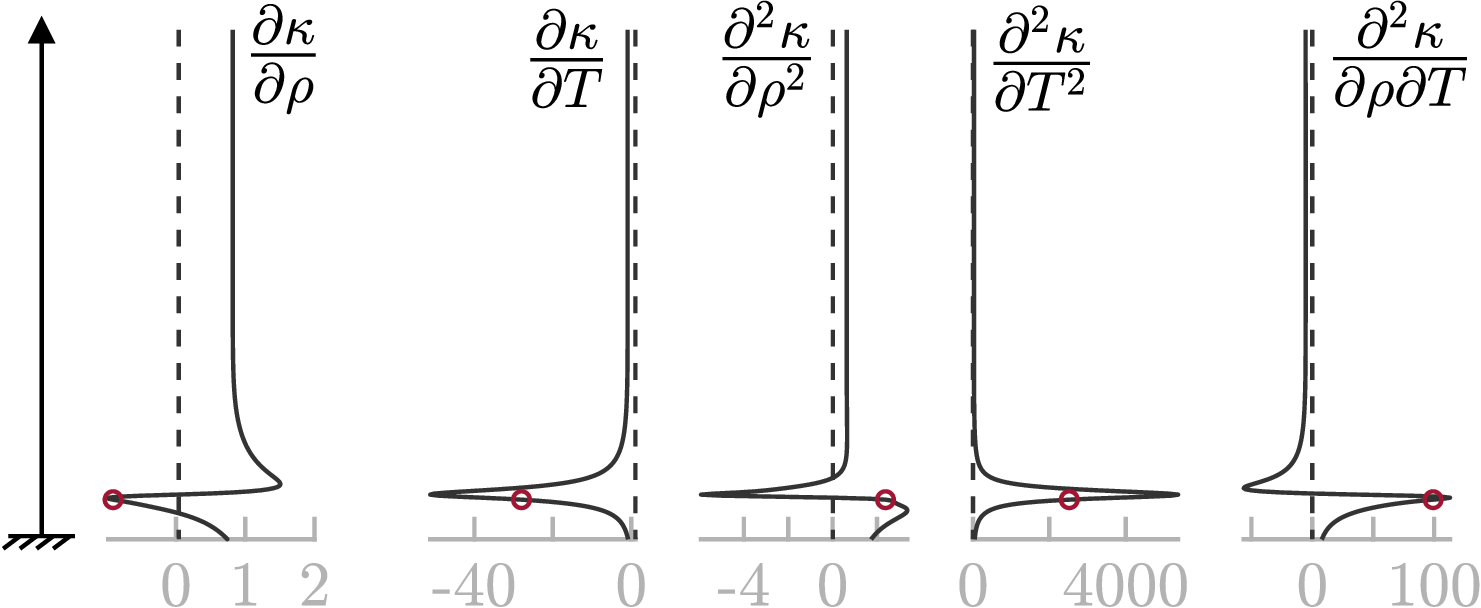}
\put(-380,82){({\it c})}
\put(-180,82){({\it d})}
\put(-375,68){$y$}
\put(-178,68){$y$}
\put(-350,75){$\overbrace{\hspace{5.5cm}}$}
\put(-152,75){$\overbrace{\hspace{5.cm}}$}
\put(-278,86){\V}
\put(-87,86){\T}
\caption{An overview of the inputs in the stability operator. Panels (a) through (d) show \B~baseflow profiles, \E~equation of state, \V~viscosity and \T~thermal conductivity. The pseudo-boiling and ideal gas regimes are plotted with solid and dashed lines, respectively (both under wall cooling). The red circle denotes the pseudo-critical point.}
\label{fig2}
\end{figure}

Considering flow instability, the flow variables $\tilde{\boldsymbol{q}}=\left(\tilde{\rho},\tilde{u},\tilde{v},\tilde{w},\tilde{T}\right)^{T}$ are decomposed into the laminar baseflow $\boldsymbol{Q}$ (steady state) and its perturbations $\boldsymbol{q}^{\prime}$:
\begin{equation}\label{eq_Q}
\tilde{\boldsymbol{q}}\left(x,y,z,t\right)=\boldsymbol{Q}\left(x,y\right)+\boldsymbol{q}^{\prime}\left(x,y,z,t\right).
\end{equation}
The stability equations are derived by subtracting the governing equations for $\tilde{\boldsymbol{q}}$ and $\boldsymbol{Q}$, both of which satisfy the Navier-Stokes equations. We investigate the perturbations in Fourier space by introducing the ansatz:
\begin{equation}\label{eq_qhat}
\boldsymbol{q}^{\prime}\left(x,y,z,t\right)=\hat{\boldsymbol{q}}\left(y\right)\exp\left(i\alpha x+i\beta z-i\omega t\right)+c.c.
\end{equation}
with $c.c.$ denoting the complex conjugate. The linearised stability equations are derived and written in a compact form: 
\begin{equation}\label{EVP}
\mathsfbi{L}\left(\boldsymbol{Q},\alpha,\beta,\omega,\Rey,\Ma\right)\hat{\boldsymbol{q}}=0.
\end{equation}
Equation \eqref{EVP} constitutes an eigenvalue problem whose dimensions are $(5\times 5)$ before spatial discretization. The detailed expressions are provided in Appendix \ref{appL}. We investigate the problem with $\boldsymbol{Q}$, $\beta$, $\omega$, $\Rey$ and $\Ma$ as inputs while $\alpha$ and $\hat{\boldsymbol{q}}$ are the eigenvalue and eigenfunction to be solved (spatial mode). We present the constituent elements of operator $\mathsfbi{L}$ in equation \eqref{eq_L}. Compared to the scalar parameters $\beta$, $\omega$, $\Rey$, and $\Ma$ (not shown in equation \eqref{eq_L}), $\boldsymbol{Q}$ includes not only the baseflow profiles ($\rho$, $u$, $v$, $T$ -- including boundary conditions, $p$) but also the equation of state, viscosity, and thermal conductivity, which depend on these profiles. In other words, the intrinsic properties of the fluid, presented in $\boldsymbol{Q}$, critically influence the stability and determine the type and outcome of the problem. 
\begin{equation}\label{eq_L}
\mathsfbi{L}=\underset{\begin{array}{l}
\spadesuit\textrm{ - baseflow profiles: }\rho,u,w,T\\[15pt]
\clubsuit\textrm{ - equation of state: }\dfrac{\partial p}{\partial\rho},\dfrac{\partial p}{\partial T},\dfrac{\partial^{2}p}{\partial\rho^{2}},\dfrac{\partial^{2}p}{\partial T^{2}},\dfrac{\partial^{2}p}{\partial\rho\partial T},\dfrac{\partial e}{\partial\rho},\dfrac{\partial e}{\partial T}\\[15pt]
\diamondsuit\textrm{ - viscosity: \ensuremath{\mu},\ensuremath{\dfrac{\partial\mu}{\partial\rho}},\ensuremath{\dfrac{\partial\mu}{\partial T}},\ensuremath{\dfrac{\partial^{2}\mu}{\partial\rho^{2}}},\ensuremath{\dfrac{\partial^{2}\mu}{\partial T^{2}}},\ensuremath{\dfrac{\partial^{2}\mu}{\partial\rho\partial T}}}\\[15pt]
\heartsuit\textrm{ - thermal conductivity: }\kappa,\dfrac{\partial\kappa}{\partial\rho},\dfrac{\partial\kappa}{\partial T},\dfrac{\partial^{2}\kappa}{\partial\rho^{2}},\dfrac{\partial^{2}\kappa}{\partial T^{2}},\dfrac{\partial^{2}\kappa}{\partial\rho\partial T}
\end{array}}{\underbrace{\left(\begin{array}{ccccc}
\spadesuit & \spadesuit & \spadesuit & \spadesuit & 0\\
\spadesuit\clubsuit\diamondsuit & \spadesuit\diamondsuit & \spadesuit\diamondsuit & \diamondsuit & \spadesuit\clubsuit\diamondsuit\\
\spadesuit\clubsuit\diamondsuit & \diamondsuit & \spadesuit\diamondsuit & \diamondsuit & \spadesuit\clubsuit\diamondsuit\\
\spadesuit\clubsuit\diamondsuit & \diamondsuit & \spadesuit\diamondsuit & \spadesuit\diamondsuit & \spadesuit\clubsuit\diamondsuit\\
\spadesuit\clubsuit\diamondsuit\heartsuit & \spadesuit\diamondsuit & \spadesuit\clubsuit\diamondsuit & \spadesuit\diamondsuit & \spadesuit\clubsuit\diamondsuit\heartsuit
\end{array}\right)}}
\end{equation}
Symbols with square brackets have been used to distinguish the four groups of inputs, with the variables associated with each group listed at the bottom of equation \eqref{eq_L}. Specifically, \B~represents baseflow profiles (density, velocities in the $x$- and $z$-directions and temperature as functions of $y$), \E~represents the equation of state (profiles of thermodynamic derivatives of pressure and internal energy), and \V~and \T~represent viscosity, thermal conductivity, and their gradients with respect to temperature and density. Compared to the Orr-Sommerfeld equation (and also to the compressible perfect/ideal-gas setup), there is an increase in the number of inputs, specifically within the three groups \E, \V, and \T. To gain an overview of all the inputs, we compare the wall-normal profiles between pseudo-boiling (solid lines) and ideal gas (dashed lines) regimes in Figure \ref{fig2}. According to the state postulate, the groups \E, \V, and \T~are functions of $\rho$ and $T$. In the stage of numerically obtaining the baseflow, $\rho$ and $T$ (and $u$ and $w$) in turn depend on the models for \E, \V, and \T~(see Section \S \ref{S2b} and the discussion in Section \S \ref{S2d}). From earlier works \citep[see reviews in][]{Guardone2024}, the non-ideal properties of a fluid can drive the baseflow to be inflectional, supporting a new inviscid mode that dominates the instability, which may significantly promote flow transition. However, it remains poorly understood how sensitive the flow stability is to the intrinsic properties of a fluid. Specifically, how important is each input, and to what degree does it influence the results? Is it feasible to ignore some of them or use idealized laws?

The answer will be pursued following the sensitivity framework. We consider a distortion $\delta\boldsymbol{Q}$ to the input $\boldsymbol{Q}$, thus equation \eqref{EVP} becomes
\begin{equation}\label{EVPd}
\mathsfbi{L}\left(\boldsymbol{Q}+\delta\boldsymbol{Q},\alpha+\delta\alpha\right)(\hat{\boldsymbol{q}}+\delta\hat{\boldsymbol{q}})=0,
\end{equation}
in which $\delta\alpha$ and $\delta\hat{\boldsymbol{q}}$ are corresponding changes in the eigenvalue and eigenfunction induced by $\delta\boldsymbol{Q}$. For brevity, we have omitted the other fixed parameters ($\beta$, $\omega$, $\Rey$ and $\Ma$). Taking Taylor's expansion of operator $\mathsfbi{L}$ yields
\begin{equation}\label{Taylor}
\mathsfbi{L}\left(\boldsymbol{Q}+\delta\boldsymbol{Q},\alpha+\delta\alpha\right)=
\mathsfbi{L}\left(\boldsymbol{Q},\alpha\right)+\frac{\partial\mathsfbi{L}}{\partial\boldsymbol{Q}}\delta\boldsymbol{Q}+\frac{\partial\mathsfbi{L}}{\partial\alpha}\delta\alpha+O\left(\delta^{2}\right),
\end{equation}
where the gradient is defined as
\begin{equation}
\frac{\partial\mathsfbi{L}}{\partial\boldsymbol{Q}}\delta\boldsymbol{Q}=\underset{s\rightarrow0}{\lim}\frac{\mathsfbi{L}\left(\boldsymbol{Q}+s\delta\boldsymbol{Q}\right)-\mathsfbi{L}\left(\boldsymbol{Q}\right)}{s}.
\end{equation}
We substitute equation \eqref{Taylor} into \eqref{EVPd} and subtract the equation for the undistorted state \eqref{EVP}. The equation at  $O(\delta)$ reads
\begin{equation}\label{SS}
\frac{\partial\mathsfbi{L}}{\partial\alpha}\hat{\boldsymbol{q}}\delta\alpha+\frac{\partial\mathsfbi{L}}{\partial\boldsymbol{Q}}\delta\boldsymbol{Q}\hat{\boldsymbol{q}}+\mathsfbi{L}\delta\hat{\boldsymbol{q}}=0.
\end{equation}
Equation \eqref{SS} forms a tripartite relation between the distortion $\delta\boldsymbol{Q}$ and the induced $\delta\alpha$ \& $\delta\hat{\boldsymbol{q}}$. Further to this relation, we define the sensitivity coefficient $\boldsymbol{S}_{\boldsymbol{Q}}$, which satisfies
\begin{equation}\label{eq_ss}
\delta\alpha=\left\langle \boldsymbol{S}_{\boldsymbol{Q}},\delta\boldsymbol{Q}\right\rangle.
\end{equation}
Here $\left\langle \right\rangle$ stands for the inner product between two vectors: $\left\langle \boldsymbol{a},\boldsymbol{b}\right\rangle =\int_{0}^{\infty}\boldsymbol{a}^{H}\boldsymbol{b}\:dy $, $H$ implies Hermitian transpose. The sensitivity coefficient, through equation \eqref{eq_ss}, measures the reactivity of the eigenvalue relative to the baseflow distortion $\delta\boldsymbol{Q}$. To find the analytical expression for $\boldsymbol{S}_{\boldsymbol{Q}}$, we seek the adjoint problem defined by the following relations:
\begin{equation}\left\langle \hat{\boldsymbol{q}}^{\dagger},\mathsfbi{L}\hat{\boldsymbol{q}}\right\rangle =\left\langle \mathsfbi{L}^{\dagger}\hat{\boldsymbol{q}}^{\dagger},\hat{\boldsymbol{q}}\right\rangle =0.
\label{EQadj}
\end{equation}
The superscript $\dagger$ is used for adjoint variables and operators. The analytical process deriving the adjoint equations has been provided in Appendix \ref{appAdj}. The inner product of $\hat{\boldsymbol{q}}^{\dagger}$ with equation \eqref{SS} eliminates its last term, giving
\begin{equation}\label{eq_deltaalpha1}
\delta\alpha=-\frac{\left\langle \hat{\boldsymbol{q}}^{\dagger},\frac{\partial\mathsfbi{L}}{\partial\boldsymbol{Q}}\delta\boldsymbol{Q}\hat{\boldsymbol{q}}\right\rangle }{\left\langle \hat{\boldsymbol{q}}^{\dagger},\frac{\partial\mathsfbi{L}}{\partial\alpha}\hat{\boldsymbol{q}}\right\rangle }.
\end{equation}
Upon normalisation of the denominator (see Appendix \ref{appNorm})
\begin{equation}
\left\langle \hat{\boldsymbol{q}}^{\dagger},\frac{\partial\mathsfbi{L}}{\partial\alpha}\hat{\boldsymbol{q}}\right\rangle =1,
\end{equation} 
Equation \eqref{eq_deltaalpha1} is simplified as
\begin{equation}\label{SS2}
\delta\alpha=-\left\langle \hat{\boldsymbol{q}}^{\dagger},\frac{\partial\mathsfbi{L}}{\partial\boldsymbol{Q}}\delta\boldsymbol{Q}\hat{\boldsymbol{q}}\right\rangle. 
\end{equation}
Comparing \eqref{SS2} and \eqref{eq_ss}, the sensitivity coefficients $\boldsymbol{S}_{\boldsymbol{Q}}$ are obtained by integrating \eqref{SS2} by parts. We provide their specific expressions in Appendix \ref{appSens} (the superscript $*$ indicates the complex conjugate of a variable) respectively for all the inputs discussed in figure \ref{fig2}. 

\subsection{Distortions of the baseflow and coupling to the linear stability}\label{S2d}
\begin{figure}
\centerline{
\includegraphics[scale=0.35]{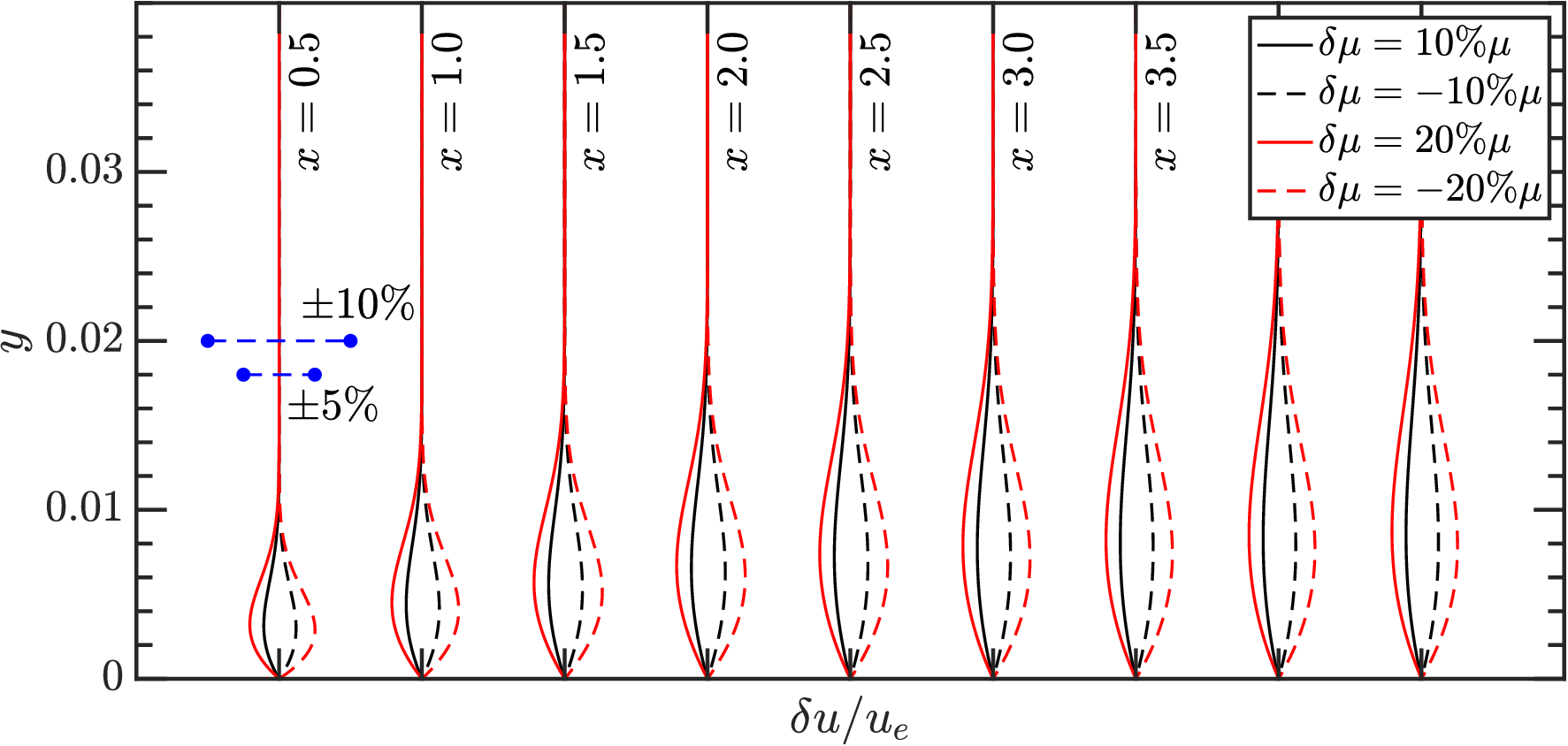}
\put(-312,140) {({\it a})}
}
\vspace{4mm}
\centerline{
\includegraphics[scale=0.35]{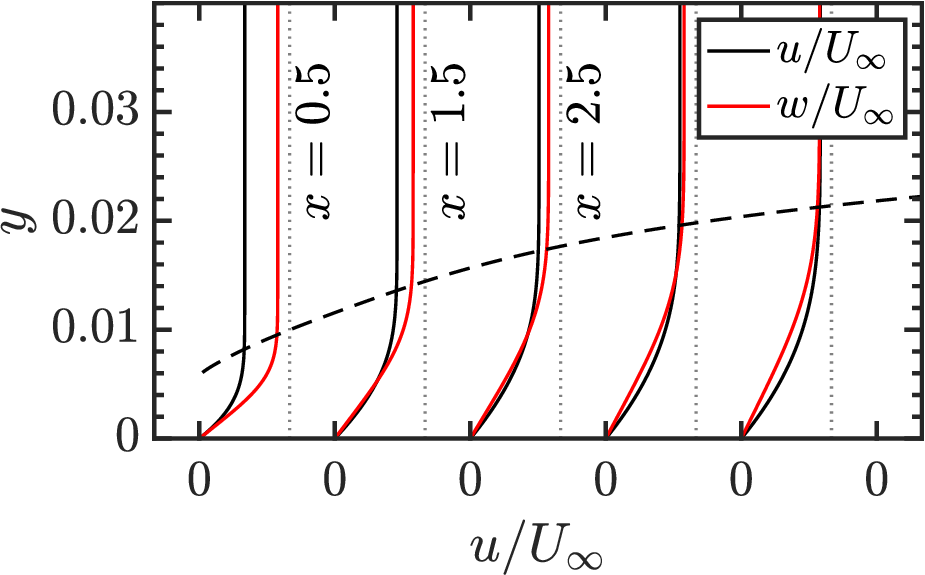}
\includegraphics[scale=0.35]{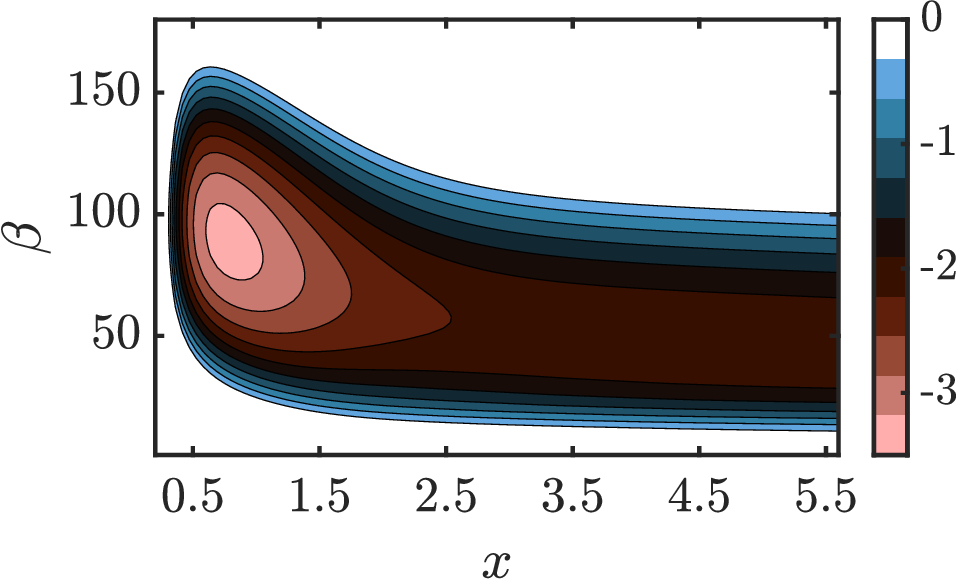}
\put(-323,92) {({\it b})}
\put(-162,92) {({\it c})}
}
\caption{Distortions of the streamwise velocity, $\delta u/u_e$, induced by viscosity alterations ($\delta \mu = \pm10\%\mu, \pm20\%\mu$) (a). The raw state baseflow profiles of $u$ and $w$ (b) and the corresponding stability diagram of the steady ($\omega=0$) crossflow instability (c) are provided. In panel (b), the dashed line stands for the boundary-layer thickness based on $0.99u/U_\infty$, and the dotted lines denote $u/U_\infty=w/U_\infty=1$. The flow is in the supercritical regime, subject to wall heating, with gas-like fluid properties. }
\label{fig_PNS_distortion}
\end{figure}

Recall Section \S \ref{S2b}, where the governing equations of the baseflow are nonlinear and the coefficient matrix in \eqref{eq_PNS} is a function of the solution vector. Here, we rewrite the equation in a more specific form:
\begin{equation}\label{PNS_distortion1}
\mathsfbi{A}\left(\boldsymbol{Q},\diamondsuit, \heartsuit, \clubsuit\right)\dfrac{\partial\boldsymbol{Q}}{\partial 
x}+\mathsfbi{B}\left(\boldsymbol{Q},\diamondsuit, \heartsuit, \clubsuit\right)\dfrac{\partial\boldsymbol{Q}}{\partial y}=\textrm{RHS}.
\end{equation}
We consider uncertainties in the fluid properties, such as distortions in viscosity $\delta\mu$ (as well as in other terms of \V, \T, and \E). We solve equation \eqref{eq_PNS} twice (with and without distorted fluid properties, respectively) to determine $\delta\boldsymbol{Q}$, which does not require small distortion amplitudes. Additionally, to understand the influence of $\delta\mu$ on the other components, take Taylor's expansion of equation \eqref{PNS_distortion1},
\begin{equation}\label{PNS_distortion2}
\begin{aligned}
\left(\mathsfbi{A}+\frac{\partial\mathsfbi{A}}{\partial\boldsymbol{Q}}\delta\boldsymbol{Q}+\frac{\partial\mathsfbi{A}}{\partial\mu}\delta\mu+O\left(\delta^{2}\right)\right)\frac{\partial\left(\boldsymbol{Q}+\delta\boldsymbol{Q}\right)}{\partial x}&+\\
\left(\mathsfbi{B}+\frac{\partial\mathsfbi{B}}{\partial\boldsymbol{Q}}\delta\boldsymbol{Q}+\frac{\partial\mathsfbi{B}}{\partial\mu}\delta\mu+O\left(\delta^{2}\right)\right)\frac{\partial\left(\boldsymbol{Q}+\delta\boldsymbol{Q}\right)}{\partial y}&=\boldsymbol{\textrm{RHS}}
\end{aligned}
\end{equation}
In the linear regime (when $\delta$ is small), by subtracting equations \eqref{PNS_distortion2} and \eqref{PNS_distortion1}, keeping terms of order $O\left(\delta\right)$:
\begin{equation}\label{PNS_distortion3}
\begin{aligned}
\mathsfbi{A}\frac{\partial\delta\boldsymbol{Q}}{\partial x}+\mathsfbi{B}\frac{\partial\delta\boldsymbol{Q}}{\partial y}=\mathrm{\delta{}_{RHS}}
\end{aligned}
\end{equation}
where
\begin{equation}
\mathrm{\delta{}_{RHS}}=-\left(\mathsfbi{A}\left(\delta\boldsymbol{Q}\right)+\mathsfbi{A}\left(\delta\mu\right)\right)\frac{\partial\boldsymbol{Q}}{\partial x}-\left(\mathsfbi{B}\left(\delta\boldsymbol{Q}\right)+\mathsfbi{B}\left(\delta\mu\right)\right)\frac{\partial\boldsymbol{Q}}{\partial y}
\end{equation}
\begin{equation}
\mathsfbi{A}\left(\delta\boldsymbol{Q}\right)=\dfrac{\partial\mathsfbi{A}}{\partial\boldsymbol{Q}}\delta\boldsymbol{Q}=\underset{s\rightarrow0}{\lim}\frac{\mathsfbi{A}\left(\boldsymbol{Q}+s\delta\boldsymbol{Q}\right)-\mathsfbi{A}\left(\boldsymbol{Q}\right)}{s},
\end{equation}
\begin{equation}
\mathsfbi{A}\left(\delta\mu\right)=\frac{\partial\mathsfbi{A}}{\partial\mu}\delta\mu.
\end{equation}
In this linear regime, $\delta\boldsymbol{Q}$ can be obtained iteratively, as in PNS, by ``lagging” the coefficients of $\mathsfbi{A}\left(\delta\boldsymbol{Q}\right)$ and $\mathsfbi{B}\left(\delta\boldsymbol{Q}\right)$.  Scrutinizing equation \eqref{PNS_distortion3}, one recognizes that a scalar deviation $\delta\mu$ can induce distortions in all components of the baseflow $\delta\boldsymbol{Q}=\left(\delta\rho, \delta u, \delta v, \delta w, \delta T\right)^{T}$ through the coupling of the distorted matrices ($\mathsfbi{A}\left(\delta\boldsymbol{Q}\right)$, $\mathsfbi{A}\left(\delta\mu\right)$, $\mathsfbi{B}\left(\delta\boldsymbol{Q}\right)$, $\mathsfbi{B}\left(\delta\mu\right)$, and the corresponding raw state $\boldsymbol{Q}$.Figure \ref{fig_PNS_distortion} provides an overview of the baseflow distortions induced by viscosity alterations: $\delta\mu=\sigma\mu(\rho,T)$. Here, $\sigma=\pm10\%$ and $\pm20\%$, respectively, leading to a bulk alteration in the $\mu(y)$ profile as given by the fluid property database. These profiles show that distortions remain within the boundary-layer thickness and their amplitudes stay largely constant moving downstream. Specifically, $\delta \mu = \pm20\%\mu$ gives rise to $(\delta u)_{\max}\approx5\% U_\infty$. The stability diagram showing the imaginary part of the eigenvalue $\alpha_i$ indicates the steady crossflow mode grows from around $x = 0.5$, reaching a maximum around $x = 1.0$ and continues growing downstream.

\begin{figure}
\centering
\includegraphics[scale=0.35]{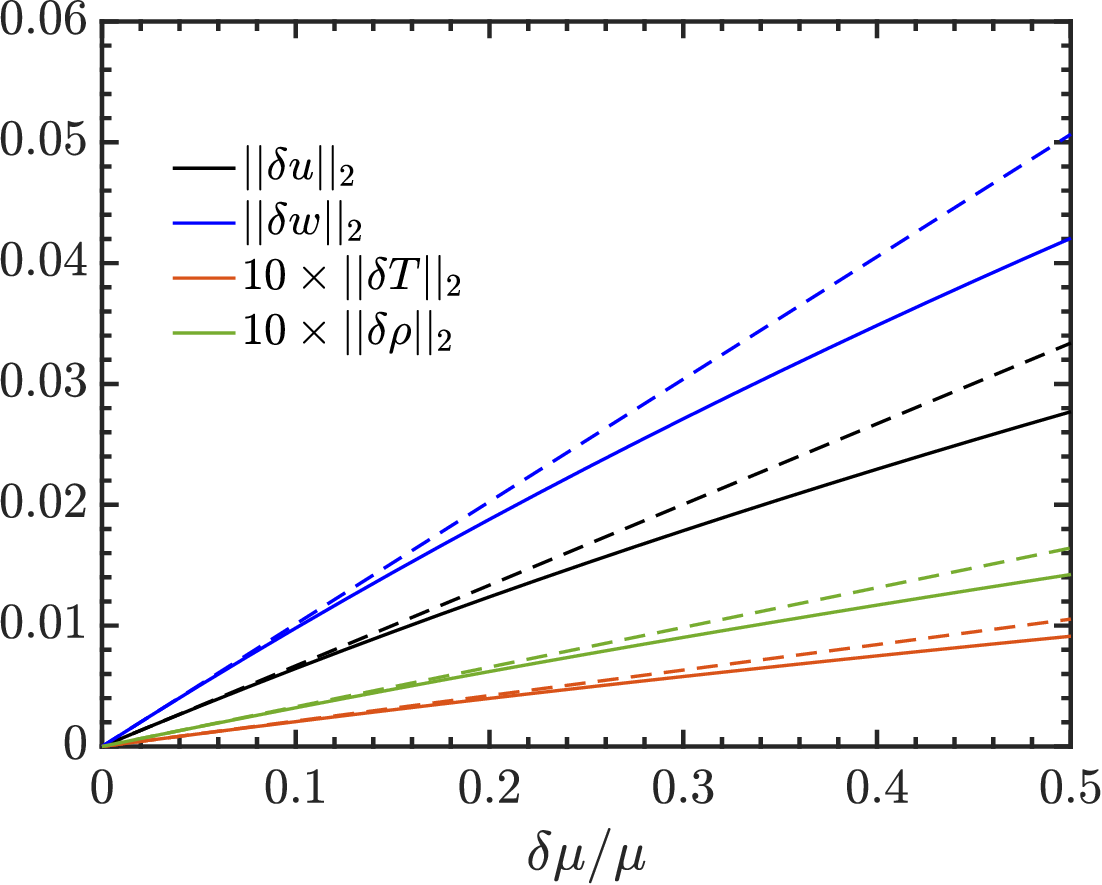} \hspace{1mm}
\includegraphics[scale=0.35]{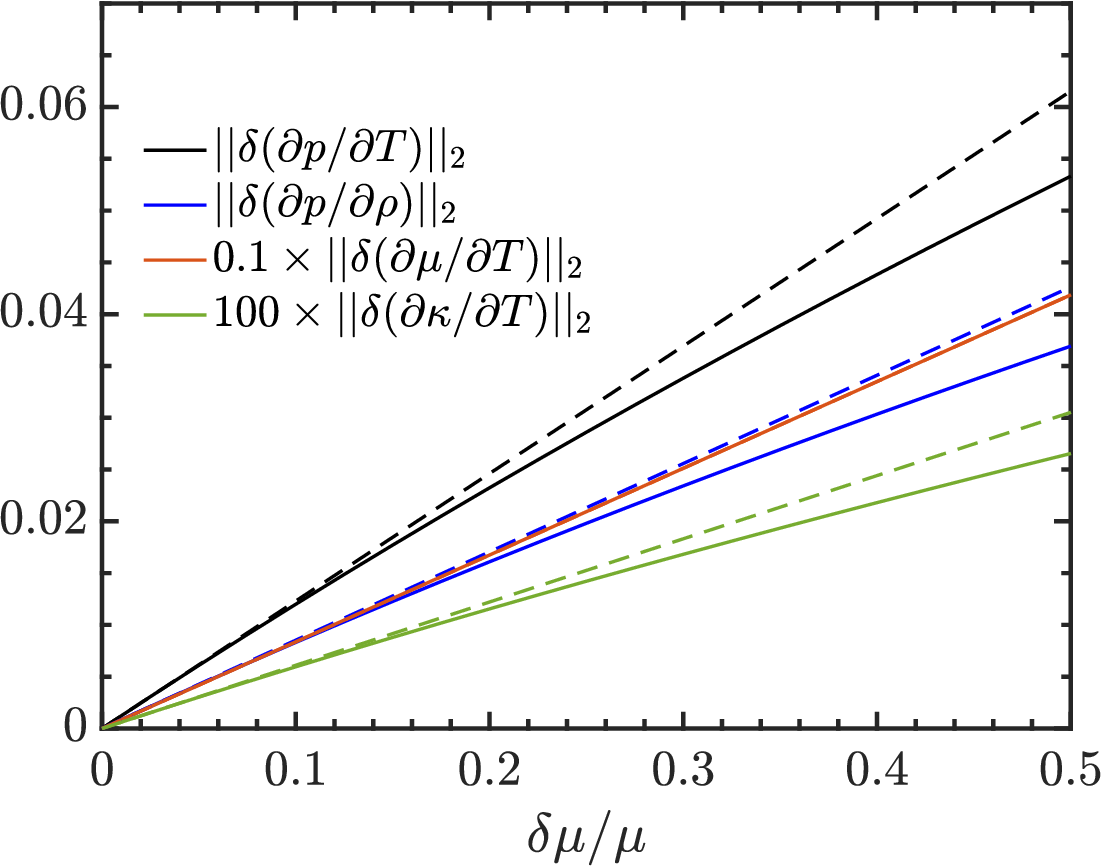}
\put(-356,133){({\it a})}
\put(-162,133){({\it b})}
\caption{Baseflow distortions induced by uncertainties in the viscosity model. The flow is in the supercritical regime. Viscosity distortion was applied in proportion to the original model, ranging from 0 to 50\%, as shown on the $x$-axis of each panel.}
\label{fig_PNS_mu}
\end{figure}

The influence of uncertainties in viscosity on the other terms of the baseflow is presented in Figure \ref{fig_PNS_mu}. Solid and dashed lines represent the actual distortion and the linear approximation, respectively. The viscosity distortion was applied in proportion to the original model: $\delta\mu = \sigma\mu(\rho,T)$ with $\sigma \in [0, 50\%]$. Linear behavior is evident when $\sigma \leq 10\%$. The influence is measured with the normalized 2-norm, with some results scaled according to the line legend (for better presentation). Here, the normalized 2-norm 
\begin{equation}\label{eqnorm}
||\boldsymbol{Q}||_2=\sqrt{\sideset{\frac{1}{N}}{_{i=1}^{N}}\sum Q_{i}^{2}},
\end{equation}
is a measure of the amplitude of the distortions accounting for the overall distortions. One may also choose to exclude the influence of the number of grid points ($N$) by using the $\infty$-norm, which considers only the maximum distortion over the wall-normal coordinates. Since the presented uncertainty is driven by viscosity, the term $\delta(\partial\mu/\partial T) \approx \sigma(\partial\mu/\partial T)$, though large, stays within the linear range. The uncertainty in viscosity induces distortions in all components of the baseflow, with considerable magnitudes in $\delta u$, $\delta w$, and terms related to pressure.

\begin{figure}
\centering
\hspace{1cm}
\includegraphics[scale=0.35]{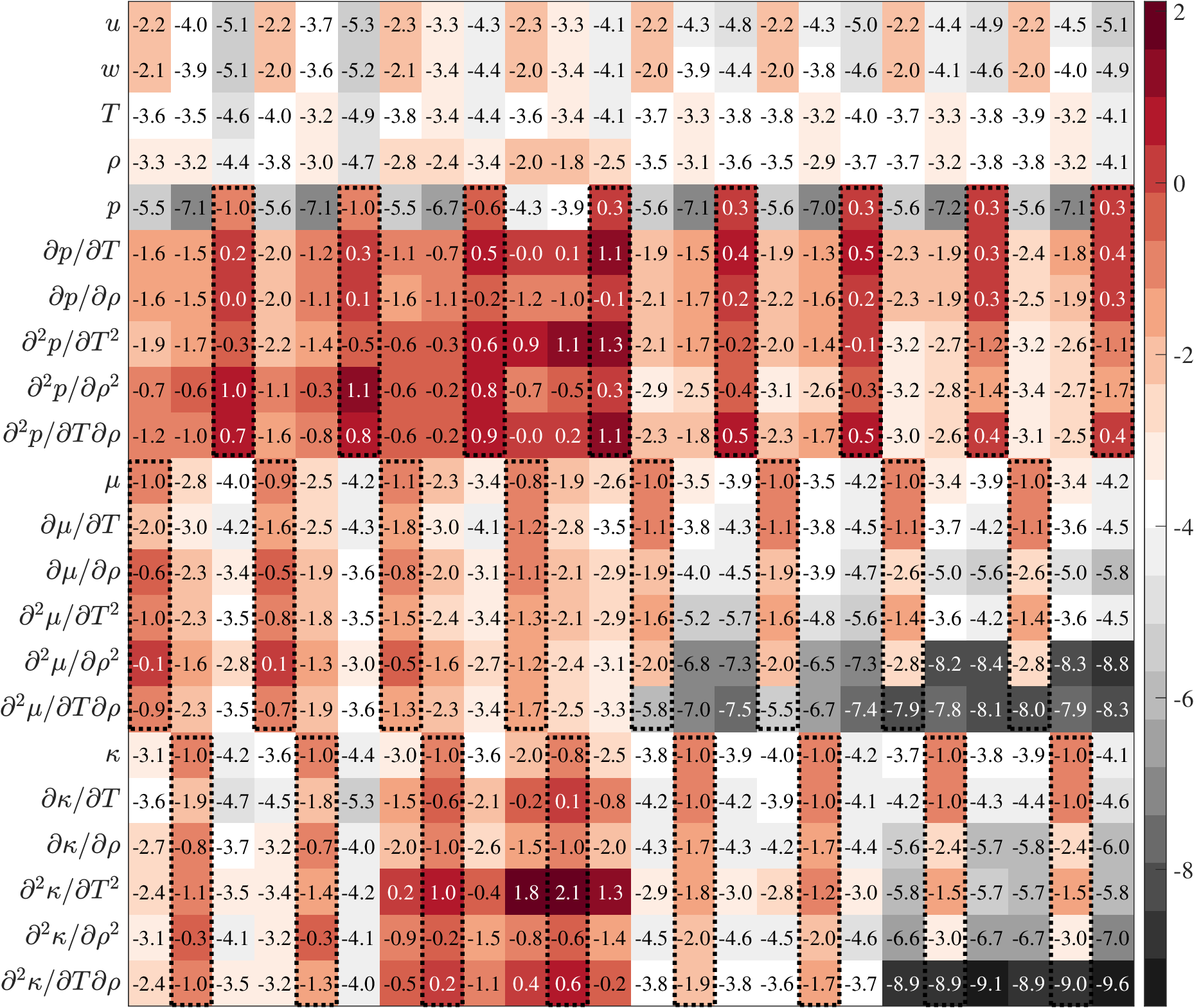}
\put(-325,237){$\left\{ \begin{array}{c}\\ \\ \\ \\ \end{array}\right.$}
\put(-325,178){$\left\{ \begin{array}{c}\\ \\ \\ \\ \\ \\  \end{array}\right.$}
\put(-325,107){$\left\{ \begin{array}{c}\\ \\ \\ \\ \\ \\  \end{array}\right.$}
\put(-325,35){$\left\{ \begin{array}{c}\\ \\ \\ \\ \\ \\  \end{array}\right.$}
\put(-338,237){$[b]$}
\put(-348,178){$[\rm EoS]$}
\put(-338,107){$[\mu]$}
\put(-338,35){$[\kappa]$}
\put(-279, -10){$\underbrace{\;\;\;\qquad\qquad\qquad}$}
\put(-254, -24){sub}
\put(-213, -10){$\underbrace{\;\;\;\qquad\qquad\qquad}$}
\put(-190, -24){trans}
\put(-147, -10){$\underbrace{\;\;\;\qquad\qquad\qquad}$}
\put(-126, -24){super}
\put(-81, -10){$\underbrace{\;\;\;\qquad\qquad\qquad}$}
\put(-60, -24){ideal}
\put(-277, -10){1\hspace{2mm}2\hspace{2.5mm}3\hspace{2mm}4\hspace{2.5mm}5\hspace{2mm}6}
\put(-212, -10){1\hspace{2mm}2\hspace{2.5mm}3\hspace{2mm}4\hspace{2.5mm}5\hspace{2mm}6}
\put(-146, -10){1\hspace{2mm}2\hspace{2.5mm}3\hspace{2mm}4\hspace{2.5mm}5\hspace{2mm}6}
\put(-80, -10){1\hspace{2mm}2\hspace{2.5mm}3\hspace{2mm}4\hspace{2.5mm}5\hspace{2mm}6}
\caption{Distortion magnitude of the baseflow components $\log(||\delta\boldsymbol{Q}||_2)$. The flow in the subcritical, transcritical, supercritical, and ideal regimes are compared. In each regime, the uncertainty is driven by $[\mu]$, $[\kappa]$, and [EoS], corresponding to columns 1-3 (with wall heating) and 4-6 (with wall cooling). The driving terms are highlighted with a rectangle of dotted lines.}
\label{fig_PNS_regimes}
\end{figure}
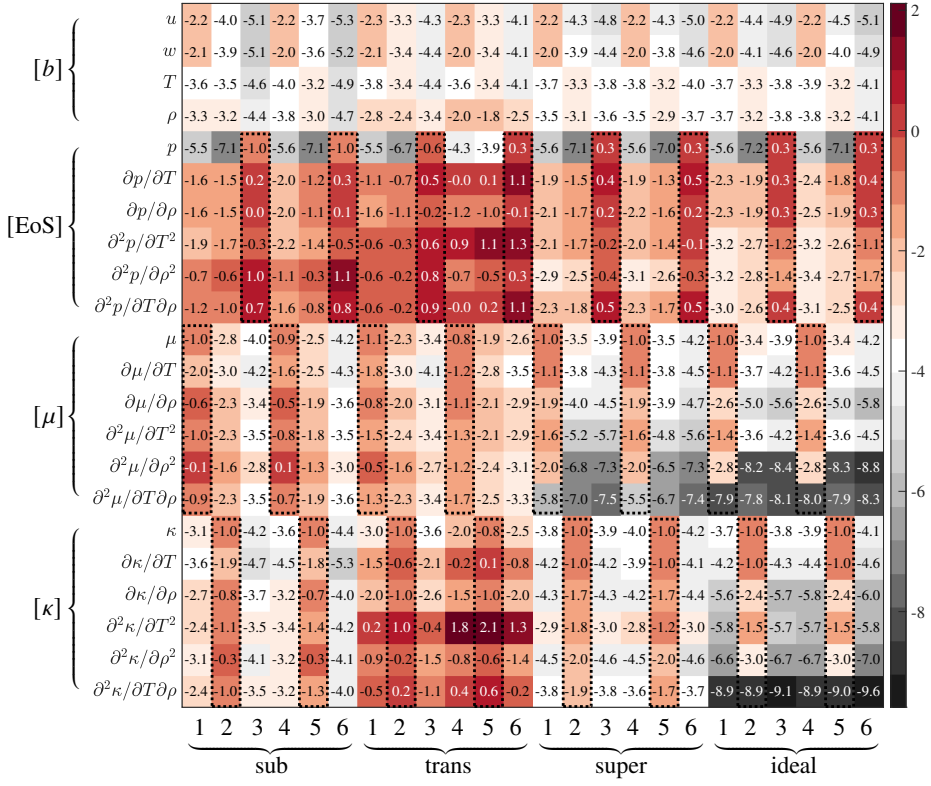

It is meaningful to compare fluid regimes, wall temperature, and sources of uncertainties in Figure \ref{fig_PNS_regimes}. The distortion magnitude $\log(||\delta\boldsymbol{Q}||_2)$ of the baseflow components is presented as a heat map. The driving uncertainty is prescribed as $\delta\mu=\sigma\mu(\rho,T)$, $\delta\kappa=\sigma\kappa(\rho,T)$, and $\delta p=\sigma p(\rho,T)$, respectively, with $\sigma=10\%$. Although the figure contains a wealth of data, key knowledge is obtained as follows:
\begin{enumerate}
\item Comparing different fluid regimes, distortions in the ideal fluid are the smallest, followed by the supercritical, subcritical, and transcritical cases, indicating an enhancing effect of non-ideal fluid properties on the induced distortion.
\item The driving uncertainty (see dotted rectangles in each column) leads to distortions in all other components of the baseflow. When the uncertainty comes from the equations of state, the driving terms remain the largest (compared to their induced distortions) across different flow regimes. Both viscosity and thermal conductivity uncertainties give rise to significant distortions in the pressure terms. The influence of thermal conductivity uncertainties is relatively smallest.
\item Uncertainties in the viscosity lead to the strongest distortion of the primary baseflow profile $[b]$, regardless of the flow regime and wall temperature. Velocities ($u$ and $w$) are more distorted by any of the studied uncertainties compared to temperature $T$ and density $\rho$.
\item Wall heating, compared to wall cooling, does not make an essential difference except for the transcritical case, where wall cooling shows larger distortions than the heating counterpart.
\end{enumerate}

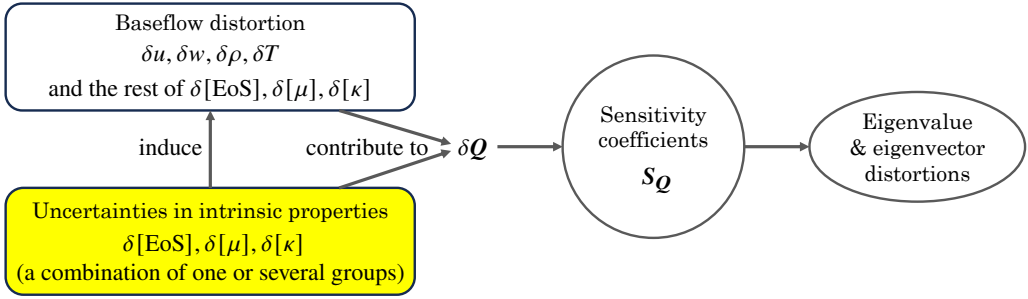
\begin{figure}
\centering
\includegraphics[scale=0.4]{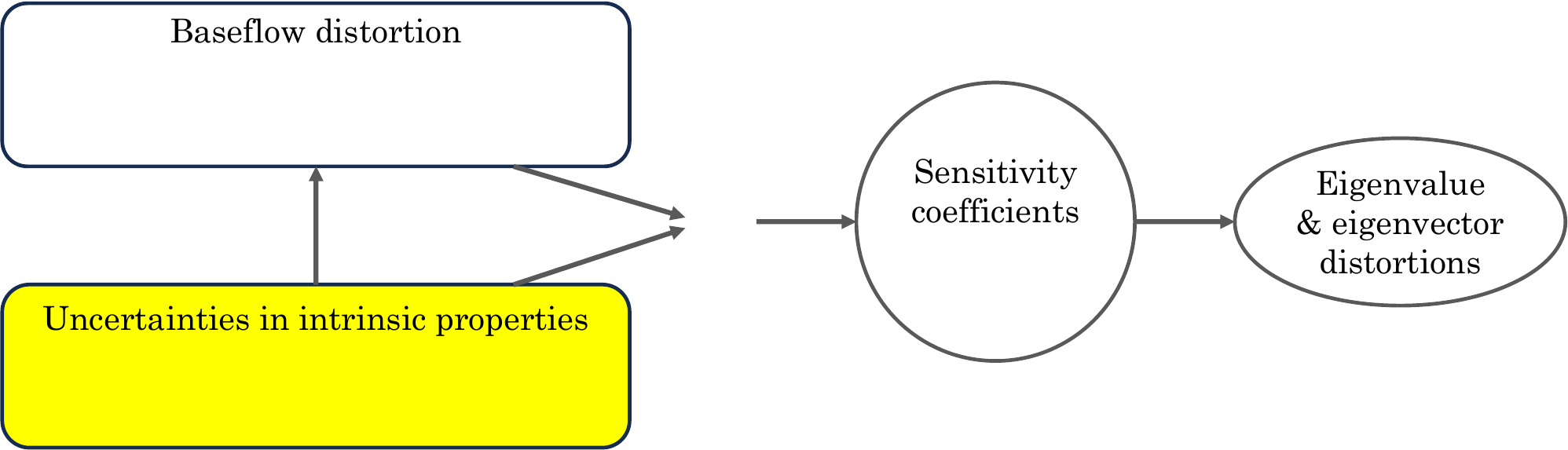}
\put(-340,17) {$\delta\E, \delta[\mu],\delta[\kappa]$}
\put(-380,6) {(a combination of one or several groups)}
\put(-333,88) {$\delta u,\delta w,\delta \rho,\delta T$}
\put(-368,75) {and the rest of $\delta\E, \delta[\mu],\delta[\kappa]$}
\put(-334,53) {induce}
\put(-271,53) {\rotatebox{0}{\textcolor{black}{contribute to}}}
\put(-214,53) {$\delta\boldsymbol{Q}$}
\put(-145,42) {$\boldsymbol{S}_{\boldsymbol{Q}}$}
\caption{Relational diagram depicting the uncertainties in intrinsic fluid properties, the baseflow distortions, and their influences on the eigenvalue problem.}
\label{fig_relation}
\end{figure}

Going through equations \eqref{eq_L} and Appendix \ref{appL}, one notes that the fluid properties are directly inputs to the stability operator \citep{guo2021sensitivity,Poulain_Content_Rigas_Garnier_Sipp_2024}. The fluid properties, the baseflow, and the stability operator form a coupled system as summarized in figure \ref{fig_relation}. The intrinsic uncertainty influences the system in an integrated manner: the primary baseflow profiles $[b]$ are distorted, which, together with the distorted fluid properties, forms inputs to the stability operator. These distortions in the eigenvalue problem are measured by the sensitivity coefficients, which will be discussed next.

\section{Results and discussions}\label{S3}
\subsection{Input distortions and the induced eigenvalue shifts}\label{S3a}

\begin{figure}
\centerline{
\includegraphics[scale=0.35]{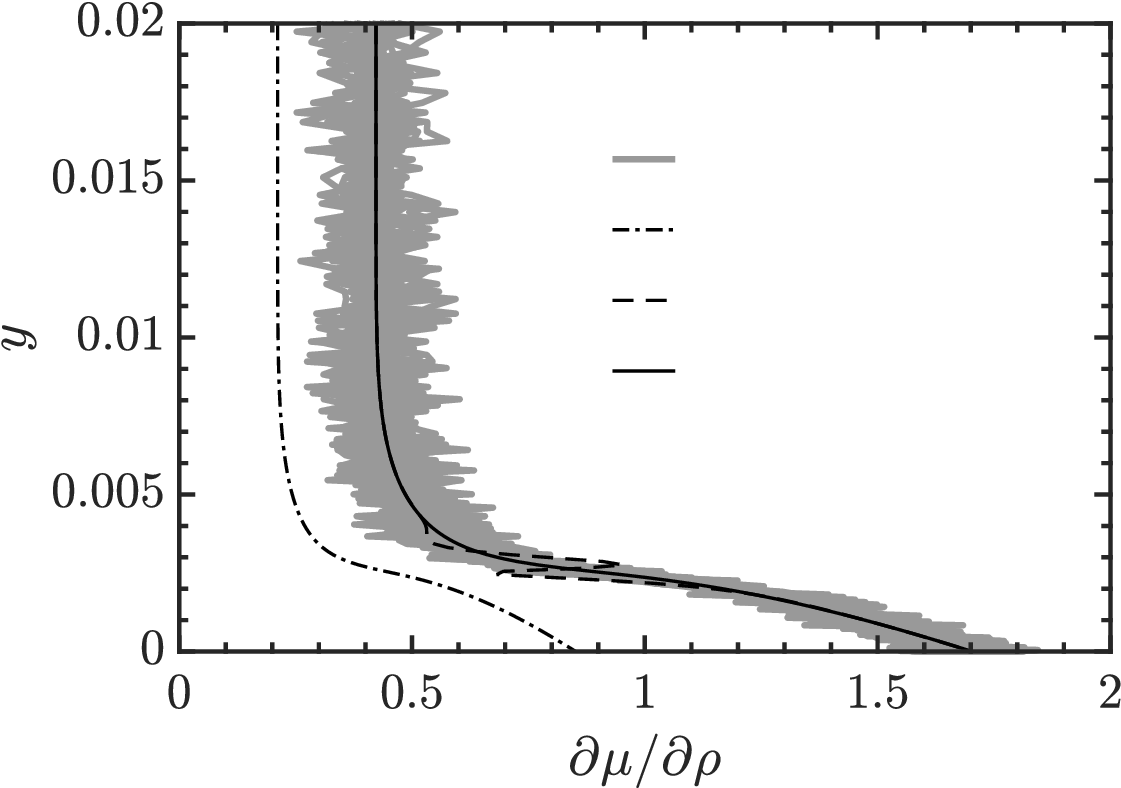}
\put(-72,104){$\boldsymbol{Q}+\delta\boldsymbol{Q}_{\rm{noise}}$}
\put(-72,92){$\boldsymbol{Q}+\delta\boldsymbol{Q}_{\rm{property}}$}
\put(-72,80) {$\boldsymbol{Q}+\delta\boldsymbol{Q}_{\rm{max}}$}
\put(-72,68) {$\boldsymbol{Q}$}
}
\caption{A portray of different distortions on $\partial\mu/\partial\rho$.}
\label{fig3}
\end{figure}

Since latent uncertainty can lead to distortions of various shapes, we begin by comparing three types of distortions (listed in Table \ref{table2}). First, white noise is assigned to assess the robustness of stability when exposed to a realistic environment, such as a laboratory experiment or a flight test. The second signal corresponds to structural sensitivity, where some inputs are biased due to an inaccurate fluid model. Third, we consider the distortion generating the maximum growth rate shift, which represents the border of sensitivity at a specific amplitude and serves as a measure of sensitivity. Figure \ref{fig3} provides a profile of $\partial\mu/\partial\rho$ in the pseudo-boiling regime with wall cooling. Unlike an ideal gas, the term $\partial\mu/\partial\rho$ is non-zero. Figure \ref{fig3} illustrates the shapes of these distortions, with their amplitudes intentionally assigned large for illustration purposes.

\begin{table}
\begin{center}\def~{\hphantom{0}}
\begin{tabular}{lll}
      Distortion type & Symbol  &  Stand for  \\[3pt]
      white noise   & $\delta\boldsymbol{Q}_{\rm{noise}}$  & random environmental perturbations \\
      structural      & $\delta\boldsymbol{Q}_{\rm{property}}$  &  distortions due to inacurate fluid models\\
      structural      & $\delta\boldsymbol{Q}_{\rm{max}}$    & distortion leading to maximum shift of the growth rate \\
\end{tabular}
\caption{{Distortions of different types.}}
\label{table2}
\end{center}
\end{table}

\begin{figure}
\centering
\includegraphics[scale=0.35]{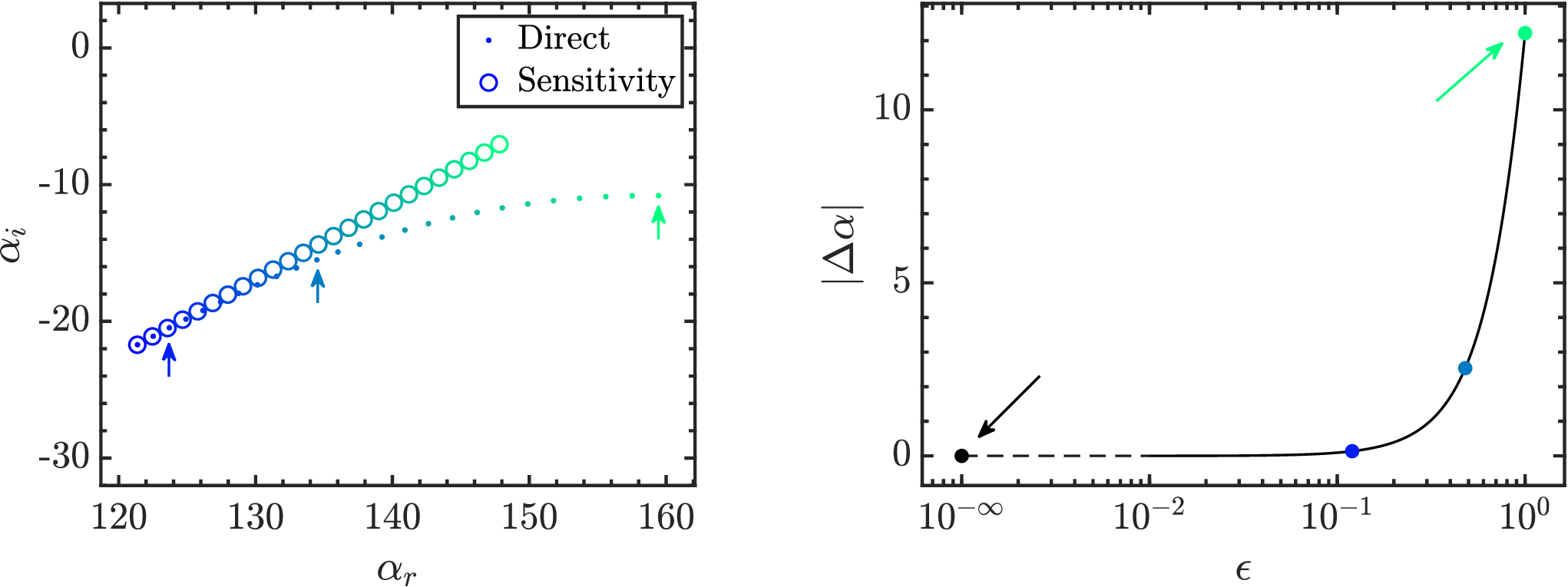}
\put(-316,118){({\it a})}
\put(-134,118){({\it b})}
\put(-212,67) {\textcolor[rgb]{0,1,0.5}{$1.00$}} 
\put(-280,53) {\textcolor[rgb]{0,0.48,0.76}{$0.52$}} 
\put(-316,39) {\textcolor[rgb]{0,0.12,0.94}{$\epsilon=0.12$}} 
\put(-81,33) {\textcolor[rgb]{0,0.12,0.94}{$\epsilon=0.12$}} 
\put(-40,51) {\textcolor[rgb]{0,0.48,0.76}{$0.52$}} 
\put(-50,118) {\textcolor[rgb]{0,1,0.5}{$\epsilon=1.00$}} 
\put(-114,48) {Fully non-ideal}
\put(-48,102) {Ideal}
\caption{Illustration of eigenvalue shift validating the sensitivity framework. (a) The eigenvalue trajectory with distorted $\partial\mu/\partial\rho$; (b) Error of eigenvalue prediction as a function of the departure parameter $\epsilon$ according to equation \eqref{eq_Qi}.}
\label{fig4}
\end{figure}
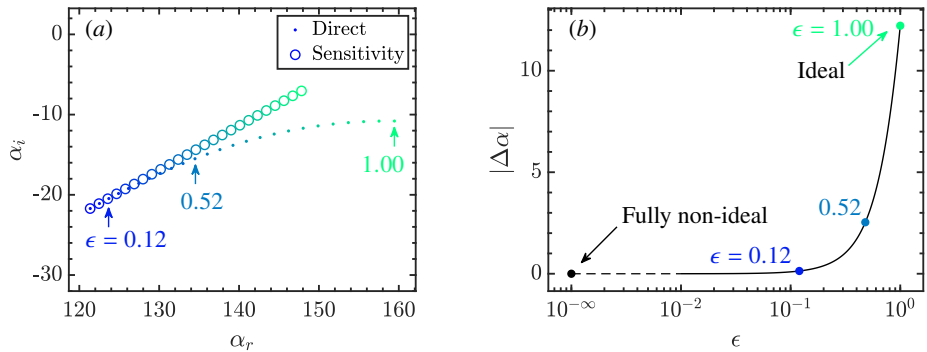


The sensitivity coefficients derived in Appendix \ref{appSens} have been validated by comparing the distorted eigenvalue using direct estimation \eqref{EVP} and the sensitivity framework \eqref{eq_ss}. Figure \ref{fig4}(a) illustrates this comparison. The case investigated involves pseudo-boiling with wall cooling. We evaluate the sensitivity of the eigenvalue at $\omega=40$, $\beta=100$, and $x=1.0$, corresponding to a typical inviscid instability mode found in a recent investigation \citep[see figure 12(b) of][]{Ren2022}. The distortion 
\begin{equation}\label{eq_Qi}
\delta\boldsymbol{Q}_{\rm{property}} |_\textrm{non-zero component} = -\epsilon\partial\mu/\partial\rho,
\end{equation}
is applied to $\partial\mu/\partial\rho$, where $0 \leq \epsilon \leq 1$ is the parameter controlling the amplitude. As $\epsilon$ increases from 0 to 1, the input $\partial\mu/\partial\rho$ is incrementally deformed by an ideal gas model until $\partial\mu/\partial\rho=0$ (fully ideal). As shown in Figure \ref{fig4}(a), the sensitivity coefficients provide an accurate prediction in the linear range, and the error (see panel b) becomes noticeable from $\epsilon=0.12$. We note that this value does however not provide a constant limit criterion for the linear regime, as the distortions can act on other input variables.

\subsection{The sensitivity coefficients and a scalar measure}\label{S3b}
\begin{figure}
\centering
\includegraphics[scale=0.4]{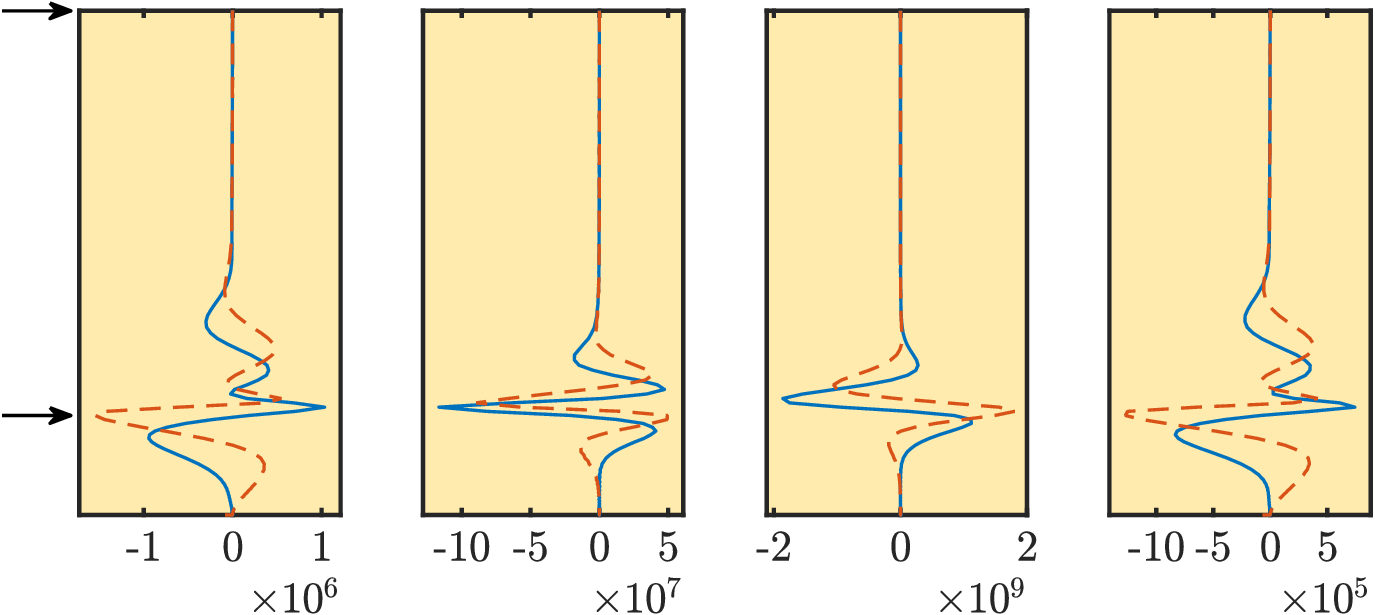}
\put(-336,118){({\it a})}
\put(-316,114){BL thickness}
\put(-316,36){Widom line}
\put(-235,100){$S_{u}$}
\put(-165,100){$S_{\rho}$}
\put(-108,100){$S_{T}$}
\put(-40,100){$S_{w}$}
\\[4mm]
\includegraphics[scale=0.4]{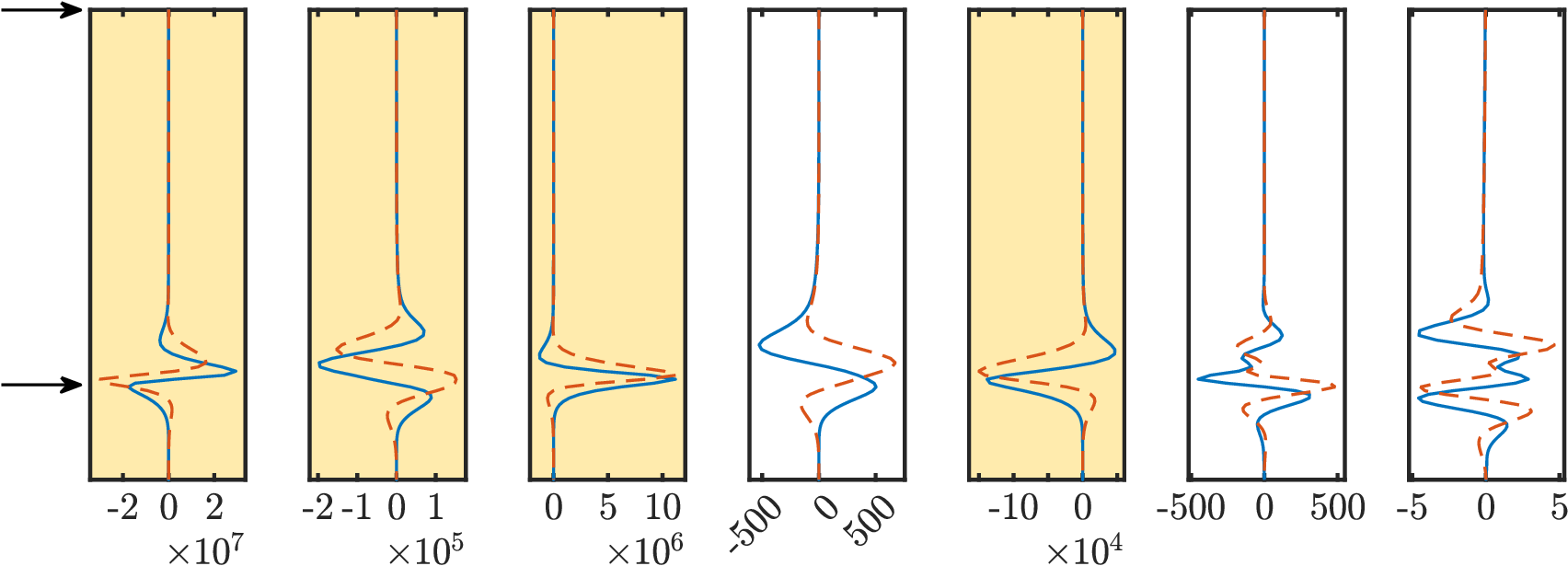}
\put(-350,115){({\it b})}
\put(-32,100){$S_{\frac{\partial e}{\partial T}}$}
\put(-78,100){$S_{\frac{\partial e}{\partial \rho}}$}
\put(-124,100){$S_{\frac{\partial^{2}p}{\partial\rho\partial T}}$}
\put(-162,100){$S_{\frac{\partial^{2}p}{\partial T^{2}}}$}
\put(-207,100){$S_{\frac{\partial^{2}p}{\partial \rho^{2}}}$}
\put(-262,100){$S_{\frac{\partial p}{\partial T}}$}
\put(-309,100){$S_{\frac{\partial p}{\partial \rho}}$}
\\[4mm]
\includegraphics[scale=0.4]{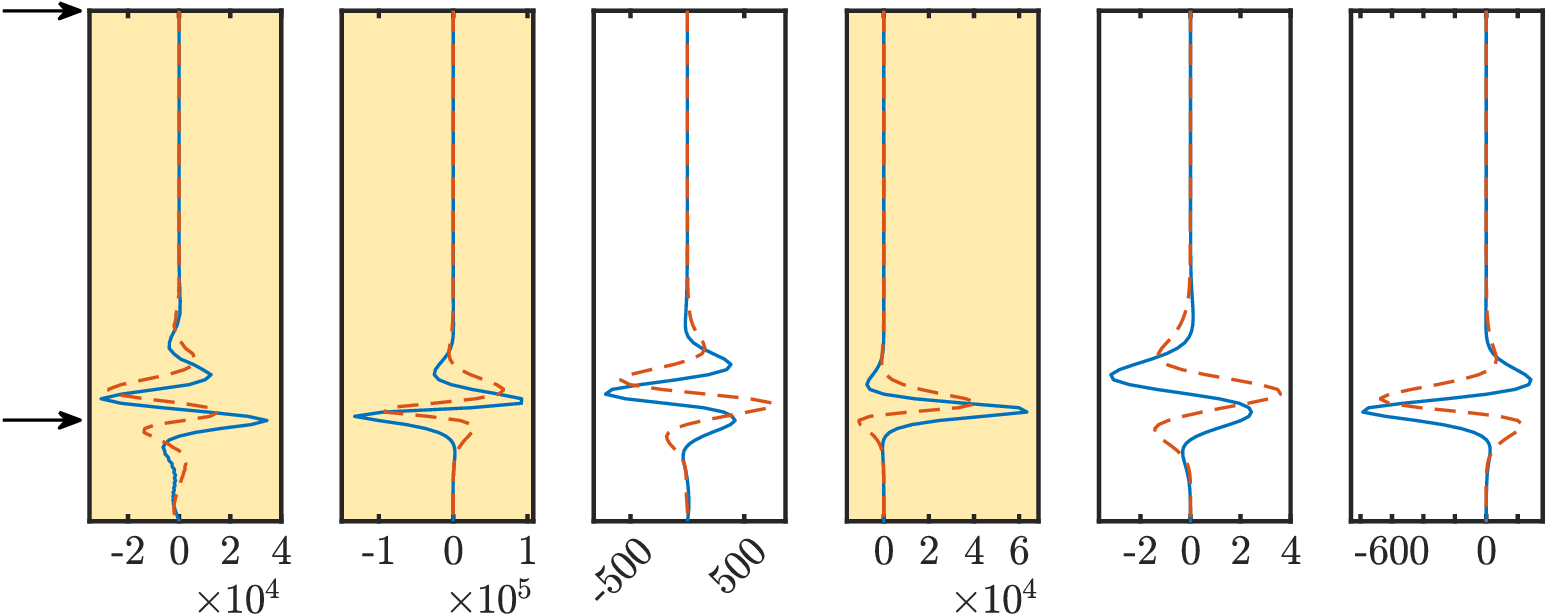}
\put(-320,115){({\it c})}
\put(-35, 100){$S_{\frac{\partial^{2}\mu}{\partial\rho\partial T}}$}
\put(-74, 100){$S_{\frac{\partial^{2}\mu}{\partial T^{2}}}$}
\put(-124,100){$S_{\frac{\partial^{2}\mu}{\partial \rho^{2}}}$}
\put(-164,100){$S_{\frac{\partial\mu}{\partial T}}$}
\put(-228,100){$S_{\frac{\partial\mu}{\partial \rho}}$}
\put(-260,100){$S_{\mu}$}
\\[4mm]
\includegraphics[scale=0.4]{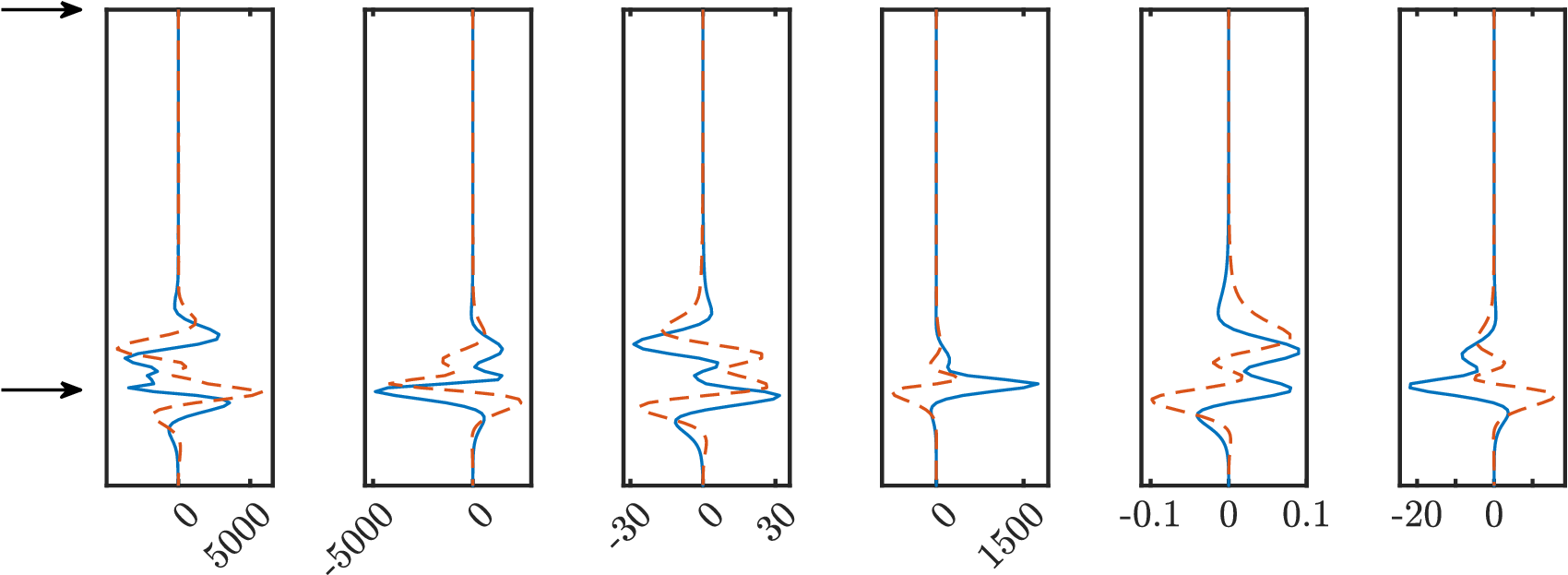}
\put(-350,115){({\it d})}
\put(-29, 100){$S_{\frac{\partial^{2}\kappa}{\partial\rho\partial T}}$}
\put(-88, 100){$S_{\frac{\partial^{2}\kappa}{\partial T^{2}}}$}
\put(-129,100){$S_{\frac{\partial^{2}\kappa}{\partial \rho^{2}}}$}
\put(-196,100){$S_{\frac{\partial\kappa}{\partial T}}$}
\put(-245,100){$S_{\frac{\partial\kappa}{\partial \rho}}$}
\put(-284,100){$S_{\kappa}$}
\caption{Assembly of sensitivity coefficients in the pseudo-boiling regime with wall cooling. Solid and dashed lines represent the real and imaginary parts respectively.}
\label{fig5}
\end{figure}
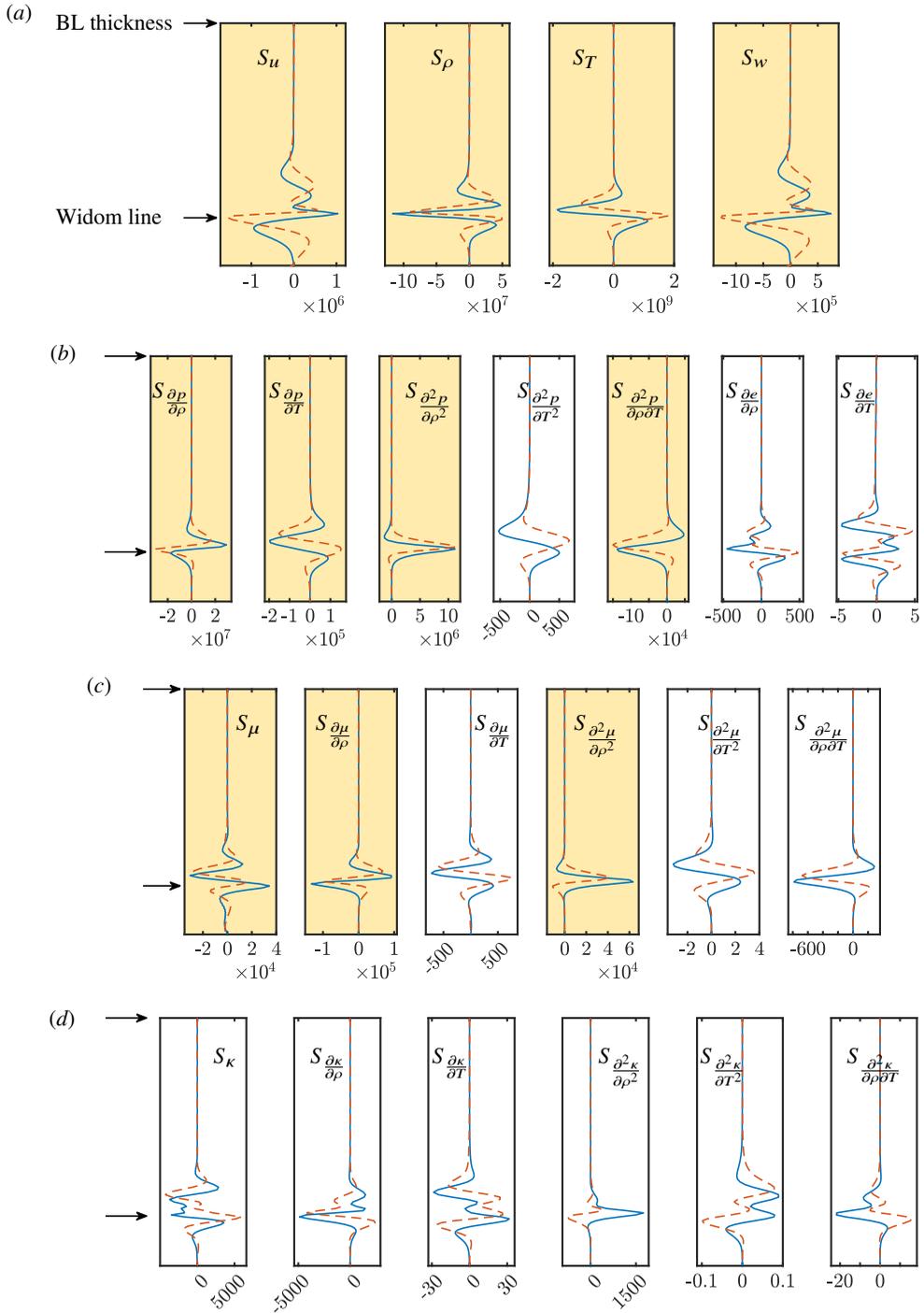

For the pseudo-boiling regime with wall cooling, we present profiles of all the sensitivity coefficients in Figure \ref{fig5}. Solid lines represent the real parts, while dashed lines represent the imaginary parts. The wave number and frequency values correspond to the case in Figure \ref{fig4}. Arrows indicate the location of the Widom line and the boundary-layer thickness $\delta_0$ ($u/u_\infty=0.99$). All panels are plotted from the wall to the height of $\delta_0$. Without exception, each profile shows shears near the Widom line and decays to zero towards the wall and upper boundaries. Distinctive differences between the profiles lie in the amplitude, spanning from $O(10^{-1})$ to $O(10^9)$. Given such a range of magnitudes on a logarithmic scale, the dominating terms are recognized. For clarity, we have applied a yellow background for values of $O(10^4)$ and higher. Before discussing different thermodynamic regimes, we propose a scalar measure for the degree of sensitivity.

Since the distortion shape can be arbitrary, the focus is placed on the distortion that leads to the maximum shift of the eigenvalue \citep{giannetti2007structural}. According to equation \eqref{eq_ss}, and accounting for the non-imaginary nature of physical distortions, the inner product reaches its maximum when $\delta\boldsymbol{Q}$ equals the real or imaginary part of $\boldsymbol{S}_{\boldsymbol{Q}}$. In other words, $\delta\boldsymbol{Q} = \pm \rm{imag}(\boldsymbol{S}_{\boldsymbol{Q}})$ results in the highest growth rate shift $\alpha_i$, while $\delta\boldsymbol{Q} = \pm \rm{real}(\boldsymbol{S}_{\boldsymbol{Q}})$ leads to the maximum shift of the chordwise wavenumber $\alpha_r$. To avoid ambiguity, we focus on the growth rate and define $\delta\boldsymbol{Q}_{{\rm max}}$ as $\pm \rm{imag}(\boldsymbol{S}_{\boldsymbol{Q}})$.

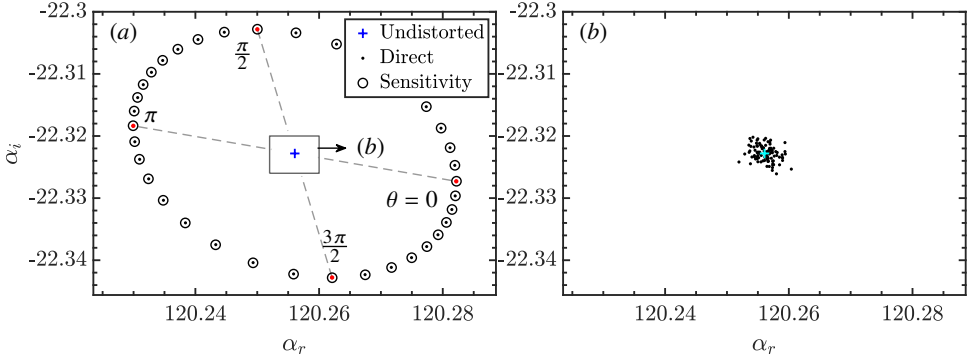
\begin{figure}
\centering
\includegraphics[scale=0.35]{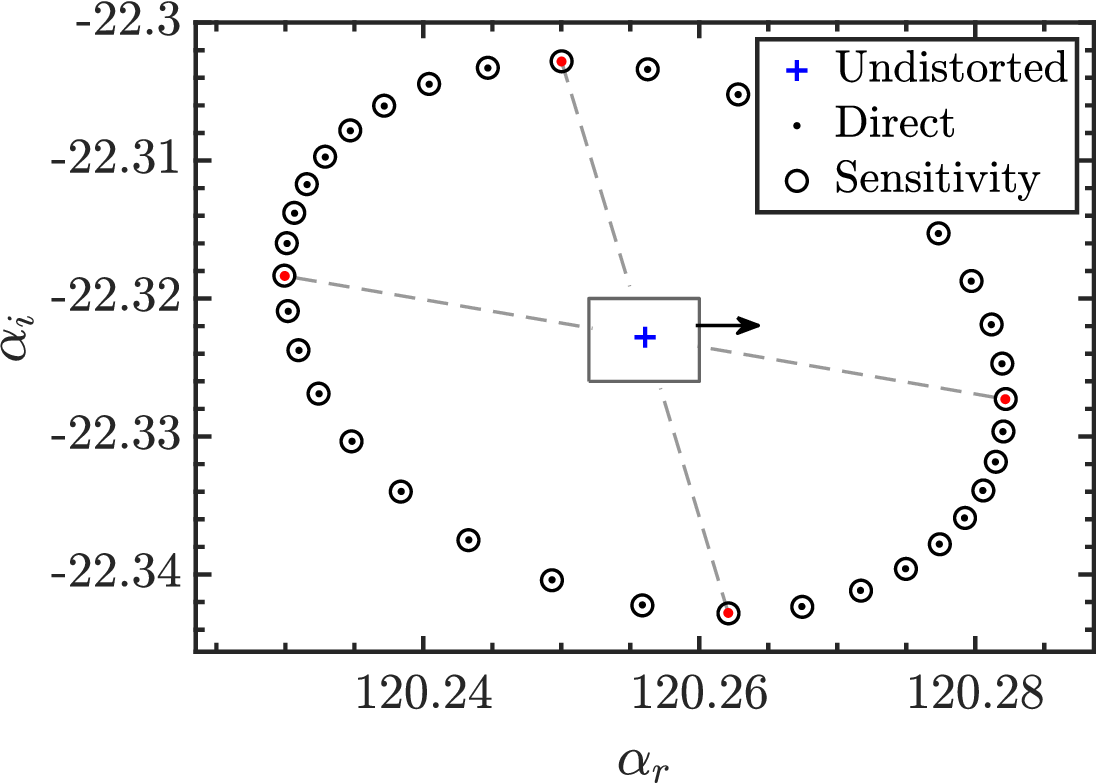}
\put(-100,110) 	{$\frac{\pi}{2}$}
\put(-133,87)  	{$\pi$}
\put(-67,38)   	{$\frac{3\pi}{2}$}
\put(-42,55)   	{$\theta=0$}
\put(-146,118){({\it a})}
\put(-53,75){({\it b})}
\includegraphics[scale=0.35]{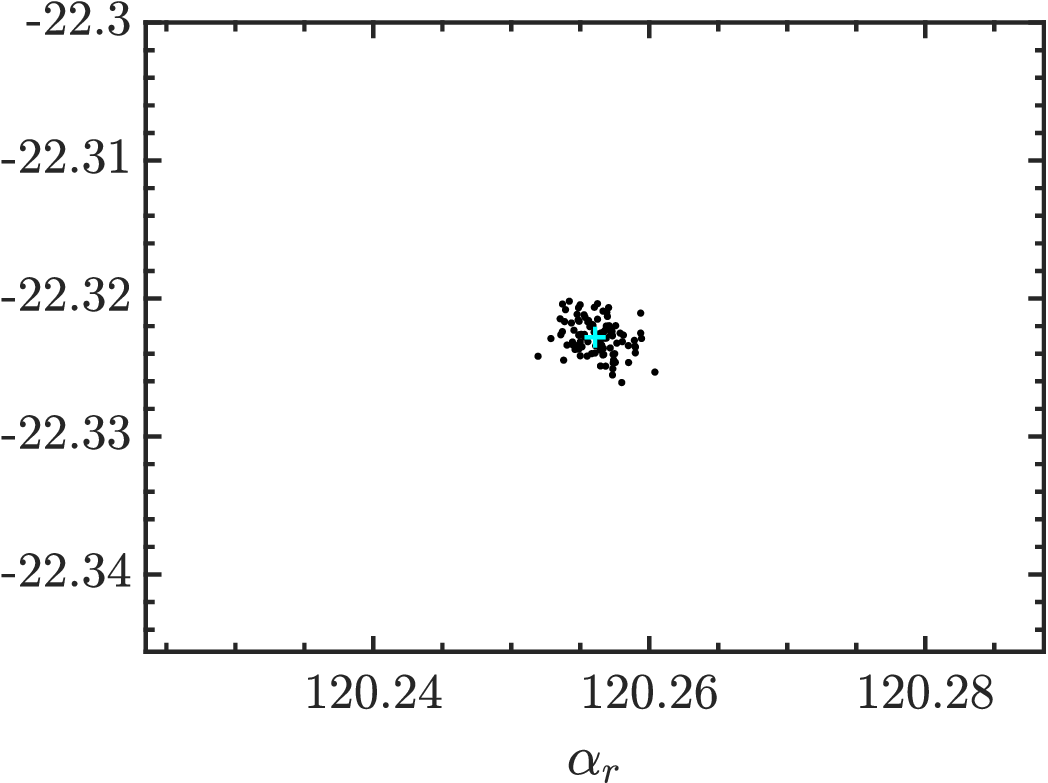}
\put(-146,118){({\it b})}
\caption{ (a) The eigenvalue shift due to $\delta\boldsymbol{Q}_{\rm{max}}$ (red dots), the term distorted is $\partial\mu/\partial\rho$; (b) Similar to panel (a) but with $\delta\boldsymbol{Q}_{\rm{noise}}$ at the same amplitude (measured with 2-norm).}
\label{fig6}
\end{figure}

To better understand the eigenvalue shift induced by $\delta\boldsymbol{Q}_{\rm{max}}$ relative to $\delta\boldsymbol{Q}_{\rm{noise}}$ of the same amplitude, we present the comparison in Figure \ref{fig6}. As seen in panel (a), the distortion
\begin{equation}
\delta\boldsymbol{Q} = \rm{real}(\boldsymbol{S}_{\boldsymbol{Q}} \exp{i\theta}),
\end{equation}
with $0\leq\theta<2\pi$, leads to an elliptic trajectory of the eigenvalue. The eigenvalues corresponding to $\delta\boldsymbol{Q}$ ($\theta=0,\pi/2,\pi,3\pi/2$) are highlighted with red dots. When the coherence of the distortions is lost and replaced with random noise, the eigenvalue shifts are significantly narrowed, as shown in panel (b), which corresponds to the central rectangle highlighted in panel (a). In both cases, we have maintained an amplitude of
\begin{equation}
\left\Vert\delta\boldsymbol{Q}\right\Vert_{2}=10^{-3}/\sqrt{N},~\mathrm{with}~N=200.
\end{equation}
In contrast to panel (a), where the eigenvalue shift is ``optimized'' along an elliptical trajectory, the introduction of random distortion causes the eigenvalue to remain closer to its undistorted value, with the displacement appearing random as well. This difference aligns with the sensitivity coefficient defined in equation \eqref{eq_ss}. Consequently, within the linear sensitivity regime, a random distortion is less likely to cause significant deviations in the eigenvalue.

\begin{figure}
\centering
\includegraphics[width=0.8\linewidth]{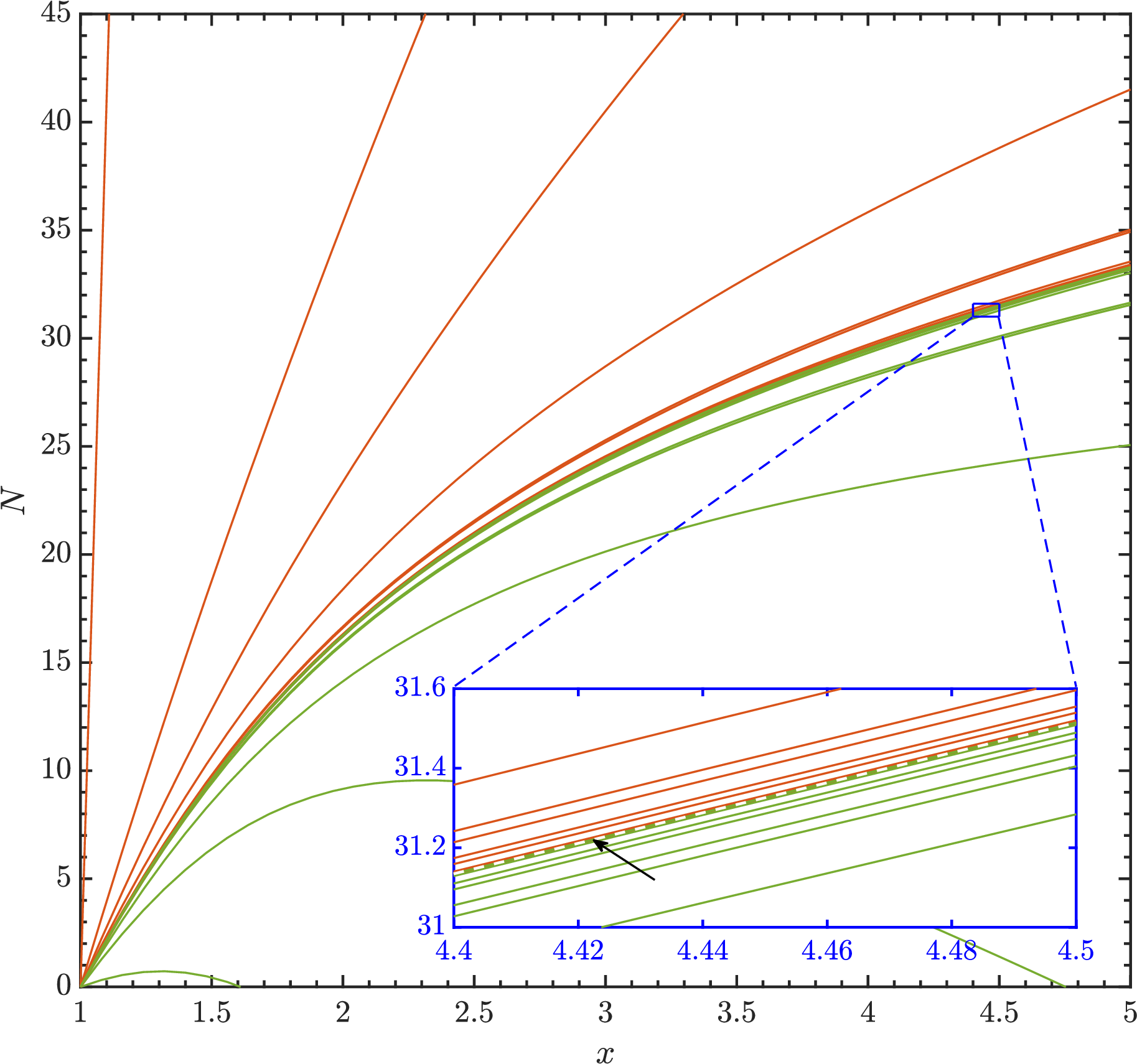}
\put(-277, 213){1}
\put(-213, 213){2}
\put(-168, 213){3}
\put(-120, 213){4}
\put(-60, 213){5, 6}
\put(-170, 103){$\overbrace{\;\qquad\qquad\qquad\qquad\qquad\qquad\qquad\qquad}$}
\put(-108, 116){$7 \cdots 23$}
\put(-131,45){\rotatebox{13}{undistorted}}
\caption{N factor influenced by $\delta\boldsymbol{Q}_{\rm{max}}$. The amplitude of an individual distortion is $\left\Vert\delta\boldsymbol{Q}\right\Vert_{2}=10^{-3}/\sqrt{200}$. The red and green curves show positive and negative shifts, respectively. The numbers $1\cdots23$ correspond to input terms (see equation \eqref{eq_L}) sorted by the impact on the N factor in descending order (see table \ref{table_Nfac}). The instability is in the pseudo-boiling regime with dominating inviscid instability.}
\label{fig7}
\end{figure}
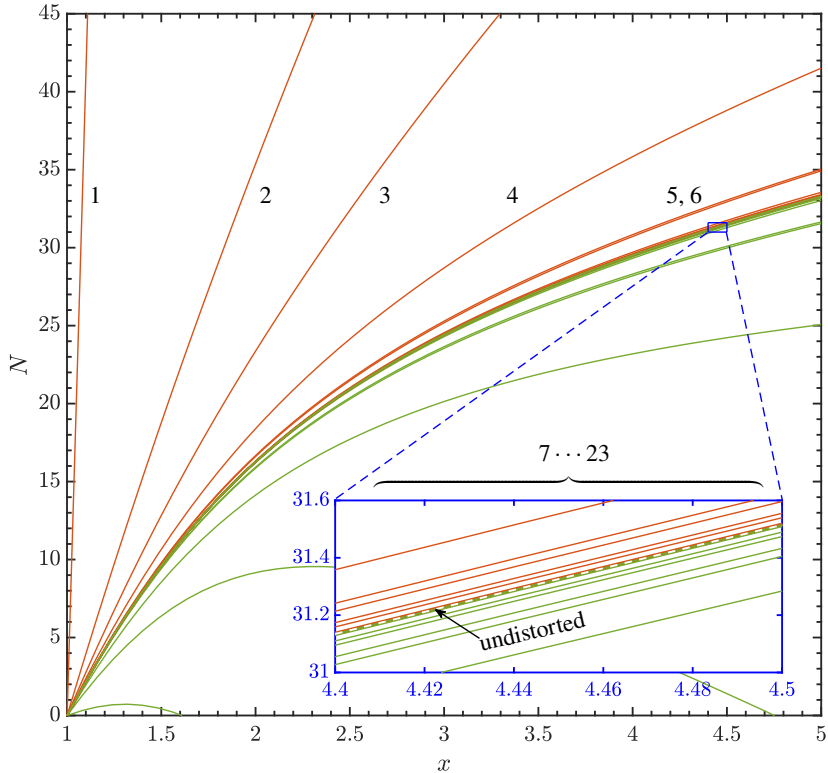

So far, we have examined the eigenvalue shift at a fixed chordwise position (corresponding to $x=1$). To account for the integral effects and gain an initial understanding of the sensitivity to different inputs, we plot the N factor (integral of the growth rate) as a function of the streamwise coordinate in Figure \ref{fig7}. Twenty-three terms corresponding to equation \eqref{eq_L} have been calculated. The distortion corresponds to $\delta\boldsymbol{Q}_{\rm{max}}$ with a 2-norm of $10^{-3}/\sqrt{200}$. Recalling the inviscid nature of the new mode (connected to the main flow), one notices that the growth rate is significantly larger than that of conventional (also largely inviscid) cross-flow modes (connected to the true cross-flow $w_s$) and reaches an N factor of 33.3 at $x=5$. Temperature distortions lead to a dramatic N factor change, followed by density and $\partial p/\partial \rho$, which agrees with the local amplitude seen in Figure \ref{fig5}. We have tabulated the top seven terms in Table \ref{table_Nfac}. If the temperature and density are correctly obtained in the baseflow stage, ensuring an accurate EOS is essential for a rational transition prediction.

\begin{table}
\begin{center}
\begin{tabular}{l@{\hskip 10mm}r@{\hskip 5mm}r@{\hskip 5mm}r@{\hskip 5mm}r}
 						&	$x = 2.04$  &  	$x=3.08$ 	& 	$x=4.04$   & 	$x=5.00$  \\[2mm]
1. $T$		 			& 	2947.1\%	  &	4504.6\%	&	5801.6\%	&	6938.3\% \\
2. $\rho$					&	119.4\%	  &	167.2\%	&	205.1\%	&     	237.1\%  \\
3. $\partial p/\partial \rho$		&	44.6\%	  &	67.7\%	&	87.1\%	&	104.2\%  \\
4. $\partial^2 p/\partial \rho^2$	&	13.3\%	  & 	17.9\%	&	21.6\%	&	24.7\%    \\	
5. $u$	 				& 	2.5\%	  &    	3.6\%	&	4.5\%	&	5.2\%    \\
6. $w$ 					&	2.2\%	  &	3.2\%	&	4.1\%	&	4.9\%    \\
7. $\partial p/\partial T$ 		&	0.4\%	  &	0.5\%	&	0.7\%	&	0.8\%	
\end{tabular}
\caption{$\delta N/N$ at four uniformly distributed observation points.}
\label{table_Nfac}
\end{center}
\end{table}

The result has also been tested as a weak function of the spanwise wavenumber $\beta$, as shown in Appendix \ref{appBeta}. Based on the above discussions, we propose a scalar measure of sensitivity:
\begin{equation}\label{measure}
M=\frac{\left|\delta\alpha_{i}\right|}{\left\Vert \delta\boldsymbol{Q}_{\mathrm{max}}\right\Vert_2 }.
\end{equation}
This measure demonstrates the maximum possible response of the growth rate to distortions scaled by its normalized norm, thereby indicating the sensitivity to certain inputs. Note that the choice of the norm is not exclusive. See also the discussion for baseflow distortions \eqref{eqnorm}. The normalized norm provides an intuitive indicator of the sensitivity amplitude which does not influence the sensitivity coefficients and the induced eigenvalue distortions. Instead of concentrating on the shape of the sensitivity coefficients, $M$ helps to compare across different cases and regimes.

\subsection{Sensitivity crossing different thermodynamic regimes}\label{S3c}

\begin{figure}
\centering
\includegraphics[scale=0.4]{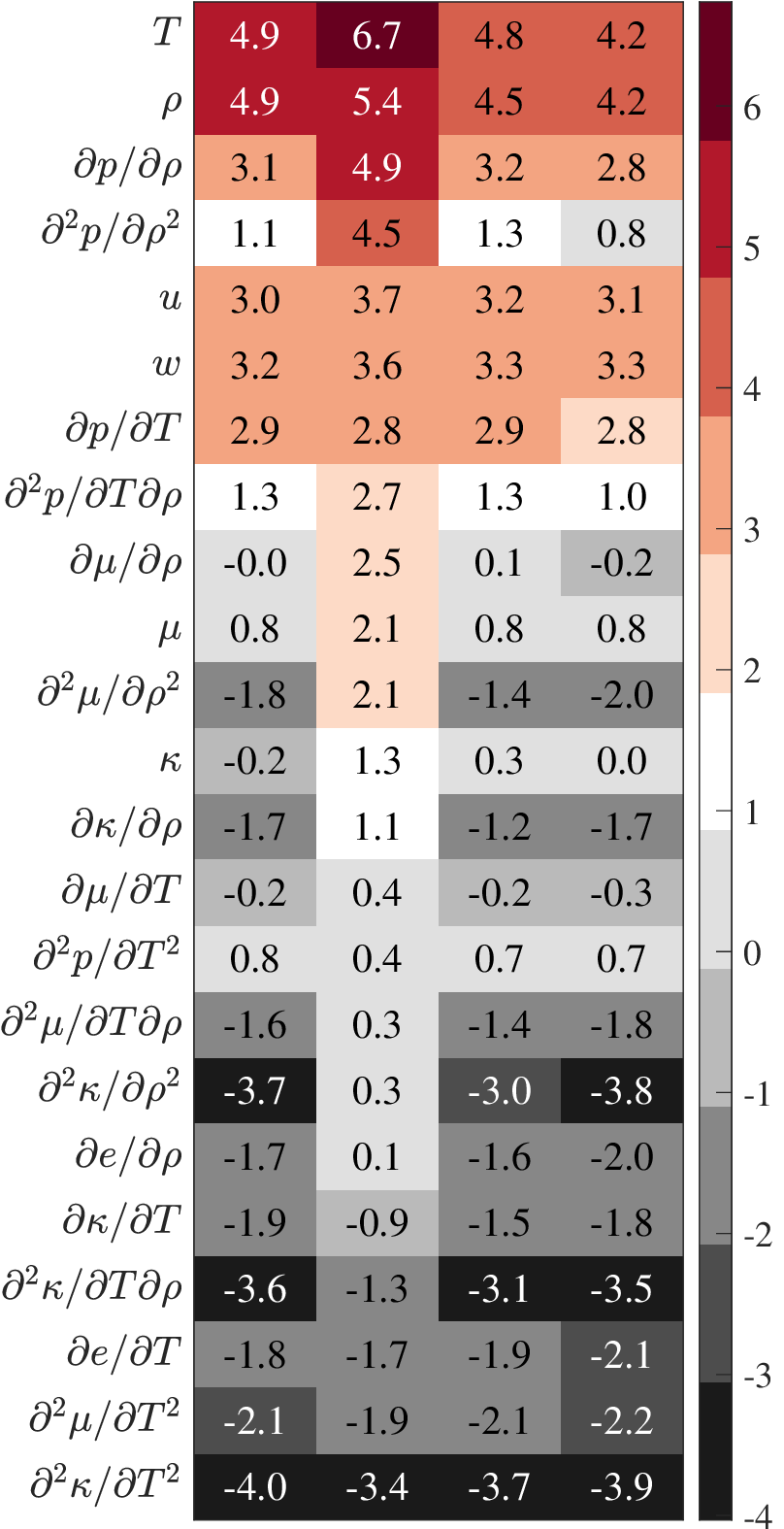}\hspace{1cm}
\includegraphics[scale=0.4,trim={0 0 0 0},clip]{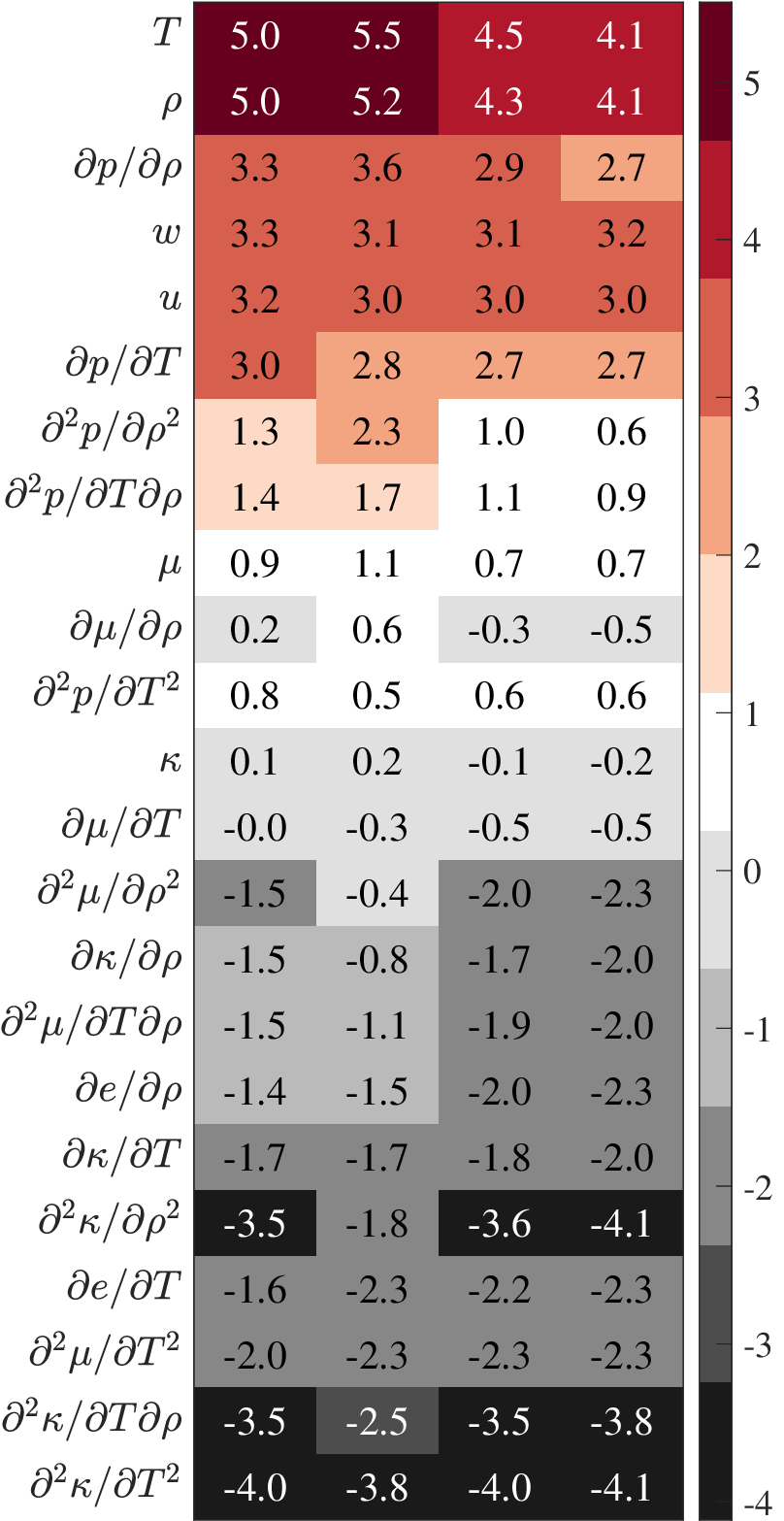}
\put(-299, -12){sub}
\put(-276, -12){trans}
\put(-250, -12){sup}
\put(-227, -12){ideal}
\put(-292, -13){$\underbrace{\;\qquad\qquad\qquad\qquad}$}
\put(-268, -30){cooling}
\put(-112, -12){sub}
\put(-89, -12){trans}
\put(-63, -12){sup}
\put(-40, -12){ideal}
\put(-105, -13){$\underbrace{\;\qquad\qquad\qquad\qquad}$}
\put(-81, -30){heating}
\caption{A comparison of sensitivity across different regimes and between different input variables. The numbers on the heatmap specify the log value of the sensitivity measure: $\log(M)$.}
\label{fig8}
\end{figure}
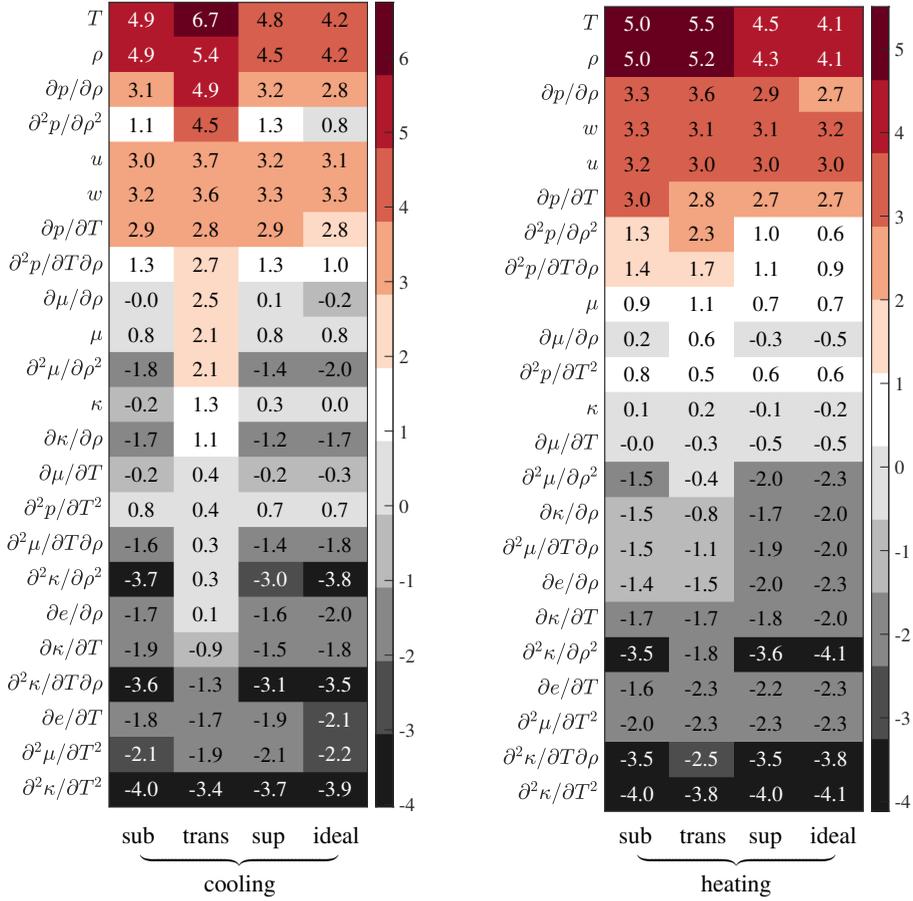


The non-ideal gas properties primarily depend on the thermodynamic regimes of the fluid. We compare the stability diagram and the sensitivity measure $M$  in Figure \ref{fig8} for regimes introduced in Table \ref{table1}, grouped into wall heating and cooling. To allow for comparison across different regimes and without loss of generality, the parameters ($x=1.0$, $\omega=15$, $\beta=80$) have been chosen for the investigations below \citep[see figure 2 of][]{ren2022non}. According their panels (b) and (c), the neutral curves and growth rates are only quantitatively different, except for the transcritical regime with wall cooling, where the instability is dominated by a new inviscid mode. In figure \ref{fig8}, the term ``distorted" is listed on the figure's left and sorted according to the sensitivity measure in the transcritical regime. Considering a representative growth rate of $O(1)$, a sensitivity measure of $M =1$ ($\log M =0$) will give $\delta\alpha_i=0.001$ at a distortion amplitude of $10^{-3}$, which starts to become visible. Note that the shape of the distortion was assumed to be the best case (recall Figure \ref{fig6}). With this estimation, the sensitivity to distortions with negative values of $\log (M)$ is minimal.

Going through the sensitivity measure of all eight cases presented in Figure \ref{fig8} implies an analogous ranking of various terms, showing that different flow regimes only moderately influence the relative position of each input of the stability operator. The temperature and density profiles are the most consequential. Subsequently, the term $\partial^2p/\partial\rho^2$ stands out only for the pseudo-boiling case with wall cooling. For the ideal regime, baseflow profiles remain more influential than other thermodynamic and transport properties. Some terms are zero (\eg $\partial \mu/\partial \rho$, $\partial \kappa/\partial \rho$) per the ideal assumption, and random or structural distortions with moderate amplitude will not invoke a significant shift in the growth rate, indicating their less active roles in the system. We notice that the transport properties (\V~and \T~profiles) are less influential for all the regimes. This implies that the ideal case will be robust provided the six key terms, $T,\rho,u,w,\partial p/\partial T, \partial p/\partial \rho$, are not distorted. On the contrary, the pseudo-boiling regimes are sensitive to more terms: $\partial^2p/\partial\rho^2, \partial^2p/\partial\rho\partial T$ for wall heating and, in addition, $\partial\mu/\partial\rho, \mu, \partial^2\mu/\partial\rho^2, \kappa$ for wall cooling.

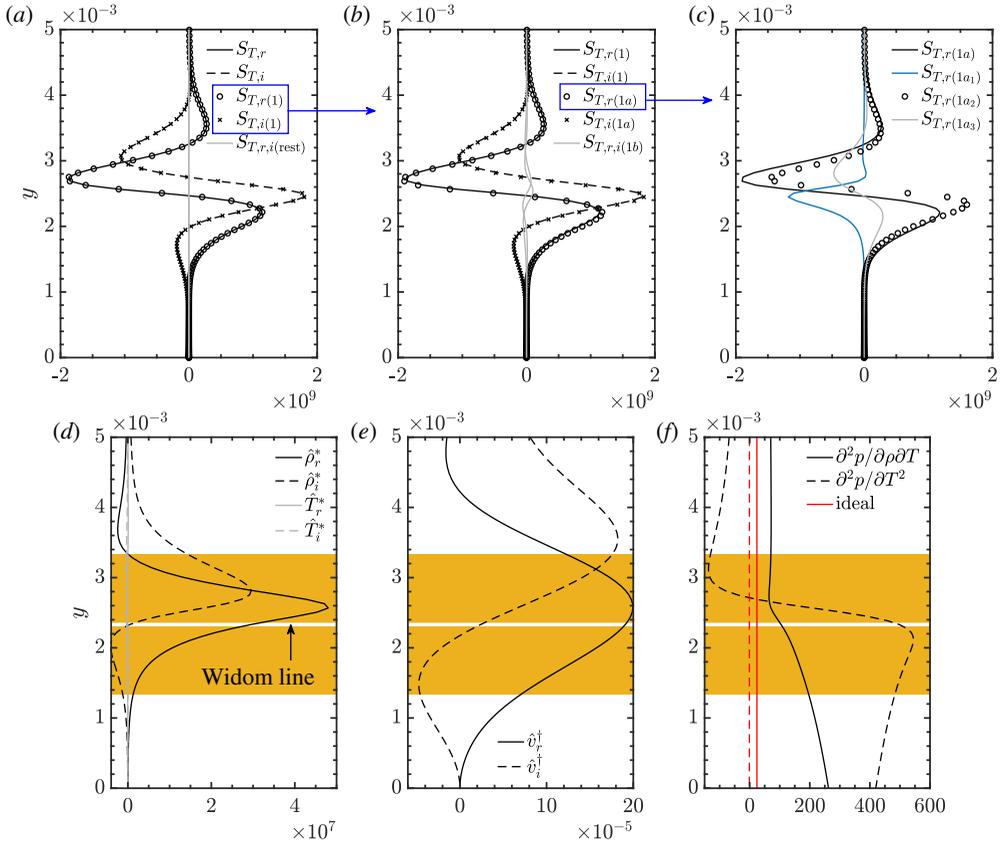
\begin{figure}
\centering
\includegraphics[scale=0.33]{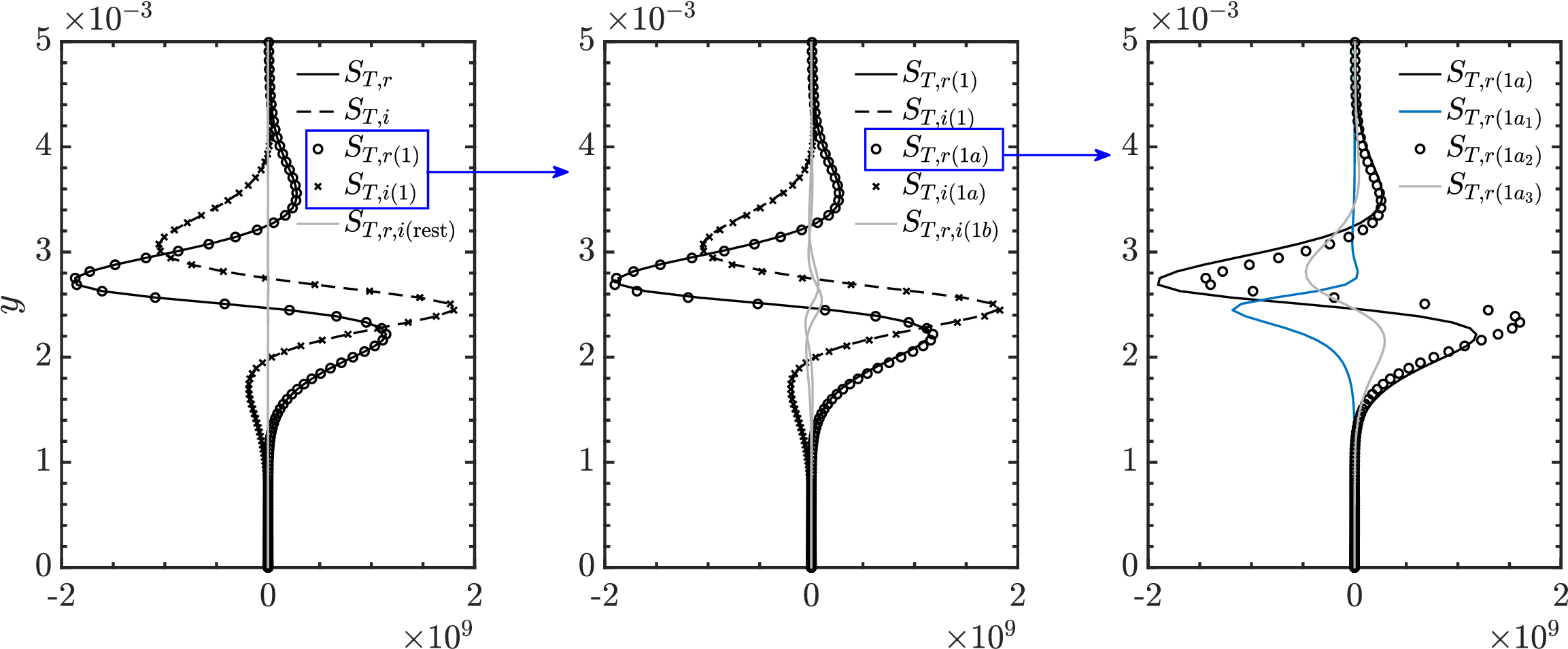}
\put(-373,145){({\it a})}
\put(-246,145){({\it b})}
\put(-116,145){({\it c})}
\\
\includegraphics[scale=0.33]{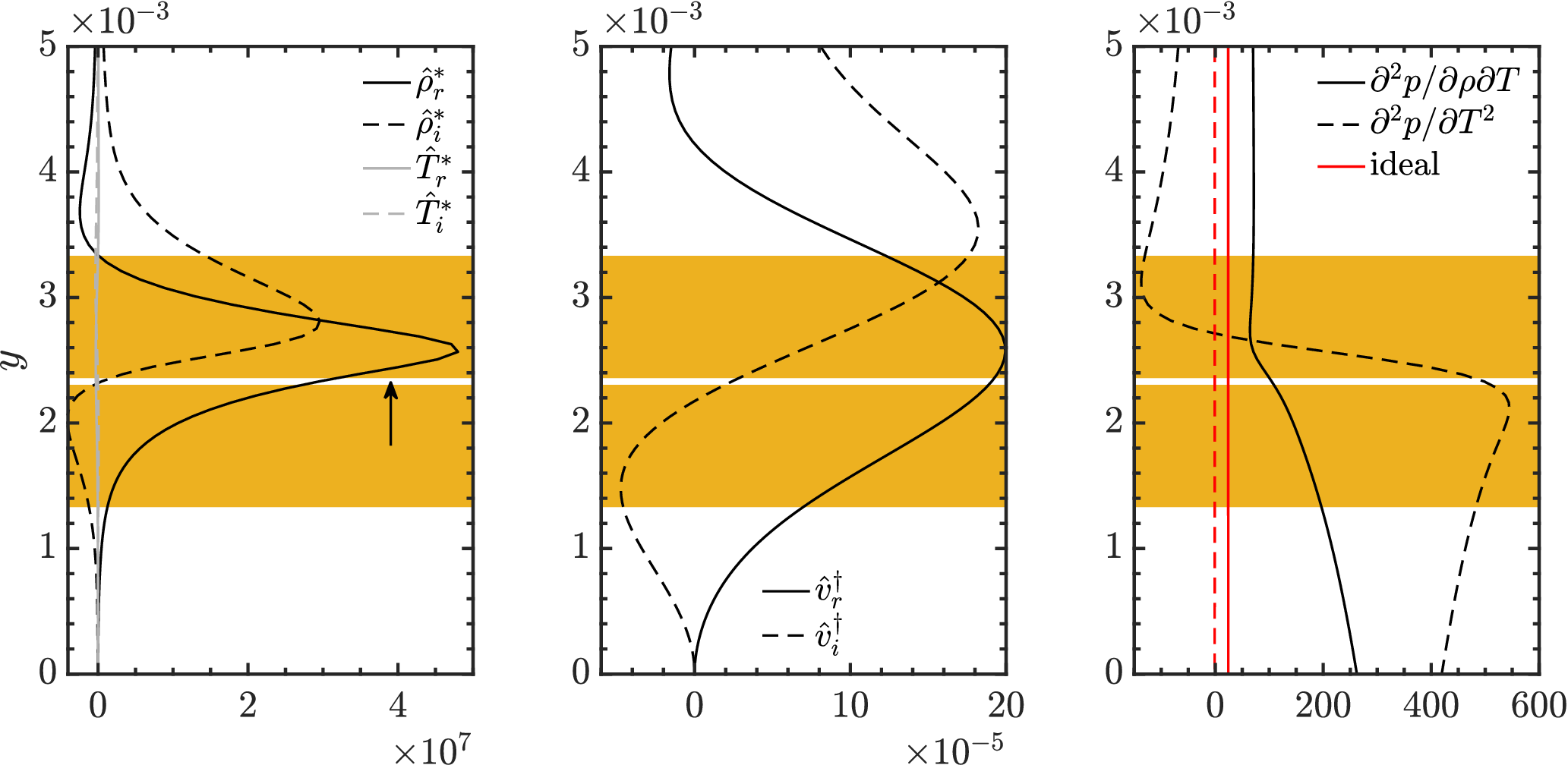}
\put(-280, 58){Widom line}
\put(-336,150){({\it d})}
\put(-224,150){({\it e})}
\put(-109,150){({\it f})}
\caption{Illustration of key terms for the sensitivity profile $S_T$ corresponding to equation \eqref{eq_ST2}. (a) $S_T$ and $S_T(1)$; (b) $S_T(1a)$ and $S_T(1b)$; (c) $S_T(1a_1)$, $S_T(1a_2)$ and $S_T(1a_3)$; \\(d -- f) The components composing $S_T(1)$.}
\label{fig9}
\end{figure}

\subsection{Magnitude range of sensitivities to various inputs}
The sensitivity profiles (figure \ref{fig5}) and their measure (figure \ref{fig8}) demonstrate a magnitude range of about $O(10^{10})$. The reason is explored here by analyzing different sensitivity profiles and their mathematical constituents. We examine the pseudo-boiling case with wall cooling. From equation \eqref{eq_ST}, the dominating terms are written as
\begin{equation}\label{eq_ST2}
\begin{aligned}
S_{T}= & \underset{S_{T}\left(1\right)}{\underbrace{\frac{\partial}{\partial y}\left(\frac{\partial^{2}p}{\partial\rho\partial T}\hat{\rho}^{*}\hat{v}^{\dagger}+\frac{\partial^{2}p}{\partial T^{2}}\hat{T}^{*}\hat{v}^{\dagger}\right)}}+\rm{visc.}\\
= & \underset{S_{T}\left(1a\right)}{\underbrace{\frac{\partial}{\partial y}\left(\frac{\partial^{2}p}{\partial\rho\partial T}\hat{\rho}^{*}\hat{v}^{\dagger}\right)}}+\underset{S_{T}\left(1b\right)}{\underbrace{\frac{\partial}{\partial y}\left(\frac{\partial^{2}p}{\partial T^{2}}\hat{T}^{*}\hat{v}^{\dagger}\right)}}+\rm{visc.}\\
= & \underset{S_{T}\left(1a_{1}\right)}{\underbrace{\frac{\partial}{\partial y}\left(\frac{\partial^{2}p}{\partial\rho\partial T}\right)\hat{\rho}^{*}\hat{v}^{\dagger}}}+\underset{S_{T}\left(1a_{2}\right)}{\underbrace{\frac{\partial^{2}p}{\partial\rho\partial T}\frac{\partial\hat{\rho}^{*}}{\partial y}\hat{v}^{\dagger}}}+\underset{S_{T}\left(1a_{3}\right)}{\underbrace{\frac{\partial^{2}p}{\partial\rho\partial T}\hat{\rho}^{*}\frac{\partial\hat{v}^{\dagger}}{\partial y}}}+S_T(1b)+\rm{visc.}
\end{aligned}
\end{equation}
with ``visc'' standing for viscous terms. The distribution of key terms are plotted in figure \ref{fig9}. As inferred from figure \ref{fig9} (a -- c), the terms $S_{T}(1)$ significantly dominate the sum of the rest terms, which are viscous (scaled by $\Rey=1.4687\times10^5$). Among $S_{T}(1)$, $S_{T}(1a)$ is much larger than $S_{T}(1b)$. We plot in panels (d -- f) the physical terms related to $S_{T}(1)$. As can be seen, the gradients of the terms  $\hat{\rho}^*$, $\hat{v}^\dagger$ and $\partial^2 p/\partial\rho\partial T$ \emph{synchronize} around the Widom line, leading to the large amplitude of $S_T$.

%
\begin{figure}
\centering
\includegraphics[scale=0.33]{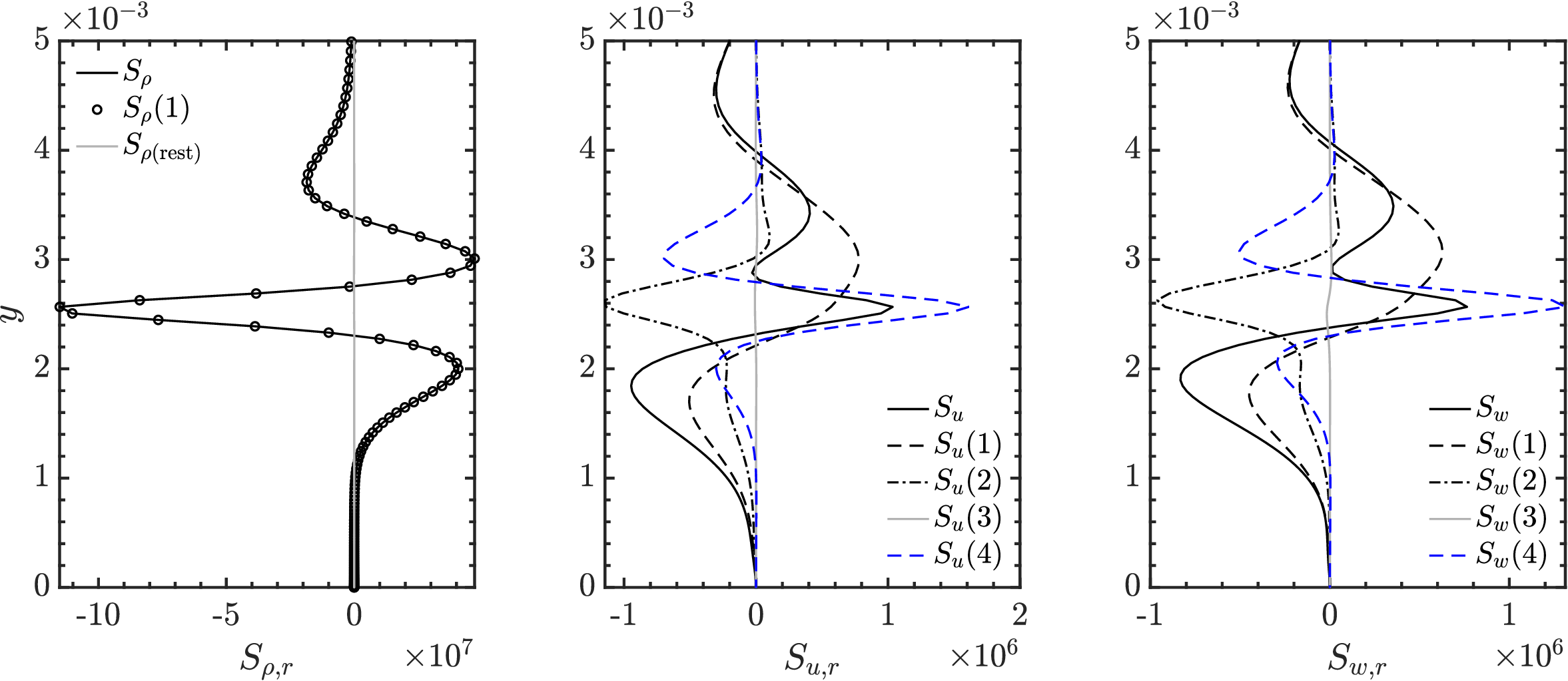}
\put(-379,160){({\it a})}
\put(-249,160){({\it b})}
\put(-119,160){({\it c})}
\caption{Sensitivity to distortions of $\rho$, $u$ and $w$ (panel a -- c). We show dominating terms (the real parts) of each profile.}
\label{fig10}
\end{figure}

Are viscous terms always negligible? The composition of $S_\rho$, $S_u$ and $S_w$ are plotted in figure \ref{fig10}. Similar to $S_T$, $S_\rho$ is dominated by an inviscid term
\begin{equation}
S_{\rho}=	\underset{S_{\rho}\left(1\right)}{\underbrace{\frac{\partial}{\partial y}\left(\frac{\partial^{2}p}{\partial\rho^{2}}\hat{\rho}^{*}\hat{v}^{\dagger}+\frac{\partial^{2}p}{\partial\rho\partial T}\hat{T}^{*}\hat{v}^{\dagger}\right)}}+\cdots
\end{equation}
Compared with equation \eqref{S_D}, the rest of the terms include viscous and inviscid terms. However, a single dominating term does not exist for $S_u$ and $S_w$. We re-write these two profiles in \eqref{eq_Su2} and \eqref{eq_Sw2}. As seen in figure \ref{fig10}, the viscous terms influence the profile in the leading order (only $S_u(3)$ \& $S_w(3)$ can be ignored),   showing the significance of viscous terms though the Reynolds number scales them. 

\begin{equation}\label{eq_Su2}
\begin{aligned}
S_{u}= & \underset{S_{u}\left(1\right)}{\underbrace{\frac{\partial}{\partial y}\left(\rho\hat{v}^{*}\hat{u}^{\dagger}\right)}}+\underset{S_{u}\left(2\right)}{\underbrace{i\alpha^{*}\left(\hat{\rho}^{*}\hat{\rho}^{\dagger}+\rho\hat{u}^{*}\hat{u}^{\dagger}+\rho\hat{v}^{*}\hat{v}^{\dagger}+\rho\hat{w}^{*}\hat{w}^{\dagger}+\rho\frac{\partial e}{\partial\rho}\hat{\rho}^{*}\hat{T}^{\dagger}+\rho\frac{\partial e}{\partial T}\hat{T}^{*}\hat{T}^{\dagger}\right)}}\\
 & +\underset{S_{u}\left(3\right)}{\underbrace{\frac{1}{Re}\frac{\partial}{\partial y}\left[2\mu\left(i\alpha^{*}\hat{v}^{*}-\frac{\partial\hat{u}^{*}}{\partial y}\right)\hat{T}^{\dagger}+\frac{\partial\mu}{\partial T}\hat{T}^{*}\left(\frac{\partial\hat{u}^{\dagger}}{\partial y}-2\frac{\partial u}{\partial y}\hat{T}^{\dagger}+i\alpha^{*}\hat{v}^{\dagger}\right)\right]}}\\
 & +\underset{S_{u}\left(4\right)}{\underbrace{\frac{1}{Re}\frac{\partial}{\partial y}\left[\frac{\partial\mu}{\partial\rho}\hat{\rho}^{*}\left(\frac{\partial\hat{u}^{\dagger}}{\partial y}-2\frac{\partial u}{\partial y}\hat{T}^{\dagger}+i\alpha^{*}\hat{v}^{\dagger}\right)\right]}}
\end{aligned}
\end{equation}

\begin{equation}\label{eq_Sw2}
\begin{aligned}
S_{w}= & \underset{S_{w}\left(1\right)}{\underbrace{\frac{\partial}{\partial y}\left(\rho\hat{v}^{*}\hat{w}^{\dagger}\right)}}+\underset{S_{w}\left(2\right)}{\underbrace{i\beta\left(\hat{\rho}^{*}\hat{\rho}^{\dagger}+\rho\hat{u}^{*}\hat{u}^{\dagger}+\rho\hat{v}^{*}\hat{v}^{\dagger}+\rho\hat{w}^{*}\hat{w}^{\dagger}+\rho\frac{\partial e}{\partial\rho}\hat{\rho}^{*}\hat{T}^{\dagger}+\rho\frac{\partial e}{\partial T}\hat{T}^{*}\hat{T}^{\dagger}\right)}}\\
 & +\underset{S_{w}\left(3\right)}{\underbrace{\frac{i\beta}{Re}\frac{\partial}{\partial y}\left(2\mu\hat{v}^{*}\hat{T}^{\dagger}+\frac{\partial\mu}{\partial\rho}\hat{\rho}^{*}\hat{v}^{\dagger}+\frac{\partial\mu}{\partial T}\hat{T}^{*}\hat{v}^{\dagger}\right)}}\\
 & +\underset{S_{w}\left(4\right)}{\underbrace{\frac{1}{Re}\frac{\partial}{\partial y}\left[\frac{\partial\mu}{\partial T}\hat{T}^{*}D\hat{w}^{\dagger}+\frac{\partial\mu}{\partial\rho}\hat{\rho}^{*}D\hat{w}^{\dagger}-2\left(\mu D\hat{w}^{*}\hat{T}^{\dagger}+\frac{\partial\mu}{\partial\rho}\frac{\partial w}{\partial y}\hat{\rho}^{*}\hat{T}^{\dagger}+\frac{\partial\mu}{\partial T}\frac{\partial w}{\partial y}\hat{T}^{*}\hat{T}^{\dagger}\right)\right]}}
\end{aligned}
\end{equation}

\begin{figure}
\centering
\includegraphics[scale=0.3]{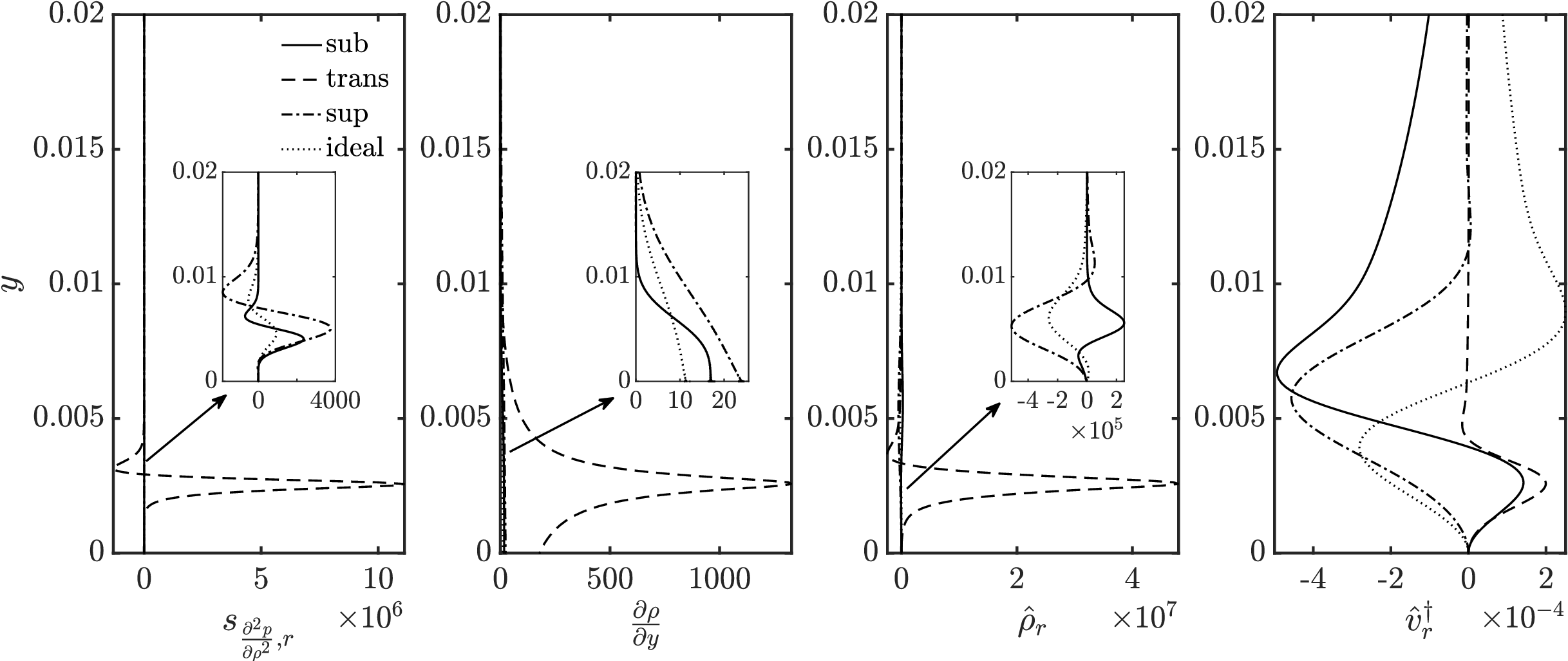}
\put(-350,154){({\it a})}
\put(-260,154){({\it b})}
\put(-172,154){({\it c})}
\put(-85,154){({\it d})}
\\
\vspace{2mm}
\includegraphics[scale=0.3]{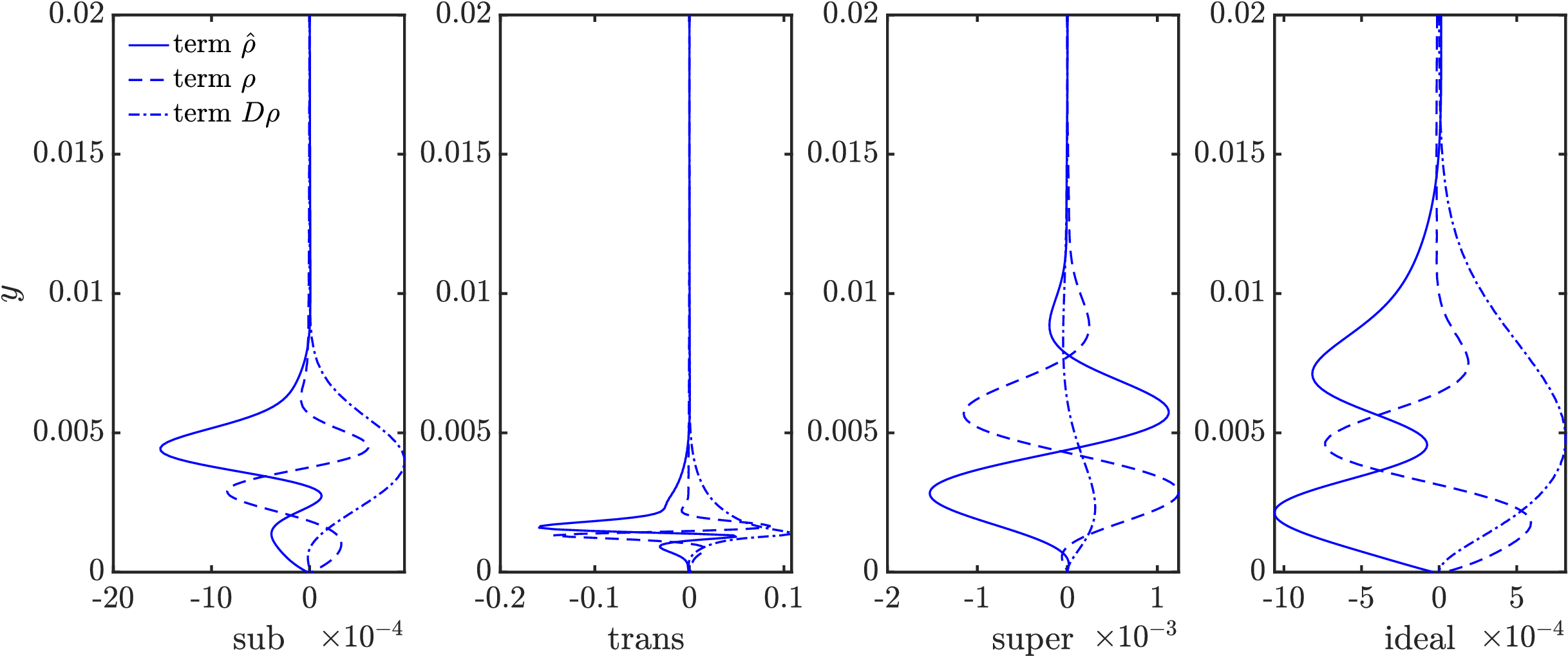}
\put(-350,154){({\it e})}
\put(-260,154){({\it f})}
\put(-172,154){({\it g})}
\put(-85,154){({\it h})}
\caption{Sensitivity (the real part) to $\partial^2 p/\partial \rho^2$ with wall cooling (a). The sensitivity equals the derivatives of the products of three terms shown in panels (b) - (d). We illustrate the balance of term $\hat{\rho}$, ${\rho}$ and $D{\rho}$ according to the continuity equation \eqref{eq_conti} in panels (e) - (h) for the four regimes considered (with wall cooling).}
\label{fig11}
\end{figure}
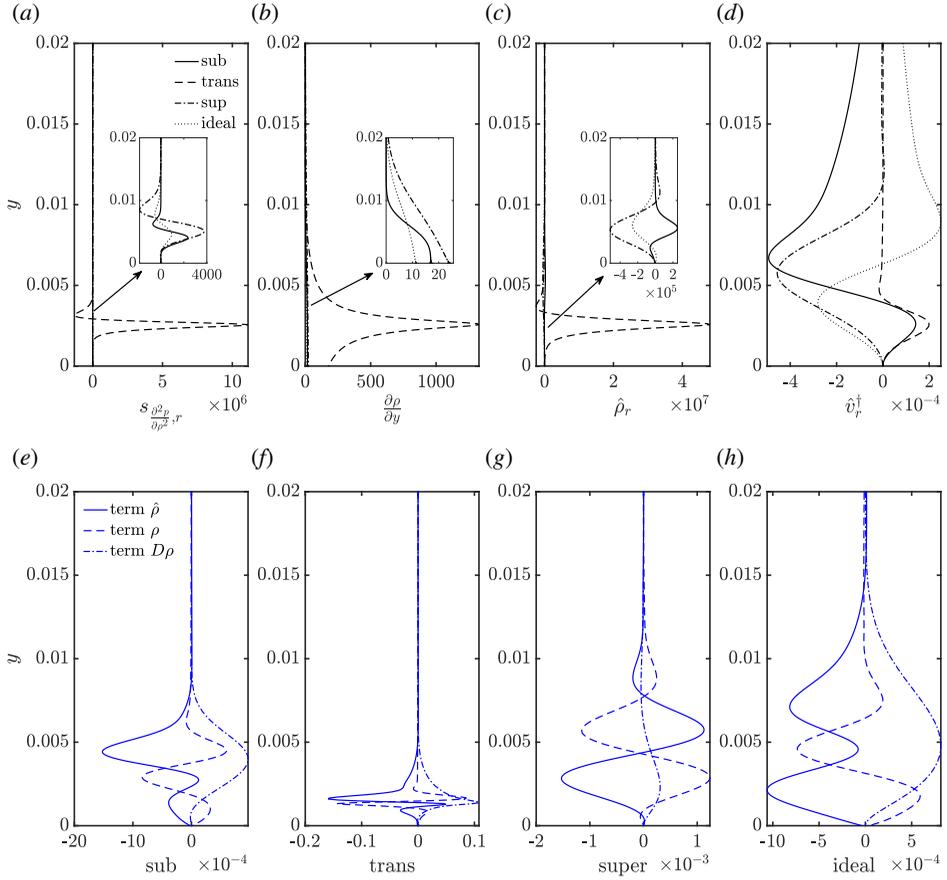

Take a column-wise comparison on figure \ref{fig8}, the sensitivity to $\partial^2 p/\partial \rho^2$ stands out in the pseudo-boiling regime with wall cooling. Recall \eqref{S_EOS}, the sensitvity coefficient amounts to a single term  $-{\partial\rho}/{\partial y}(\hat{\rho}^{*}\hat{v}^{\dagger})$. In figure \ref{fig11}, we compare the real parts of these terms across different regimes. From panel (a), we observe that the terms in the pseudo-boiling case is four orders of magnitudes larger than in other regimes, in accordance with figure \ref{fig8}. Panels (b) to (d) unveil that the differences are due to $\partial\rho/\partial y$ and $\hat{\rho}$. Both terms are largely two orders larger than the rest of the cases. In fact, according to the continuity equation, 
\begin{equation}\label{eq_conti}
\underset{\mathrm{term\:}D\rho}{\underbrace{\hat{v}\frac{\partial\rho}{\partial y}}}+\underset{\mathrm{term\:}\hat{\rho}}{\underbrace{\left(i\alpha u-i\omega+i\beta w\right)\hat{\rho}}}+\underset{\mathrm{term\:}\rho}{\underbrace{\left(i\alpha\hat{u}+D\hat{v}+i\beta\hat{w}\right)\rho}}=0
\put(-195,-28){\rotatebox{13}{$\longrightarrow$}}
\put(-198,-35){\rotatebox{13}{\scriptsize (leads to)}}
\end{equation}
the term $D\rho$ is balanced by terms $\rho$ and $\hat{\rho}$ . As seen from panel (e-h) of figure \ref{fig11}, all three terms are in the same order of magnitude regardless of the thermodynamic regime. Simultaneously, juxtaposing the four panels reveals that the pseudo-boiling regime has a significantly larger amplitude of the three balancing terms according to equation \eqref{eq_conti}, clarifying the differences observed in panel (a). The non-ideal gas behaviour constitutes a quadratic effect, magnifying the influence of $\partial\rho/\partial y$ on sensitivities into its square (through the product with term $\hat{\rho}$). At the same time, we notice that the thermodynamic derivatives (\eg $\partial^2p/(\partial\rho\partial T)$ and $\partial^2p/\partial T^2$) and their wall-normal gradients are significantly larger in the pseudo-boiling regime (see figure \ref{fig9}f). Multiplication with $\hat{\rho}$ leads to tremendous sensitivity to temperature and density distortions.

\begin{figure}
\centering
\includegraphics[scale=0.35]{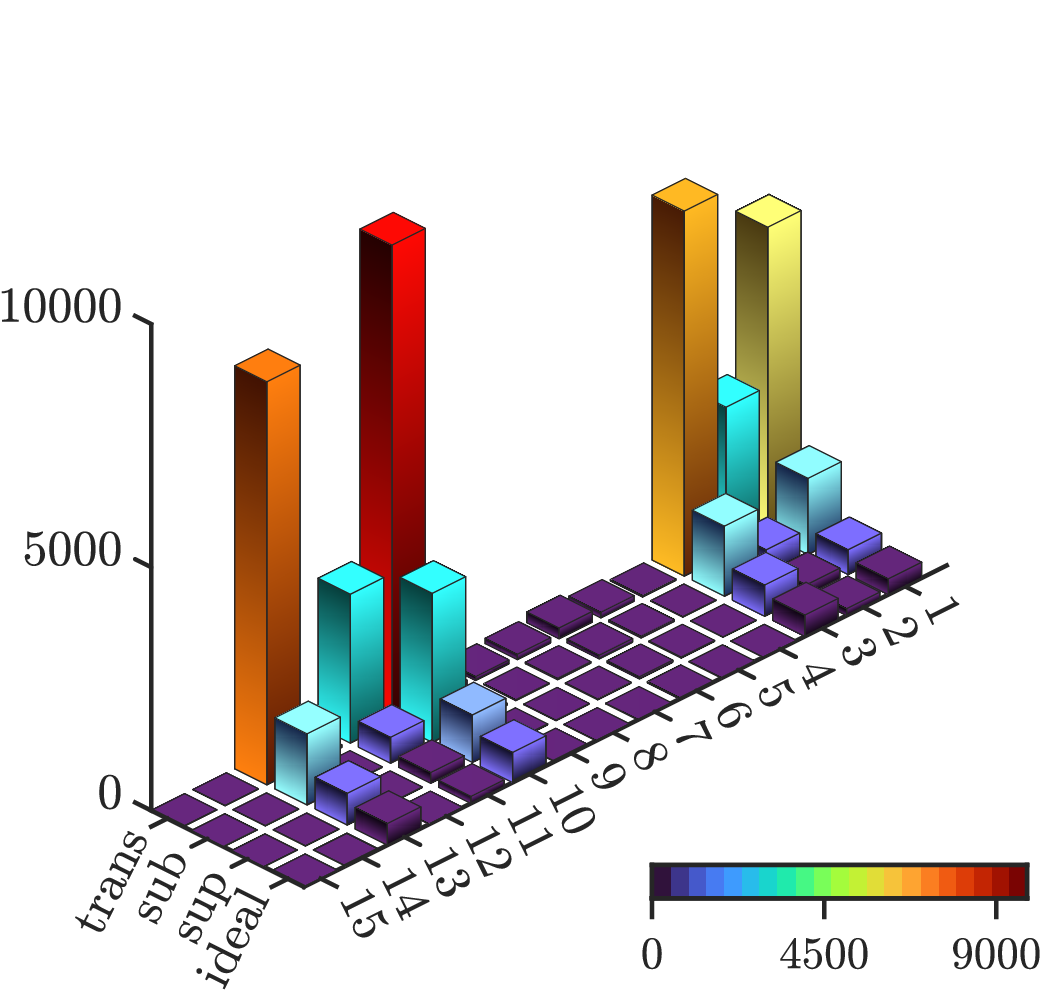} \hspace{5mm}
\includegraphics[scale=0.35]{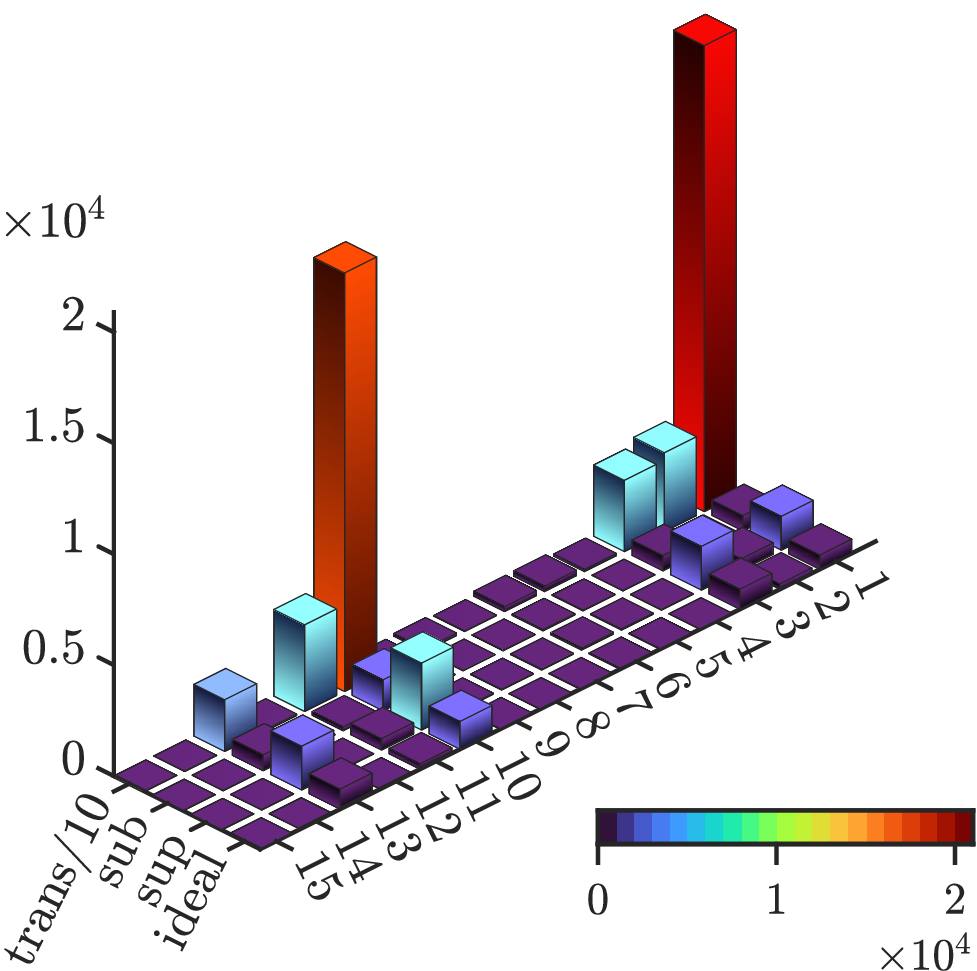}
\put(-360,140){({\it a})}
\put(-165,140){({\it b})}
\caption{Exposition of sensitivity terms for $\partial \mu/\partial \rho$ with wall heating (a) and cooling (b). The terms corresponding to \eqref{S_mu_D} have been labeled as 1 -- 15 on the plot. The height of each pillar stands for the 2-norm of a term.}
\label{fig12}
\end{figure}

As another example, the sensitivity to $\partial\mu/\partial\rho$ is studied according to equation \eqref{S_mu_D}. The sensitivity coefficients contain 15 terms, which are presented in figure \ref{fig12}, comparing across different thermodynamic regimes. Since the amplitude for the pseudo-boiling case is more prominent in the wall cooling case, we have scaled their amplitudes with a factor of 10, as indicated on the axis of panel (b). With wall cooling, we find term 1 and 10 stand out in the transcritical case, say, $\frac{\partial u}{\partial y}D\hat{\rho}^{*}\hat{u}^{\dagger}$ and $\frac{\partial w}{\partial y}D\hat{\rho}^{*}\hat{w}^{\dagger}$, both terms are led by $D\hat{\rho}^{*}$. This again attributes to the variation of the density near the Widom line. We also noticed that some terms (4 -- 9 and 12, 14 and 15) remain small for all the cases.

Taking an overview of the sensitivity coefficients, it is unambiguous that the sensitivity can be uniformly written as:
\begin{equation}
{\rm Sensitivity}=\mathlarger{\mathlarger{\Sigma}}
\colorbox{Gray!30}{wave number}
\cdot
\colorbox{green!10}{$\dfrac{1}{\Rey,\Pran\Ec}$} 
\cdot
\colorbox{RedOrange!10}{$\left(D,D^2\right)q$} 
\cdot
\colorbox{red!20}{$\left(D,D^2\right)\hat{q}^*$} 
\cdot
\colorbox{blue!10}{$\left(D,D^2\right)\hat{q}^\dagger$} 
\put(-258, 27){optional, profile dependent}
\put(-281, 14){$\overbrace{\hspace{5.8cm}}$}
\put(-161, -8){$\underbrace{\hspace{3.6cm}}$}
\put(-163, -26){quadratic effect, $D\rho \rightarrow \hat{\rho}$}
\put(-144, -39){(Pseudo-boiling)}
\end{equation}
where $q$, $\hat{q}^*$ and $\hat{q}^\dagger$ stand for the baseflow, the eigenfunction (complex conjugate) and the adjoint eigenfunction inclusive of their spatial gradients. The sensitivity is fully viscous for the viscosity and thermal conductivity while mixed for the baseflow and EOS profiles, inducing a significant decrease of sensitivity to viscosity and thermal conductivity terms shown in figure \ref{fig8}. 

This subsection examined the sensitivity both term-wise and case-wise. The impact of all inputs differs significantly, with a magnitude range of about $O(10^{10})$. In the pseudo-boiling regime, the wall-normal gradient of baseflow profiles  (\eg $\rho$, $\partial^2p/\partial\rho\partial T$ and $\partial^2p/\partial T^2$) and the density perturbation $\hat{\rho}$ become large simultaneously following the balancing relation of the continuity equation. The overall sensitivity thus increases with a boost due to a quadratic effect in certain terms (\eg sensitivity to $\partial^2p/\partial \rho^2$) that remain small in other regimes. Special care must be taken in modeling, experiments, or applications, as a slight deviation may lead to a significant shift in the transition location.

\subsection{Fluid model relaxation}\label{S3d}

\begin{figure}
\centering
\includegraphics[scale=0.35]{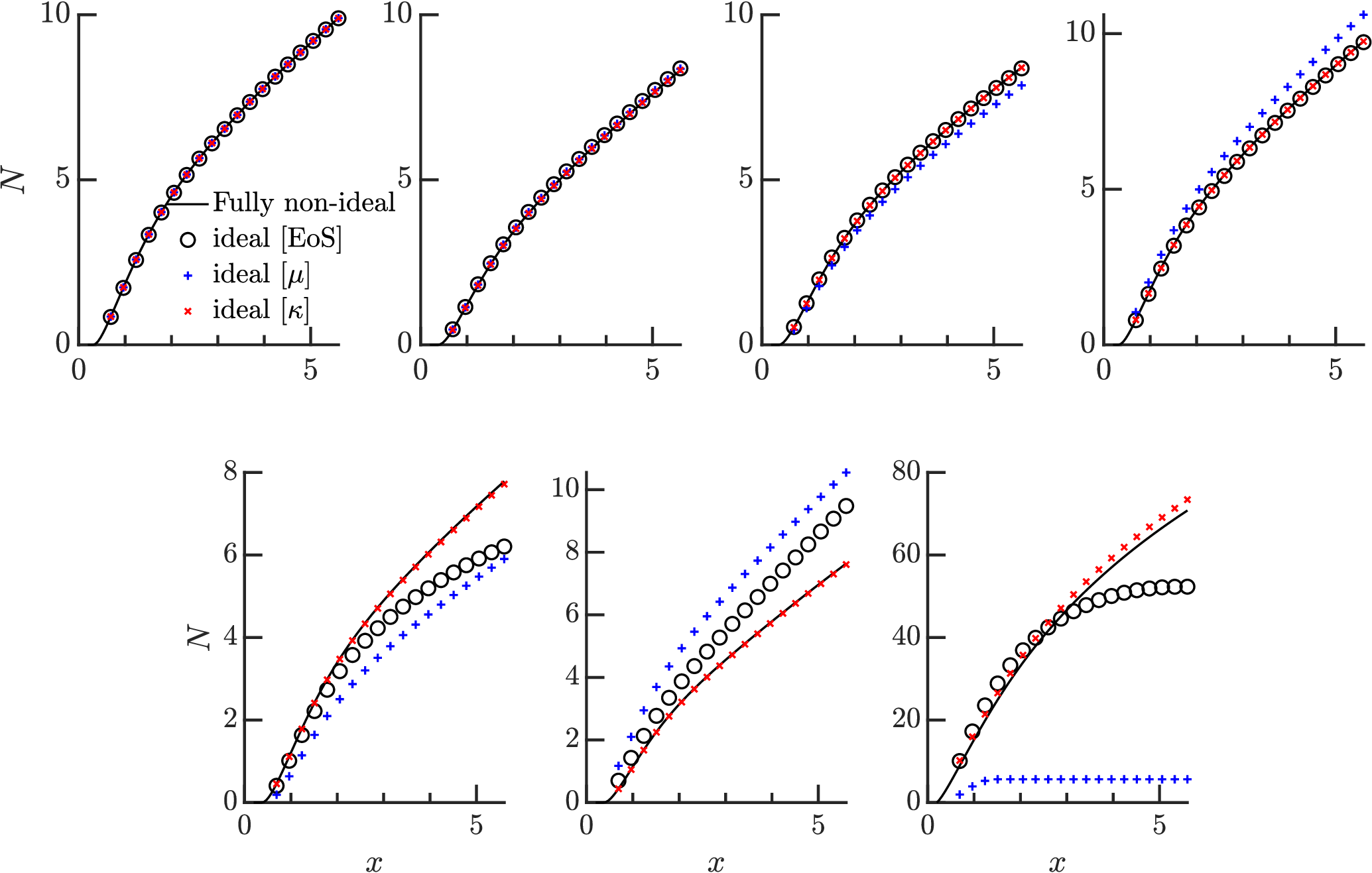}
\put(-324,200){wall-}
\put(-324,190){heating}
\put(-238,200){wall-cooling}
\put(-151,200){wall-heating}
\put(-63,200){wall-}
\put(-63,190){cooling}
\put(-281,92){wall-heating}
\put(-194,92){wall-cooling}
\put(-107,92){wall-cooling}
\put(-107,82){$\omega=20$}
\put(-312, 225){$\overbrace{\hspace{4cm}}$}
\put(-140, 225){$\overbrace{\hspace{4cm}}$}
\put(-250, 105){$\overbrace{\hspace{6cm}}$}
\put(-268,236){gas-like}
\put(-103,236){liquid-like}
\put(-190,115){pseudo-boiling}
\caption{Comparison of N factors using non-ideal and ideal models (for \E, \V~and \T~respectively). The thermodynamic regime (gas-like, liquid-like, pseudo-boiling) has been signified for relevant panels. All cases are subject to $\beta=80$, $\omega=0$ (steady CF mode), except the last panel with $\beta=80$, $\omega=20$ (inviscid TS mode).}
\label{figure18}
\end{figure}
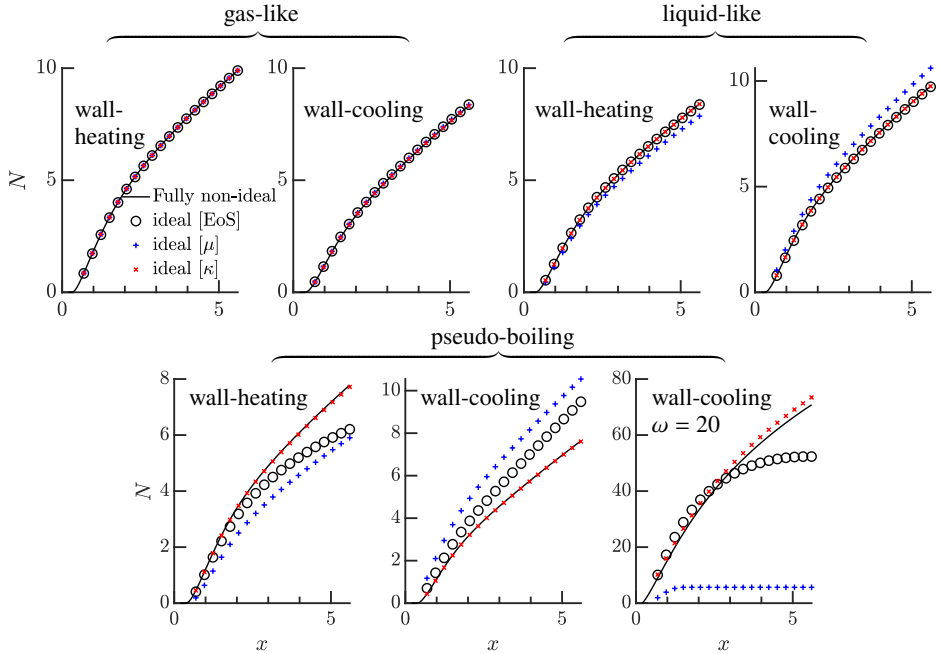

Based on the above discussions and towards an effective engineering prediction, one naturally raises the question of how accurate it is to use an empirical law for specific inputs in different thermodynamic regimes. Considering the dependence of the baseflow and stability operator on fluid properties (recall Figure \ref{fig_relation}), the outcome depends on two factors: the departure of the input from its actual value and the eigenvalue sensitivity to it. The error is thus proportional to the non-ideality of the fluid regime. For example, if one uses ideal gas laws for all the fluid properties, the non-ideal gas behavior will be totally ignored, and the difference will be significant in the highly non-ideal regime. An extreme example is in the transcritical regime for wall cooling, where the calculated mode will be a cross-flow mode rather than the 2-D inviscid mode, whose growth rate is much larger.

On the other hand, if the baseflow has been correctly solved numerically or measured in experiments, while in the stage of stability analysis the inputs related to the EOS or transport properties are supplied with ideal-gas models, how reliable are the results? This is the classic issue of using a “standard” simplified stability solver on complex-physics baseflow profiles. This also means that no distortion amplitudes need to be specified as before. We supplement the discussion by analyzing the N-factors based on stability calculations. The biased inputs are subject to ideal models for EOS ($p=\rho RT$), viscosity (Sutherland's law), and thermal conductivity (Sutherland's law), as shown in Figure \ref{figure18}. Recall the sensitivity measure presented in Figure \ref{fig8}. Provided the primary baseflow profiles (group \B~in \eqref{eq_L}) are correct, in the gas-like regime, the sensitivity is only noteworthy for terms $\partial p/\partial T$ and $\partial p/\partial \rho$. Both inputs can be satisfactorily characterized using the ideal gas equation of state, not leading to a visible difference in the N-factor. Furthermore, the liquid-like regime shows concerns for \V~terms, while the pseudo-boiling regime requires that \E~and \V~terms be adequately accounted for. In particular, the inviscid TS mode additionally requires \T~terms, as the N-factor and the absolute error are much more considerable.

Table \ref{table_sum} summarizes the \emph{idealizable} fluid models for stability analysis, indicating for each fluid regime which terms can be safely taken from the simple ideal model when correct baseflow profiles are supplied. This means that the induced distortions and the corresponding sensitivity are weak enough not to induce a notable change in the N factor. Table \ref{table_sum} provides a first estimation of standard stability solvers in predicting flow transition in different fluid regimes, where only the supercritical regime (gas-like), approximately similar to an ideal gas, is applicable to ideal-gas models. This assessment is qualitative and assumes a low Mach number configuration; the error can increase or decrease when the flow temperature range approaches or recedes from the Widom line, where viscous heating effects may become significant at higher Mach numbers.

\begin{table}
\begin{center}\def~{\hphantom{0}}
\begin{tabular}{c@{\hskip 5mm}c@{\hskip 5mm}c@{\hskip 5mm}c}
{fluid regime} & {wall temperature}  & instability mode & idealizable models   \\ [5mm]
subcritical	& heating/cooling	& CF mode	& \E,~\T    \\
transcritical 	&heating		& CF mode & \T   \\
transcritical 	&cooling		& inviscid TS mode & none    \\
supercritical 	&heating/cooling   & CF mode         	& \E,~\V,~\T 
\end{tabular}
\caption{A summary of \emph{idealizable} fluid models for linear stability analysis, provided correct baseflow profiles ($u$, $w$, $\rho$, $T$) are supplied.}
\label{table_sum}                               
\end{center}
\end{table}

\section{Conclusions}\label{S4}
The investigation examines the sensitivity of three-dimensional boundary-layer modal instability to intrinsic uncertainties in fluid properties, including the equation of state (EoS) and transport properties. We adopt a representative fluid, supercritical CO2 at 80 bar (static pressure), which exhibits significant non-ideal gas behavior crossing the Widom line. The flow matches the redesigned DLR experiment with identical pressure coefficient distribution (strong favorable pressure gradient), Reynolds number, and a low Mach number. Distinguishable thermodynamic regimes—liquid-like, pseudo-boiling, gas-like, and ideal gas—emerge by prescribing respective temperature boundary conditions. The mechanisms for linear instability found recently \citep{Ren2022} demonstrate a changeover of the dominant mode from conventional cross-flow (CF) instability to inviscid Tollmien–Schlichting (TS) type in the pseudo-boiling regime with wall cooling, despite the strong favorable pressure gradient.

The sensitivity of linear stability is a response of a coupled system, including the acquisition of the laminar baseflow and the eigenvalue problem. Uncertainties in any one or combination of terms of the EoS, viscosity, and thermal conductivity give rise to distortions in primary baseflow profiles ($u$, $w$, $\rho$, $T$) and all profiles related to fluid properties needed for the stability analysis. The shape and amplitude of these distortions are discussed and found to be dependent on the source of uncertainty and fluid regimes. 

Further, these distortions are measured by the sensitivity coefficients, whose product indicates the alterations of the eigenvalue. The sensitivity coefficients are theoretically formulated for each input of the stability operator using the adjoint equations, indicating the eigenvalue shift when performing the inner product with corresponding physical distortions. The adjoint equations have been validated using the bi-orthogonal relationship, and all sensitivity coefficients have been validated against linear stability calculations with distorted inputs. Considering that an actual distortion can be arbitrary (e.g., random noise) or structural (e.g., model deviation), we propose a unified measure for the degree of sensitivity that describes the maximum possible eigenvalue shift. It is worth noting that the adjoint-equation-based linear sensitivity provides details to understand the system's response to the bias of specific inputs, thereby directing the most effective control strategies. The method also efficiently predicts growth rate shifts due to different distortions. On the other hand, one may circumvent the details and obtain a stability prediction directly with distorted inputs, which is less effective. However, the amplitude of the distortion can be considerable in this case.

The sensitivity has been investigated for all inputs of the stability operator and compared horizontally across different thermodynamic regimes. A range of its amplitude spanning the order of $10^{10}$ for different inputs is found and attributed to the scaling of the Reynolds number for viscosity-related terms and different wall-normal gradients of the baseflow and perturbation profiles. The sensitivity is significantly larger (by 1-2 orders of magnitude) in the pseudo-boiling regime. In particular, a quadratic effect has been uncovered. The boost of $\partial\rho/\partial y$ results in an equivalent rise for $\hat{\rho}$ through a balancing relation of the continuity equation. The product of $\partial\rho/\partial y$ and $\hat{\rho}$ gives rise to a quadratic sensitivity increase to $\partial^2 p/\partial\rho^2$ across the Widom line. Meanwhile, the tremendous sensitivity to temperature and density terms — especially regarding the respective direct baseflow profiles — is also due to large thermodynamic derivatives (e.g., $\partial^2 p/(\partial\rho\partial T)$ and $\partial^2 p/\partial T^2$) and their wall-normal gradients.

Towards practical applications for transition prediction, the N factor deviation has been investigated when certain groups of terms are given following an ideal-gas assumption, based on the use of the correct, non-ideal baseflow profiles. In other words, a standard instability solver is applied to the fully non-ideal baseflow. In particular, we show that in the gas-like regime, using the idealized equation of state (EOS), viscosity, and thermal conductivity do not provoke noticeable differences. The liquid-like regime requires viscosity to be correctly modeled, while the pseudo-boiling regime needs all elements to be correct except for the thermal conductivity for the CF mode. Accurately predicting the inviscid TS mode's growth rate requires proper modeling of all the above elements --- employing an ideal-fluid solver to the correct baseflow profiles gives a meaningless instability result.

The present research focused on the sensitivity of flow stability to intrinsic uncertainties in fluid properties. This question arises from the ever-growing industrial demand and the ongoing upgrades and improvements of fluid property databases. The different sensitivity behaviors found in this research indicate that an accurate model describing the fluid is essential in highly non-ideal regimes, particularly for the accurate description of the new dominating inviscid mode.
 

\backsection[Acknowledgements]{JR acknowledges helpful discussions with Drs. J. Park (Coventry University), Y. Xi, S. Fu (Tsinghua University), P.C. Boldini and R. Pecnik (TU Delft).}

\backsection[Funding]{This work was supported by National Natural Science Foundation of China (Grant Nos.12372215 and 92152109), the Aeronautical Science Fund (Grant No. 2024M005072001) and the Alexander von Humboldt foundation.}

\backsection[Declaration of interests]{The authors report no conflict of interest.}

\backsection[Data availability statement]{The data that support the findings of this study are available from the first and corresponding author upon reasonable request.}

\backsection[Author ORCID]
{\\Jie Ren~     		\href{https://orcid.org/0000-0001-8448-4361}{https://orcid.org/0000-0001-8448-4361}; 
 \\Yongxiang Wu	\href{https://orcid.org/0000-0002-9653-1935}{https://orcid.org/0000-0002-9653-1935};
 \\Xuerui Mao       	\href{https://orcid.org/0000-0002-8452-5773}{https://orcid.org/0000-0002-8452-5773};
 \\Markus Kloker 	\href{https://orcid.org/0000-0002-5352-7442}{https://orcid.org/0000-0002-5352-7442}}

\appendix
\section{The stability operator $\mathsfbi{L}$}\label{appL}
The specific expressions of the linear stability operator introduced in \eqref{eq_L} are given below.
\begin{equation}
\mathsfbi{L}_{1,1}=-i\alpha u+i\omega-i\beta w,\;
\mathsfbi{L}_{1,2}=-i\alpha\rho,\;
\mathsfbi{L}_{1,3}=-\rho D-\frac{\partial\rho}{\partial y},\;
\mathsfbi{L}_{1,4}=-i\beta\rho.
\end{equation}

\begin{equation}
\mathsfbi{L}_{2,1}=-i\alpha\frac{\partial p}{\partial\rho}+\frac{1}{Re}\frac{\partial\mu}{\partial\rho}\frac{\partial u}{\partial y}D+\frac{1}{Re}\frac{\partial\mu}{\partial\rho}\frac{\partial^{2}u}{\partial y^{2}}+\frac{1}{Re}\frac{\partial u}{\partial y}\left(\frac{\partial^{2}\mu}{\partial\rho^{2}}\frac{\partial\rho}{\partial y}+\frac{\partial^{2}\mu}{\partial\rho\partial T}\frac{\partial T}{\partial y}\right).
\end{equation}

\begin{equation}
\mathsfbi{L}_{2,2}=-\frac{\alpha^{2}\left(2\mu+\lambda\right)}{Re}-i\alpha\rho u+i\omega\rho+\frac{1}{Re}\frac{\partial\mu}{\partial y}D-i\beta\rho w+\frac{\mu}{Re}D^{2}-\frac{\beta^{2}\mu}{Re}.
\end{equation}

\begin{equation}
\mathsfbi{L}_{2,3}=\frac{i\alpha\left(\mu+\lambda\right)D}{Re}+\frac{i\alpha}{Re}\frac{\partial\mu}{\partial y}-\rho\frac{\partial u}{\partial y},\;
\mathsfbi{L}_{2,4}=-\frac{\alpha\beta\left(\mu+\lambda\right)}{Re}.
\end{equation}

\begin{equation}
\mathsfbi{L}_{2,5}=-i\alpha\frac{\partial p}{\partial T}+\frac{1}{Re}\frac{\partial\mu}{\partial T}\frac{\partial u}{\partial y}D+\frac{1}{Re}\frac{\partial\mu}{\partial T}\frac{\partial^{2}u}{\partial y^{2}}+\frac{1}{Re}\frac{\partial u}{\partial y}\left(\frac{\partial^{2}\mu}{\partial T^{2}}\frac{\partial T}{\partial y}+\frac{\partial^{2}\mu}{\partial T\partial\rho}\frac{\partial\rho}{\partial y}\right).
\end{equation}

\begin{equation}
\mathsfbi{L}_{3,1}=\frac{i\alpha}{Re}\frac{\partial\mu}{\partial\rho}\frac{\partial u}{\partial y}-\frac{\partial p}{\partial\rho}D+i\beta\frac{1}{Re}\frac{\partial\mu}{\partial\rho}\frac{\partial w}{\partial y}-\frac{\partial^{2}p}{\partial\rho^{2}}\frac{\partial\rho}{\partial y}-\frac{\partial^{2}p}{\partial\rho\partial T}\frac{\partial T}{\partial y}.
\end{equation}

\begin{equation}
\mathsfbi{L}_{3,2}=\frac{i\alpha\left(\mu+\lambda\right)D}{Re}+\frac{i\alpha}{Re}\frac{\partial\lambda}{\partial y}.
\end{equation}

\begin{equation}
\mathsfbi{L}_{3,3}=-\frac{\alpha^{2}\mu}{Re}-i\alpha\rho u+i\omega\rho+\frac{1}{Re}\frac{\partial\left(2\mu+\lambda\right)}{\partial y}D-i\beta\rho w+\frac{2\mu+\lambda}{Re}D^{2}-\frac{\beta^{2}\mu}{Re}.
\end{equation}

\begin{equation}
\mathsfbi{L}_{3,4}=i\beta\frac{1}{Re}\frac{\partial\lambda}{\partial y}+\frac{i\beta\left(\mu+\lambda\right)D}{Re}.
\end{equation}

\begin{equation}
\mathsfbi{L}_{3,5}=\frac{i\alpha}{Re}\frac{\partial\mu}{\partial T}\frac{\partial u}{\partial y}-\frac{\partial p}{\partial T}D+i\beta\frac{1}{Re}\frac{\partial\mu}{\partial T}\frac{\partial w}{\partial y}-\frac{\partial^{2}p}{\partial T^{2}}\frac{\partial T}{\partial y}-\frac{\partial^{2}p}{\partial\rho\partial T}\frac{\partial\rho}{\partial y}.
\end{equation}

\begin{equation}
\mathsfbi{L}_{4,1}=\frac{1}{Re}\frac{\partial\mu}{\partial\rho}\frac{\partial w}{\partial y}D-i\beta\frac{\partial p}{\partial\rho}+\frac{1}{Re}\frac{\partial\mu}{\partial\rho}\frac{\partial^{2}w}{\partial y^{2}}+\frac{1}{Re}\frac{\partial w}{\partial y}\left(\frac{\partial^{2}\mu}{\partial\rho^{2}}\frac{\partial\rho}{\partial y}+\frac{\partial^{2}\mu}{\partial\rho\partial T}\frac{\partial T}{\partial y}\right).
\end{equation}

\begin{equation}
\mathsfbi{L}_{4,2}=-\frac{\alpha\beta\left(\mu+\lambda\right)}{Re},\;
\mathsfbi{L}_{4,3}=i\beta\frac{1}{Re}\frac{\partial\mu}{\partial y}+\frac{i\beta\left(\mu+\lambda\right)D}{Re}-\rho\frac{\partial w}{\partial y}.
\end{equation}

\begin{equation}
\mathsfbi{L}_{4,4}=-\frac{\alpha^{2}\mu}{Re}-i\alpha\rho u+i\omega\rho+\frac{1}{Re}\frac{\partial\mu}{\partial y}D-i\beta\rho w+\frac{\mu}{Re}D^{2}-\frac{\beta^{2}\left(2\mu+\lambda\right)}{Re}.
\end{equation}

\begin{equation}
\mathsfbi{L}_{4,5}=\frac{1}{Re}\frac{\partial\mu}{\partial T}\frac{\partial w}{\partial y}D-i\beta\frac{\partial p}{\partial T}+\frac{1}{Re}\frac{\partial\mu}{\partial T}\frac{\partial^{2}w}{\partial y^{2}}+\frac{1}{Re}\frac{\partial w}{\partial y}\left(\frac{\partial^{2}\mu}{\partial T^{2}}\frac{\partial T}{\partial y}+\frac{\partial^{2}\mu}{\partial T\partial\rho}\frac{\partial\rho}{\partial y}\right).
\end{equation}

\begin{equation}
\begin{aligned}\mathsfbi{L}_{5,1}= & -i\alpha\rho u\frac{\partial e}{\partial\rho}+i\omega\rho\frac{\partial e}{\partial\rho}+\frac{1}{RePrEc}\frac{\partial\kappa}{\partial\rho}\frac{\partial T}{\partial y}D-i\beta\rho w\frac{\partial e}{\partial\rho}\\
 & +\frac{1}{RePrEc}\left(\frac{\partial\kappa}{\partial\rho}\frac{\partial^{2}T}{\partial y^{2}}+\frac{\partial^{2}\kappa}{\partial\rho^{2}}\frac{\partial\rho}{\partial y}\frac{\partial T}{\partial y}+\frac{\partial^{2}\kappa}{\partial\rho\partial T}\left(\frac{\partial T}{\partial y}\right)^{2}\right)\\
 & +\frac{1}{Re}\frac{\partial\mu}{\partial\rho}\left(\left(\frac{\partial u}{\partial y}\right)^{2}+\left(\frac{\partial w}{\partial y}\right)^{2}\right).
\end{aligned}
\end{equation}

\begin{equation}
\mathsfbi{L}_{5,2}=-i\alpha p+\frac{2\mu}{Re}\frac{\partial u}{\partial y}D.
\end{equation}

\begin{equation}
\mathsfbi{L}_{5,3}=\frac{2i\alpha\mu}{Re}\frac{\partial u}{\partial y}-pD+i\beta\frac{2\mu}{Re}\frac{\partial w}{\partial y}-\rho\frac{\partial e}{\partial y}.
\end{equation}

\begin{equation}
\mathsfbi{L}_{5,4}=\frac{2\mu}{Re}\frac{\partial w}{\partial y}D-i\beta p.
\end{equation}

\begin{equation}
\begin{aligned}
\mathsfbi{L}_{5,5}= &-\frac{\alpha^{2}\kappa}{RePrEc}-i\alpha\rho u\frac{\partial e}{\partial T}+i\omega\rho\frac{\partial e}{\partial T}+\frac{1}{RePrEc}\left(\frac{\partial\kappa}{\partial y}+\frac{\partial\kappa}{\partial T}\frac{\partial T}{\partial y}\right)D\\
&-i\beta\rho w\frac{\partial e}{\partial T}+\frac{\kappa}{RePrEc}D^{2}-\frac{\kappa\beta^{2}}{RePrEc}\\
&+\frac{1}{RePrEc}\left(\frac{\partial\kappa}{\partial T}\frac{\partial^{2}T}{\partial y^{2}}+\frac{\partial^{2}\kappa}{\partial T^{2}}\left(\frac{\partial T}{\partial y}\right)^{2}+\frac{\partial^{2}\kappa}{\partial\rho\partial T}\frac{\partial T}{\partial y}\frac{\partial\rho}{\partial y}\right)\\
&+\frac{1}{Re}\frac{\partial\mu}{\partial T}\left(\left(\frac{\partial u}{\partial y}\right)^{2}+\left(\frac{\partial w}{\partial y}\right)^{2}\right).
\end{aligned}
\end{equation}

\section{Derivation of the adjoint equations}\label{appAdj}
We re-write the stability operator based on the order of wall-normal derivative:
\begin{equation}
\mathsfbi{L}=\mathsfbi{L}_{0}+\mathsfbi{L}_{1}D+\mathsfbi{L}_{2}D^{2}
\end{equation}
Upon integration by parts,
\begin{equation}
\left\langle \hat{\boldsymbol{q}}^{\dagger},\mathsfbi{L}\hat{\boldsymbol{q}}\right\rangle =	\left\langle \left(\mathsfbi{L}_{0}^{H}-\mathsfbi{L}_{1}^{H}D-\frac{\partial\mathsfbi{L}_{1}^{H}}{\partial y}+\mathsfbi{L}_{2}^{H}D^{2}+2\frac{\partial\mathsfbi{L}_{2}^{H}}{\partial y}D+\frac{\partial^{2}\mathsfbi{L}_{2}^{H}}{\partial y^{2}}\right)\hat{\boldsymbol{q}}^{\dagger},\hat{\boldsymbol{q}}\right\rangle +\rm{B.T.}
\end{equation}
$D=\partial/\partial y$,  B.T. stands for boundary terms produced during integration: 
\begin{equation}
\rm{B.T.}=\underset{=0}{\underbrace{\left.\hat{\boldsymbol{q}}^{\dagger H}\mathsfbi{L}_{1}\hat{\boldsymbol{q}}\right|_{0}^{\infty}}}+\underset{\textrm{need }\hat{u}^{\dagger}=\hat{v}^{\dagger}=\hat{w}^{\dagger}=\hat{T}^{\dagger}=0}{\underbrace{\left.\hat{\boldsymbol{q}}^{\dagger H}\mathsfbi{L}_{2}D\hat{\boldsymbol{q}}\right|_{0}^{\infty}}}\underset{=0}{\underbrace{-\left.D\hat{\boldsymbol{q}}^{\dagger H}\mathsfbi{L}_{2}\hat{\boldsymbol{q}}\right|_{0}^{\infty}-\left.\hat{\boldsymbol{q}}^{\dagger H}D\mathsfbi{L}_{2}\hat{\boldsymbol{q}}\right|_{0}^{\infty}}}
\end{equation}
To make sure the B.T. vanishes, we specify the boundary conditions for adjoint vectors:
\begin{equation}
\hat{u}^{\dagger}=\hat{v}^{\dagger}=\hat{w}^{\dagger}=\hat{T}^{\dagger}=0,\qquad (y=0,\;\infty)
\end{equation}
According to the definition of the adjoint equations \eqref{EQadj}, the following relation holds.
\begin{equation}\label{EQorth1}
\mathsfbi{L}^{\dagger}\left(\alpha^{\dagger}\right)\hat{\boldsymbol{q}}^{\dagger}=\mathsfbi{L}^{\dagger}\left(\alpha^{*}\right)\hat{\boldsymbol{q}}^{\dagger}\Leftrightarrow\alpha^{\dagger}=\alpha^{*}
\end{equation}
In addition, the bi-orthogonal relationship is obtained from the definition of the adjoint equations. Examine $\hat{\boldsymbol{q}}_{i}$ and $\hat{\boldsymbol{q}}_{j}^{\dagger}$,
\begin{equation}
\left\langle \mathsfbi{L}^{\dagger}\left(\alpha_{j}^{\dagger}\right)\hat{\boldsymbol{q}}_{j}^{\dagger},\hat{\boldsymbol{q}}_{i}\right\rangle =\left\langle \hat{\boldsymbol{q}}_{j}^{\dagger},\mathsfbi{L}\left(\alpha_{j}\right)\hat{\boldsymbol{q}}_{i}\right\rangle =\left\langle \hat{\boldsymbol{q}}_{j}^{\dagger},\mathsfbi{L}\left(\alpha_{i}\right)\hat{\boldsymbol{q}}_{i}\right\rangle =0
\end{equation}
gives
\begin{equation}\label{EQorth2}
\left\langle \hat{\boldsymbol{q}}_{j}^{\dagger},\left[\mathsfbi{L}\left(\alpha_{i}\right)-\mathsfbi{L}\left(\alpha_{j}\right)\right]\hat{\boldsymbol{q}}_{i}\right\rangle =0
\end{equation}
\eqref{EQorth1} and \eqref{EQorth2} have been used to validate the adjoint equations derived.
	
\section{Normalisation of the eigenvector and its adjoint}\label{appNorm}
Following discussions in \S \ref{S2c}, we normalise $\hat{\boldsymbol{q}}$ and $\hat{\boldsymbol{q}}^{\dagger}$ such that the following term equal to unity.
\begin{align}
\begin{split}
\left\langle \hat{\boldsymbol{q}}^{\dagger},\frac{\partial\mathsfbi{L}}{\partial\alpha}\hat{\boldsymbol{q}}\right\rangle = & \int_{0}^{\infty}\hat{\boldsymbol{q}}^{\dagger H}\frac{\partial\mathsfbi{L}}{\partial\alpha}\hat{\boldsymbol{q}}\:dy\\
=\int_{0}^{\infty} & \hat{\rho}^{\dagger*}\left(-iu\hat{\rho}-i\rho\hat{u}\right)\:dy\\
+\int_{0}^{\infty} & \hat{u}^{\dagger*}\left(-i\frac{\partial p}{\partial\rho}\hat{\rho}+i\rho u\hat{u}-i\frac{\partial p}{\partial T}\hat{T}\right)\:dy\\
+\int_{0}^{\infty} & \hat{u}^{\dagger*}\left(-\frac{2\alpha\left(2\mu+\lambda\right)}{Re}\hat{u}+\frac{i\left(\mu+\lambda\right)D\hat{v}}{Re}+\frac{i}{Re}\frac{\partial\mu}{\partial y}\hat{v}-\frac{\beta\left(\mu+\lambda\right)}{Re}\hat{w}\right)\:dy\\
+\int_{0}^{\infty} & \hat{v}^{\dagger*}\left(\frac{i}{Re}\frac{\partial\mu}{\partial\rho}\frac{\partial u}{\partial y}\hat{\rho}+\frac{i\left(\mu+\lambda\right)D\hat{u}}{Re}+\frac{i}{Re}\frac{\partial\lambda}{\partial y}\hat{u}\right)\:dy\\
+\int_{0}^{\infty} & \hat{v}^{\dagger*}\left(-\frac{2\alpha\mu}{Re}\hat{v}-i\rho u\hat{v}+\frac{i}{Re}\frac{\partial\mu}{\partial T}\frac{\partial u}{\partial y}\hat{T}\right)\:dy\\
+\int_{0}^{\infty} & \hat{w}^{\dagger*}\left(-\frac{\beta\left(\mu+\lambda\right)}{Re}\hat{u}-\frac{2\alpha\mu}{Re}\hat{w}-i\rho u\hat{w}\right)\:dy\\
+\int_{0}^{\infty} & \hat{T}^{\dagger*}\left(-i\rho u\frac{\partial e}{\partial\rho}\hat{\rho}-ip\hat{u}+\frac{2i\mu}{Re}\frac{\partial u}{\partial y}\hat{v}-\frac{2\alpha\kappa}{RePrEc}\hat{T}-i\rho u\frac{\partial e}{\partial T}\hat{T}\right)\:dy
\end{split}
\end{align}	
	
\section{Sensitivity coefficients}\label{appSens}
\begin{align}
\begin{split}
\left\langle \boldsymbol{S}_{\boldsymbol{Q}},\delta\boldsymbol{Q}\right\rangle = & \left(\begin{array}{c}
S_{u}\\
S_{w}\\
S_{\rho}\\
S_{T}\\
S_{p}
\end{array}\right)^{H}\left(\begin{array}{c}
\delta_{u}\\
\delta_{w}\\
\delta_{\rho}\\
\delta_{T}\\
\delta_{p}
\end{array}\right)+\left(\begin{array}{c}
S_{\frac{\partial p}{\partial\rho}}\\
S_{\frac{\partial p}{\partial T}}\\
S_{\frac{\partial^{2}p}{\partial\rho^{2}}}\\
S_{\frac{\partial^{2}p}{\partial T^{2}}}\\
S_{\frac{\partial^{2}p}{\partial\rho\partial T}}
\end{array}\right)^{H}\left(\begin{array}{c}
\delta_{\frac{\partial p}{\partial\rho}}\\
\delta_{\frac{\partial p}{\partial T}}\\
\delta_{\frac{\partial^{2}p}{\partial\rho^{2}}}\\
\delta_{\frac{\partial^{2}p}{\partial T^{2}}}\\
\delta_{\frac{\partial^{2}p}{\partial\rho\partial T}}
\end{array}\right)+\left(\begin{array}{c}
S_{\frac{\partial e}{\partial\rho}}\\
S_{\frac{\partial e}{\partial T}}
\end{array}\right)^{H}\left(\begin{array}{c}
\delta_{\frac{\partial e}{\partial\rho}}\\
\delta_{\frac{\partial e}{\partial T}}
\end{array}\right)+\\
 & \left(\begin{array}{c}
S_{\mu}\\
S_{\frac{\partial\mu}{\partial\rho}}\\
S_{\frac{\partial\mu}{\partial T}}\\
S_{\frac{\partial^{2}\mu}{\partial\rho^{2}}}\\
S_{\frac{\partial^{2}\mu}{\partial T^{2}}}\\
S_{\frac{\partial^{2}\mu}{\partial\rho\partial T}}
\end{array}\right)^{H}\left(\begin{array}{c}
\delta_{\mu}\\
\delta_{\frac{\partial\mu}{\partial\rho}}\\
\delta_{\frac{\partial\mu}{\partial T}}\\
\delta_{\frac{\partial^{2}\mu}{\partial\rho^{2}}}\\
\delta_{\frac{\partial^{2}\mu}{\partial T^{2}}}\\
\delta_{\frac{\partial^{2}\mu}{\partial\rho\partial T}}
\end{array}\right)+\left(\begin{array}{c}
S_{\kappa}\\
S_{\frac{\partial\kappa}{\partial\rho}}\\
S_{\frac{\partial\kappa}{\partial T}}\\
S_{\frac{\partial^{2}\kappa}{\partial\rho^{2}}}\\
S_{\frac{\partial^{2}\kappa}{\partial T^{2}}}\\
S_{\frac{\partial^{2}\kappa}{\partial\rho\partial T}}
\end{array}\right)^{H}\left(\begin{array}{c}
\delta_{\kappa}\\
\delta_{\frac{\partial\kappa}{\partial\rho}}\\
\delta_{\frac{\partial\kappa}{\partial T}}\\
\delta_{\frac{\partial^{2}\kappa}{\partial\rho^{2}}}\\
\delta_{\frac{\partial^{2}\kappa}{\partial T^{2}}}\\
\delta_{\frac{\partial^{2}\kappa}{\partial\rho\partial T}}
\end{array}\right)
\end{split}
\end{align}

\begin{align}
\begin{split}
S_{u}= & \frac{\partial}{\partial y}\left(\rho\hat{v}^{*}\hat{u}^{\dagger}\right)\\
 & +i\alpha^{*}\left(\hat{\rho}^{*}\hat{\rho}^{\dagger}+\rho\hat{u}^{*}\hat{u}^{\dagger}+\rho\hat{v}^{*}\hat{v}^{\dagger}+\rho\hat{w}^{*}\hat{w}^{\dagger}+\rho\frac{\partial e}{\partial\rho}\hat{\rho}^{*}\hat{T}^{\dagger}+\rho\frac{\partial e}{\partial T}\hat{T}^{*}\hat{T}^{\dagger}\right)\\
 & +\frac{2}{Re}\frac{\partial}{\partial y}\left[\mu\left(i\alpha^{*}\hat{v}^{*}-\frac{\partial\hat{u}^{*}}{\partial y}\right)\hat{T}^{\dagger}\right]\\
 & +\frac{1}{Re}\frac{\partial}{\partial y}\left[\frac{\partial\mu}{\partial T}\hat{T}^{*}\left(\frac{\partial\hat{u}^{\dagger}}{\partial y}-2\frac{\partial u}{\partial y}\hat{T}^{\dagger}+i\alpha^{*}\hat{v}^{\dagger}\right)\right]\\
 & +\frac{1}{Re}\frac{\partial}{\partial y}\left[\frac{\partial\mu}{\partial\rho}\hat{\rho}^{*}\left(\frac{\partial\hat{u}^{\dagger}}{\partial y}-2\frac{\partial u}{\partial y}\hat{T}^{\dagger}+i\alpha^{*}\hat{v}^{\dagger}\right)\right]
\end{split}
\end{align}

\begin{align}
\begin{split}
S_{w}= & \frac{\partial}{\partial y}\left(\rho\hat{v}^{*}\hat{w}^{\dagger}\right)+i\beta\left(\hat{\rho}^{*}\hat{\rho}^{\dagger}+\rho\hat{u}^{*}\hat{u}^{\dagger}+\rho\hat{v}^{*}\hat{v}^{\dagger}+\rho\hat{w}^{*}\hat{w}^{\dagger}+\rho\frac{\partial e}{\partial\rho}\hat{\rho}^{*}\hat{T}^{\dagger}+\rho\frac{\partial e}{\partial T}\hat{T}^{*}\hat{T}^{\dagger}\right)\\
 & +\frac{i\beta}{Re}\frac{\partial}{\partial y}\left(2\mu\hat{v}^{*}\hat{T}^{\dagger}+\frac{\partial\mu}{\partial\rho}\hat{\rho}^{*}\hat{v}^{\dagger}+\frac{\partial\mu}{\partial T}\hat{T}^{*}\hat{v}^{\dagger}\right)\\
 & +\frac{1}{Re}\frac{\partial}{\partial y}\left(\frac{\partial\mu}{\partial T}\hat{T}^{*}D\hat{w}^{\dagger}+\frac{\partial\mu}{\partial\rho}\hat{\rho}^{*}D\hat{w}^{\dagger}\right)\\
 & -\frac{2}{Re}\frac{\partial}{\partial y}\left(\mu D\hat{w}^{*}\hat{T}^{\dagger}+\frac{\partial\mu}{\partial\rho}\frac{\partial w}{\partial y}\hat{\rho}^{*}\hat{T}^{\dagger}+\frac{\partial\mu}{\partial T}\frac{\partial w}{\partial y}\hat{T}^{*}\hat{T}^{\dagger}\right)
\end{split}
\end{align}

\begin{align}\label{eq_ST}
\begin{split}
S_{T}= & \frac{\partial}{\partial y}\left(\frac{\partial^{2}p}{\partial\rho\partial T}\hat{\rho}^{*}\hat{v}^{\dagger}+\frac{\partial^{2}p}{\partial T^{2}}\hat{T}^{*}\hat{v}^{\dagger}\right)\\
 & -\frac{1}{Re}\frac{\partial}{\partial y}\left(\frac{\partial u}{\partial y}\frac{\partial^{2}\mu}{\partial\rho\partial T}\hat{\rho}^{*}\hat{u}^{\dagger}+\frac{\partial u}{\partial y}\frac{\partial^{2}\mu}{\partial T^{2}}\hat{T}^{*}\hat{u}^{\dagger}+\frac{\partial w}{\partial y}\frac{\partial^{2}\mu}{\partial\rho\partial T}\hat{\rho}^{*}\hat{w}^{\dagger}+\frac{\partial w}{\partial y}\frac{\partial^{2}\mu}{\partial T^{2}}\hat{T}^{*}\hat{w}^{\dagger}\right)\\
 & -\frac{1}{RePrEc}\frac{\partial}{\partial y}\left(\frac{\partial\kappa}{\partial\rho}D\hat{\rho}^{*}\hat{T}^{\dagger}+\frac{\partial^{2}\kappa}{\partial\rho^{2}}\frac{\partial\rho}{\partial y}\hat{\rho}^{*}\hat{T}^{\dagger}+2\frac{\partial^{2}\kappa}{\partial\rho\partial T}\frac{\partial T}{\partial y}\hat{\rho}^{*}\hat{T}^{\dagger}\right)\\
 & -\frac{1}{RePrEc}\frac{\partial}{\partial y}\left(\frac{\partial\kappa}{\partial T}D\hat{T}^{*}\hat{T}^{\dagger}+\frac{\partial^{2}\kappa}{\partial\rho\partial T}\frac{\partial\rho}{\partial y}\hat{T}^{*}\hat{T}^{\dagger}+2\frac{\partial^{2}\kappa}{\partial T^{2}}\frac{\partial T}{\partial y}\hat{T}^{*}\hat{T}^{\dagger}\right)\\
 & +\frac{1}{RePrEc}\frac{\partial^{2}}{\partial y^{2}}\left(\frac{\partial\kappa}{\partial\rho}\hat{\rho}^{*}\hat{T}^{\dagger}+\frac{\partial\kappa}{\partial T}\hat{T}^{*}\hat{T}^{\dagger}\right)
\end{split}
\end{align}

\begin{align}\label{S_D}
\begin{split}
S_{\rho}= & i\alpha^{*}\hat{u}^{*}\hat{\rho}^{\dagger}-\left(D\hat{v}^{*}\right)\hat{\rho}^{\dagger}+\frac{\partial}{\partial y}\left(\hat{v}^{*}\hat{\rho}^{\dagger}\right)+i\beta\hat{w}^{*}\hat{\rho}^{\dagger}\\
 & -\frac{\partial u}{\partial y}\hat{v}^{*}\hat{u}^{\dagger}-\frac{\partial w}{\partial y}\hat{v}^{*}\hat{w}^{\dagger}-\frac{\partial e}{\partial y}\hat{v}^{*}\hat{T}^{\dagger}\\
 & +\frac{\partial}{\partial y}\left(\frac{\partial^{2}p}{\partial\rho^{2}}\hat{\rho}^{*}\hat{v}^{\dagger}+\frac{\partial^{2}p}{\partial\rho\partial T}\hat{T}^{*}\hat{v}^{\dagger}\right)\\
 & +\left(i\alpha^{*}u-i\omega+i\beta w\right)\left(\hat{u}^{*}\hat{u}^{\dagger}+\hat{v}^{*}\hat{v}^{\dagger}+\hat{w}^{*}\hat{w}^{\dagger}+\frac{\partial e}{\partial\rho}\hat{\rho}^{*}\hat{T}^{\dagger}+\frac{\partial e}{\partial T}\hat{T}^{*}\hat{T}^{\dagger}\right)+\\
 & -\frac{1}{Re}\frac{\partial}{\partial y}\left(\frac{\partial u}{\partial y}\frac{\partial^{2}\mu}{\partial\rho^{2}}\hat{\rho}^{*}\hat{u}^{\dagger}+\frac{\partial u}{\partial y}\frac{\partial^{2}\mu}{\partial T\partial\rho}\hat{T}^{*}\hat{u}^{\dagger}+\frac{\partial w}{\partial y}\frac{\partial^{2}\mu}{\partial\rho^{2}}\hat{\rho}^{*}\hat{w}^{\dagger}+\frac{\partial w}{\partial y}\frac{\partial^{2}\mu}{\partial T\partial\rho}\hat{T}^{*}\hat{w}^{\dagger}\right)\\
 & -\frac{1}{RePrEc}\frac{\partial}{\partial y}\left(\frac{\partial^{2}\kappa}{\partial\rho^{2}}\frac{\partial T}{\partial y}\hat{\rho}^{*}\hat{T}^{\dagger}+\frac{\partial^{2}\kappa}{\partial\rho\partial T}\frac{\partial T}{\partial y}\hat{T}^{*}\hat{T}^{\dagger}\right)
\end{split}
\end{align}

\begin{equation}
S_{p}=\left(i\alpha^{*}\hat{u}^{*}-D\hat{v}^{*}+i\beta\hat{w}^{*}\right)\hat{T}^{\dagger}
\end{equation}

\begin{equation}\label{S_EOS}
\left. \begin{array}{l}
\displaystyle
S_{\frac{\partial p}{\partial\rho}}=i\alpha^{*}\hat{\rho}^{*}\hat{u}^{\dagger}-D\hat{\rho}^{*}\hat{v}^{\dagger}+i\beta\hat{\rho}^{*}\hat{w}^{\dagger} \\[10pt]
\displaystyle
S_{\frac{\partial p}{\partial T}}=i\alpha^{*}\hat{T}^{*}\hat{u}^{\dagger}-D\hat{T}\hat{v}^{\dagger}+i\beta\hat{T}^{*}\hat{w}^{\dagger} \\[10pt]
\displaystyle
S_{\frac{\partial^{2}p}{\partial\rho^{2}}}=-\frac{\partial\rho}{\partial y}\hat{\rho}^{*}\hat{v}^{\dagger}\\[10pt]
\displaystyle
S_{\frac{\partial^{2}p}{\partial T^{2}}}=-\frac{\partial T}{\partial y}\hat{T}^{*}\hat{v}^{\dagger}\\[10pt]
\displaystyle
S_{\frac{\partial^{2}p}{\partial\rho\partial T}}=-\left(\frac{\partial T}{\partial y}\hat{\rho}^{*}+\frac{\partial\rho}{\partial y}\hat{T}^{*}\right)\hat{v}^{\dagger}\\[10pt]
\displaystyle
S_{\frac{\partial e}{\partial\rho}}=-\rho\frac{\partial\rho}{\partial y}\hat{v}^{*}\hat{T}^{\dagger}+\left(i\alpha^{*}\rho u-i\omega\rho+i\beta\rho w\right)\hat{\rho}^{*}\hat{T}^{\dagger}\\[10pt]
\displaystyle
S_{\frac{\partial e}{\partial T}}=-\rho\frac{\partial T}{\partial y}\hat{v}^{*}\hat{T}^{\dagger}+\left(i\alpha^{*}\rho u-i\omega\rho+i\beta\rho w\right)\hat{T}^{*}\hat{T}^{\dagger}
\end{array} \right\}
\end{equation}

\begin{align}
\begin{split}
S_{\mu}= & \frac{1}{Re}\left(-\frac{4}{3}\alpha^{*2}\hat{u}^{*}\hat{u}^{\dagger}+D^{2}\hat{u}^{*}\hat{u}^{\dagger}-\beta^{2}\hat{u}^{*}\hat{u}^{\dagger}-\frac{1}{3}i\alpha^{*}D\hat{v}^{*}\hat{u}^{\dagger}-\frac{1}{3}\alpha^{*}\beta\hat{w}^{*}\hat{u}^{\dagger}\right)\\
 & +\frac{1}{Re}\left(-\frac{1}{3}i\alpha^{*}D\hat{u}^{*}\hat{v}^{\dagger}-\alpha^{*2}\hat{v}^{*}\hat{v}^{\dagger}+\frac{4}{3}D^{2}\hat{v}^{*}\hat{v}^{\dagger}-\beta^{2}\hat{v}^{*}\hat{v}^{\dagger}-\frac{1}{3}i\beta D\hat{w}^{*}\hat{v}^{\dagger}\right)\\
 & +\frac{1}{Re}\left(-\frac{1}{3}\alpha^{*}\beta\hat{u}^{*}\hat{w}^{\dagger}-\frac{1}{3}i\beta D\hat{v}^{*}\hat{w}^{\dagger}-\alpha^{*2}\hat{w}^{*}\hat{w}^{\dagger}+D^{2}\hat{w}^{*}\hat{w}^{\dagger}-\frac{4}{3}\beta^{2}\hat{w}^{*}\hat{w}^{\dagger}\right)\\
 & +\frac{2}{Re}\left(\frac{\partial u}{\partial y}D\hat{u}^{*}\hat{T}^{\dagger}-i\alpha^{*}\frac{\partial u}{\partial y}\hat{v}^{*}\hat{T}^{\dagger}-i\beta\frac{\partial w}{\partial y}\hat{v}^{*}\hat{T}^{\dagger}+\frac{\partial w}{\partial y}D\hat{w}^{*}\hat{T}^{\dagger}\right)
\end{split}
\end{align}

\begin{align}\label{S_mu_D}
\begin{split}
S_{\frac{\partial\mu}{\partial\rho}}= & \frac{1}{Re}\left(\underset{1}{\underbrace{\frac{\partial u}{\partial y}D\hat{\rho}^{*}\hat{u}^{\dagger}}}+\underset{2}{\underbrace{\frac{\partial^{2}u}{\partial y^{2}}\hat{\rho}^{*}\hat{u}^{\dagger}}}+\underset{3}{\underbrace{\frac{\partial\rho}{\partial y}D\hat{u}^{*}\hat{u}^{\dagger}}}+\underset{4}{\underbrace{-i\alpha^{*}\frac{\partial\rho}{\partial y}\hat{v}^{*}\hat{u}^{\dagger}}}+\underset{5}{\underbrace{-i\alpha^{*}\frac{\partial u}{\partial y}\hat{\rho}^{*}\hat{v}^{\dagger}}}\right)\\
 & +\frac{1}{Re}\left(\underset{6}{\underbrace{-i\beta\frac{\partial w}{\partial y}\hat{\rho}^{*}\hat{v}^{\dagger}}}+\underset{7}{\underbrace{\frac{2}{3}i\alpha^{*}\frac{\partial\rho}{\partial y}\hat{u}^{*}\hat{v}^{\dagger}}}+\underset{8}{\underbrace{\frac{4}{3}\frac{\partial\rho}{\partial y}D\hat{v}^{*}\hat{v}^{\dagger}}}+\underset{9}{\underbrace{\frac{2}{3}i\beta\frac{\partial\rho}{\partial y}\hat{w}^{*}\hat{v}^{\dagger}}}+\underset{10}{\underbrace{\frac{\partial w}{\partial y}D\hat{\rho}^{*}\hat{w}^{\dagger}}}\right)\\
 & +\frac{1}{Re}\left(\underset{11}{\underbrace{\frac{\partial^{2}w}{\partial y^{2}}\hat{\rho}^{*}\hat{w}^{\dagger}}}+\underset{12}{\underbrace{-i\beta\frac{\partial\rho}{\partial y}\hat{v}^{*}\hat{w}^{\dagger}}}+\underset{13}{\underbrace{\frac{\partial\rho}{\partial y}D\hat{w}^{*}\hat{w}^{\dagger}}}+\underset{14}{\underbrace{\left(\frac{\partial u}{\partial y}\right)^{2}\hat{\rho}^{*}\hat{T}^{\dagger}}}+\underset{15}{\underbrace{\left(\frac{\partial w}{\partial y}\right)^{2}\hat{\rho}^{*}\hat{T}^{\dagger}}}\right)
\end{split}
\end{align}

\begin{align}
\begin{split}
S_{\frac{\partial\mu}{\partial T}}= & \frac{1}{Re}\left(\frac{\partial T}{\partial y}D\hat{u}^{*}\hat{u}^{\dagger}-i\alpha^{*}\frac{\partial T}{\partial y}\hat{v}^{*}\hat{u}^{\dagger}+\frac{\partial u}{\partial y}D\hat{T}^{*}\hat{u}^{\dagger}+\frac{\partial^{2}u}{\partial y^{2}}\hat{T}^{*}\hat{u}^{\dagger}\right)\\
 & +\frac{1}{Re}\left(\frac{2}{3}i\alpha^{*}\frac{\partial T}{\partial y}\hat{u}^{*}\hat{v}^{\dagger}+\frac{4}{3}\frac{\partial T}{\partial y}D\hat{v}^{*}\hat{v}^{\dagger}+\frac{2}{3}i\beta\frac{\partial T}{\partial y}\hat{w}^{*}\hat{v}^{\dagger}\right)\\
 & +\frac{1}{Re}\left(-i\alpha^{*}\frac{\partial u}{\partial y}\hat{T}^{*}\hat{v}^{\dagger}-i\beta\frac{\partial w}{\partial y}\hat{T}^{*}\hat{v}^{\dagger}-i\beta\frac{\partial T}{\partial y}\hat{v}^{*}\hat{w}^{\dagger}+\frac{\partial T}{\partial y}D\hat{w}^{*}\hat{w}^{\dagger}\right)\\
 & +\frac{1}{Re}\left(+\frac{\partial w}{\partial y}D\hat{T}^{*}\hat{w}^{\dagger}+\frac{\partial^{2}w}{\partial y^{2}}\hat{T}^{*}\hat{w}^{\dagger}+\left(\frac{\partial u}{\partial y}\right)^{2}\hat{T}^{*}\hat{T}^{\dagger}+\left(\frac{\partial w}{\partial y}\right)^{2}\hat{T}^{*}\hat{T}^{\dagger}\right)
\end{split}
\end{align}

\begin{align}
\begin{split}
S_{\frac{\partial^{2}\mu}{\partial\rho^{2}}}=\frac{1}{Re}\left(\frac{\partial u}{\partial y}\frac{\partial\rho}{\partial y}\hat{\rho}^{*}\hat{u}^{\dagger}+\frac{\partial w}{\partial y}\frac{\partial\rho}{\partial y}\hat{\rho}^{*}\hat{w}^{\dagger}\right)
\end{split}
\end{align}

\begin{align}
\begin{split}
S_{\frac{\partial^{2}\mu}{\partial T^{2}}}=\frac{1}{Re}\left(\frac{\partial u}{\partial y}\frac{\partial T}{\partial y}\hat{T}^{*}\hat{u}^{\dagger}+\frac{\partial w}{\partial y}\frac{\partial T}{\partial y}\hat{T}^{*}\hat{w}^{\dagger}\right)
\end{split}
\end{align}

\begin{align}
\begin{split}
S_{\frac{\partial^{2}\mu}{\partial\rho\partial T}}=\frac{1}{Re}\left(\frac{\partial u}{\partial y}\frac{\partial T}{\partial y}\hat{\rho}^{*}\hat{u}^{\dagger}+\frac{\partial u}{\partial y}\frac{\partial\rho}{\partial y}\hat{T}^{*}\hat{u}^{\dagger}+\frac{\partial w}{\partial y}\frac{\partial T}{\partial y}\hat{\rho}^{*}\hat{w}^{\dagger}+\frac{\partial w}{\partial y}\frac{\partial\rho}{\partial y}\hat{T}^{*}\hat{w}^{\dagger}\right)
\end{split}
\end{align}

\begin{equation}\label{S_ka}
\left. \begin{array}{l}
\displaystyle
S_{\kappa}=\frac{1}{RePrEc}\left(-\alpha^{*2}+D^{2}-\beta^{2}\right)\hat{T}^{*}\hat{T}^{\dagger} \\[10pt]
\displaystyle
S_{\frac{\partial\kappa}{\partial\rho}}=\frac{1}{RePrEc}\left(\frac{\partial T}{\partial y}D\hat{\rho}^{*}\hat{T}^{\dagger}+\frac{\partial^{2}T}{\partial y^{2}}\hat{\rho}^{*}\hat{T}^{\dagger}+\frac{\partial\rho}{\partial y}D\hat{T}^{*}\hat{T}^{\dagger}\right) \\[10pt]
\displaystyle
S_{\frac{\partial\kappa}{\partial T}}=\frac{1}{RePrEc}\left(2\frac{\partial T}{\partial y}D\hat{T}^{*}\hat{T}^{\dagger}+\frac{\partial^{2}T}{\partial y^{2}}\hat{T}^{*}\hat{T}^{\dagger}\right)\\[10pt]
\displaystyle
S_{\frac{\partial^{2}\kappa}{\partial\rho^{2}}}=\frac{1}{RePrEc}\frac{\partial\rho}{\partial y}\frac{\partial T}{\partial y}\hat{\rho}^{*}\hat{T}^{\dagger}\\[10pt]
\displaystyle
S_{\frac{\partial^{2}\kappa}{\partial T^{2}}}=\frac{1}{RePrEc}\left(\frac{\partial T}{\partial y}\right)^{2}\hat{T}^{*}\hat{T}^{\dagger}\\[10pt]
\displaystyle
S_{\frac{\partial^{2}\kappa}{\partial\rho\partial T}}=\frac{1}{RePrEc}\left(\left(\frac{\partial T}{\partial y}\right)^{2}\hat{\rho}^{*}\hat{T}^{\dagger}+\frac{\partial T}{\partial y}\frac{\partial\rho}{\partial y}\hat{T}^{*}\hat{T}^{\dagger}\right)
\end{array} \right\}
\end{equation}

\section{Influence of the wavenumber on sensitivity}\label{appBeta}
\begin{figure}
\centering
\includegraphics[width=1.0\linewidth]{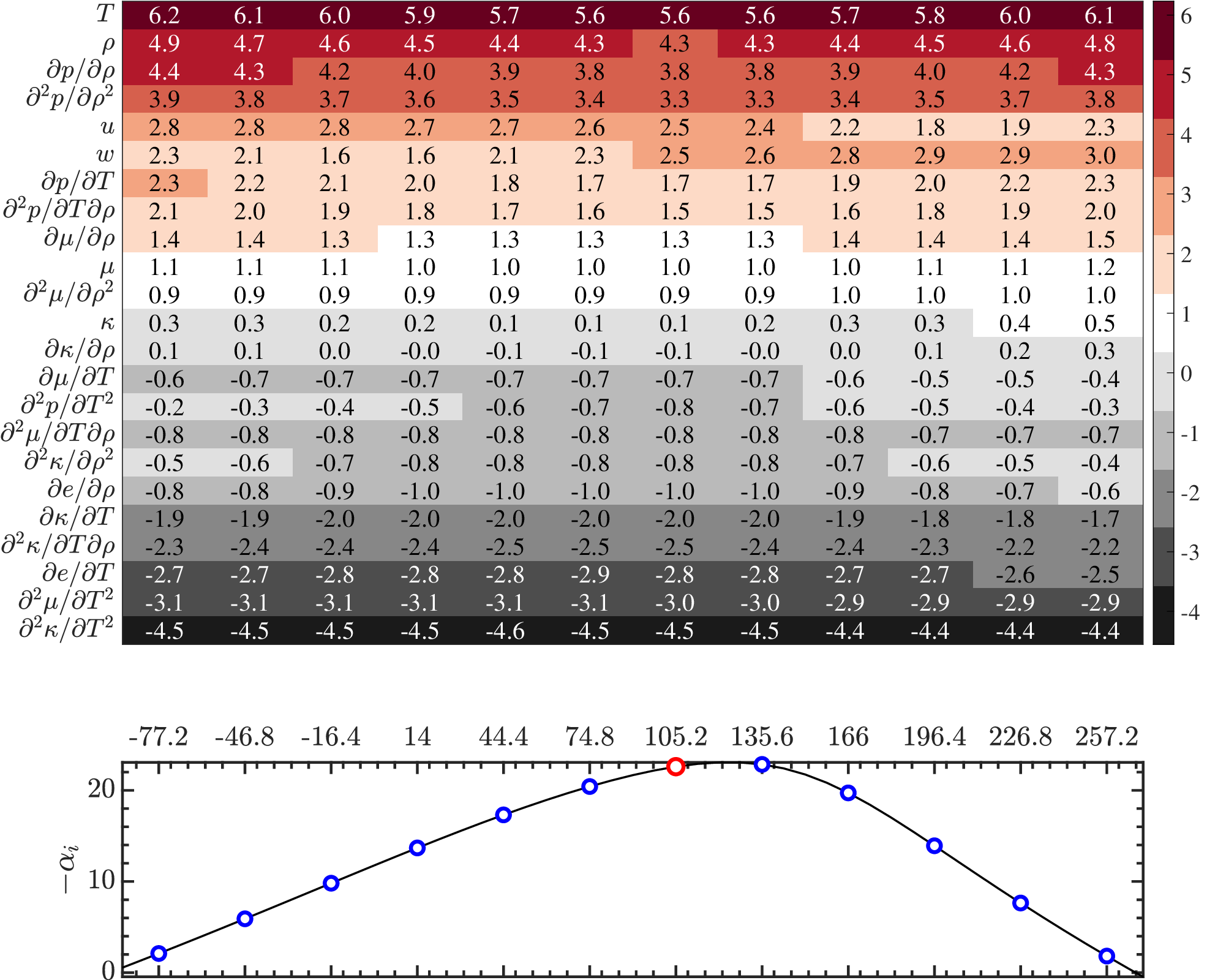}
\put(-183, 91){$\beta$}
\put(-366,318){({\it a})}
\put(-366,78){({\it b})}
\caption{(a) Sensitivity measure $\log(M)$ as a function of the wavenumber $\beta$. (b) Growth rate versus $\beta$. The fluid is in the pseudo-boiling regime with $x=1$, $\omega=40$ (inviscid TS mode).}
\label{figa}
\end{figure}

We provide an example of the influence of the spanwise wavenumber on the sensitivity. Figure \ref{figa} shows that the sensitivity measure is only mildly influenced by $\beta$, with the rank of input terms largely unaffected. The fluid is in the pseudo-boiling regime. 
\FloatBarrier

\bibliography{main}
\bibliographystyle{jfm}

\end{document}